\documentstyle[emulateapj]{article}

\newcommand{\etal}{et~al.\ }
\newcommand{\PVdblt}{{\rm P}\kern 0.1em{\sc v}~$\lambda\lambda 1117, 1128$}
\newcommand{\CaIIdblt}{{\rm Ca}\kern 0.1em{\sc ii}~$\lambda\lambda 3934, 3969$}
\newcommand{\CIVdblt}{{\rm C}\kern 0.1em{\sc iv}~$\lambda\lambda 1548, 1550$}
\newcommand{\MgIIdblt}{{\rm Mg}\kern 0.1em{\sc ii}~$\lambda\lambda 2796, 2803$}
\newcommand{\NVdblt}{{\rm N}\kern 0.1em{\sc v}~$\lambda\lambda 1238, 1242$}
\newcommand{\SVIdblt}{{\rm S}\kern 0.1em{\sc vi}~$\lambda\lambda 933, 944$}
\newcommand{\OVIdblt}{{\rm O}\kern 0.1em{\sc vi}~$\lambda\lambda 1031, 1037$}
\newcommand{\SiIIdblt}{{\rm Si}\kern 0.1em{\sc ii}~$\lambda\lambda 1190, 1193$}
\newcommand{\SiIVdblt}{{\rm Si}\kern 0.1em{\sc iv}~$\lambda\lambda 1393, 1402$}
\newcommand{\PV}{\hbox{{\rm P}\kern 0.1em{\sc v}}}
\newcommand{\AlI}{\hbox{{\rm Al}\kern 0.1em{\sc i}}}
\newcommand{\AlII}{\hbox{{\rm Al}\kern 0.1em{\sc ii}}}
\newcommand{\AlIII}{{\hbox{\rm Al}\kern 0.1em{\sc iii}}}
\newcommand{\CaII}{\hbox{{\rm Ca}\kern 0.1em{\sc ii}}}
\newcommand{\CII}{\hbox{{\rm C}\kern 0.1em{\sc ii}}}
\newcommand{\CIIe}{\hbox{{\rm C$^{\ast}$}\kern 0.1em{\sc ii}}}
\newcommand{\CIII}{\hbox{{\rm C}\kern 0.1em{\sc iii}}}
\newcommand{\CIV}{\hbox{{\rm C}\kern 0.1em{\sc iv}}}
\newcommand{\CV}{\hbox{{\rm C}\kern 0.1em{\sc v}}}
\newcommand{\HI}{\hbox{{\rm H}\kern 0.1em{\sc i}}}
\newcommand{\HII}{\hbox{{\rm H}\kern 0.1em{\sc ii}}}
\newcommand{\Lya}{\hbox{{\rm Ly}\kern 0.1em$\alpha$}}
\newcommand{\Lyb}{\hbox{{\rm Ly}\kern 0.1em$\beta$}}
\newcommand{\Lyg}{\hbox{{\rm Ly}\kern 0.1em$\gamma$}}
\newcommand{\Lyd}{\hbox{{\rm Ly}\kern 0.1em$\delta$}}
\newcommand{\Lye}{\hbox{{\rm Ly}\kern 0.1em$\epsilon$}}
\newcommand{\Lyphi}{\hbox{{\rm Ly}\kern 0.1em$\phi$}}
\newcommand{\Lyfive}{\hbox{{\rm Ly}\kern 0.1em$5$}}
\newcommand{\Lysix}{\hbox{{\rm Ly}\kern 0.1em$6$}}
\newcommand{\Lyseven}{\hbox{{\rm Ly}\kern 0.1em$7$}}
\newcommand{\Lyeight}{\hbox{{\rm Ly}\kern 0.1em$8$}}
\newcommand{\Lynine}{\hbox{{\rm Ly}\kern 0.1em$9$}}
\newcommand{\Lyten}{\hbox{{\rm Ly}\kern 0.1em$10$}}
\newcommand{\Lyeleven}{\hbox{{\rm Ly}\kern 0.1em$11$}}
\newcommand{\HeI}{\hbox{{\rm He}\kern 0.1em{\sc i}}}
\newcommand{\HeII}{\hbox{{\rm He}\kern 0.1em{\sc ii}}}
\newcommand{\FeI}{\hbox{{\rm Fe}\kern 0.1em{\sc i}}}
\newcommand{\FeII}{\hbox{{\rm Fe}\kern 0.1em{\sc ii}}}
\newcommand{\FeIII}{\hbox{{\rm Fe}\kern 0.1em{\sc iii}}}
\newcommand{\MnII}{\hbox{{\rm Mn}\kern 0.1em{\sc ii}}}
\newcommand{\MgI}{\hbox{{\rm Mg}\kern 0.1em{\sc i}}}
\newcommand{\MgII}{\hbox{{\rm Mg}\kern 0.1em{\sc ii}}}
\newcommand{\MgIII}{\hbox{{\rm Mg}\kern 0.1em{\sc iii}}}
\newcommand{\NI}{\hbox{{\rm N}\kern 0.1em{\sc i}}}
\newcommand{\NII}{\hbox{{\rm N}\kern 0.1em{\sc ii}}}
\newcommand{\NIII}{\hbox{{\rm N}\kern 0.1em{\sc iii}}}
\newcommand{\NV}{\hbox{{\rm N}\kern 0.1em{\sc v}}}
\newcommand{\OVI}{\hbox{{\rm O}\kern 0.1em{\sc vi}}}
\newcommand{\OI}{\hbox{{\rm O}\kern 0.1em{\sc i}}}
\newcommand{\OII}{\hbox{[{\rm O}\kern 0.1em{\sc ii}]}}
\newcommand{\OIV}{\hbox{{\rm O}\kern 0.1em{\sc iv}]}}
\newcommand{\SI}{{\rm S}\kern 0.1em{\sc i}}
\newcommand{\SIV}{{\rm S}\kern 0.1em{\sc iv}}
\newcommand{\SVI}{{\rm S}\kern 0.1em{\sc vi}}
\newcommand{\SiI}{\hbox{{\rm Si}\kern 0.1em{\sc i}}}
\newcommand{\SiII}{\hbox{{\rm Si}\kern 0.1em{\sc ii}}}
\newcommand{\SiIII}{\hbox{{\rm Si}\kern 0.1em{\sc iii}}}
\newcommand{\SiIV}{\hbox{{\rm Si}\kern 0.1em{\sc iv}}}
\newcommand{\SII}{\hbox{{\rm S}\kern 0.1em{\sc ii}}}
\newcommand{\SIII}{\hbox{{\rm S}\kern 0.1em{\sc iii}}}
\newcommand{\NaI}{\hbox{{\rm Na}\kern 0.1em{\sc i}}}
\newcommand{\TiII}{\hbox{{\rm Ti}\kern 0.1em{\sc ii}}}
\newcommand{\kms}{\hbox{km~s$^{-1}$}}
\newcommand{\cmsq}{\hbox{cm$^{-2}$}}

\begin{document}

\lefthead{CHURCHILL \& VOGT}
\righthead{{\MgII} KINEMATICS}
\slugcomment{The Astonomical Journal (2001) 122, in press  (August)}

\title{The Kinematics of Intermediate Redshift
M\lowercase{g}~{\sc ii} Absorbers\altaffilmark{1}}

\author{Christopher W. Churchill\altaffilmark{2}}
\affil{Department of Astronomy and Astrophysics \\
       The Pennsylvania State University \\
       University Park, PA 16802 \\
       {\it cwc@astro.psu.edu}}
\and

\author{Steven S. Vogt}
\affil{UCO/Lick Observatories \\
       Board of Studies in Astronomy and Astrophysics \\
       University of California, Santa Cruz, CA 96054 \\
       {\it vogt@ucolick.org}}

\altaffiltext{1}{Based  in  part   on  observations  obtained  at  the
W.~M. Keck Observatory, which  is operated as a scientific partnership
among Caltech, the University of California, and NASA. The Observatory
was made possible by the  generous financial support of the W.~M. Keck
Foundation.}
\altaffiltext{2}{Work   partially    conducted   while   at   UCO/Lick
Observatories and the Board  of Studies in Astronomy and Astrophysics,
University of California, Santa Cruz, CA 96054}

\begin{abstract}

We  present  23  quasar   absorption  line  systems  selected  by  the
{\MgIIdblt} doublet with $W_{r}(2796)  > 0.3$~{\AA} over the redshift
range  $0.4   \leq  z  \leq   1.2$.   The  kinematics   and  ``profile
morphologies''  are  studied  at  $\simeq 6$~{\kms}  resolution  to  a
$5~\sigma$  equivalent  width detection  threshold  of $W_{r}(2796)  =
0.015$~{\AA}.  We are thus sensitive to very weak ``clouds'' isolated
in velocity.  The absorption profiles were segregated into ``kinematic
subsystems'', and the  properties (velocities, velocity widths, column
densities, etc.)   were measured directly  from the data and  by Voigt
profile modeling.

Most  absorbers are  characterized by  a dominant  kinematic subsystem
with $W_{r}(2796) > 0.2$~{\AA}  and velocity spreads ranging from $10$
to  $50$~{\kms}   in  proportion  to  the   system  equivalent  width.
Additional kinematic  subsystems have velocities  separations as large
as $400$~{\kms}  relative to  the dominant subsystem.   The equivalent
widths   and  velocity   spreads  of   these  weaker   subsystems  are
anti--correlated  with  their velocities  and  their equivalent  width
distribution  turns down from  a power  law below  $W_{r}(2796) \simeq
0.08$~{\AA}.  These ``moderate''  and ``high velocity'' subsystems are
inferred to  have sub--Lyman limit  {\HI}, and therefore  are probably
not higher redshift analogues to Galactic high velocity clouds (HVCs).
 
Weak subsystems  are asymmetrically distributed in  velocity such that
they are either all blueshifted  or all redshifted with respect to the
dominant  subsystem.   This  implies,  that although  the  ``kinematic
morphologies'' vary greatly on a case--by--case basis, a given line of
sight  is apparently  probing  a well  defined  spatial and  kinematic
structure.  We investigate  a simple kinematic model that  relies on a
rotating  disk  to  explain   the  observed  asymmetries.   There  are
systematic differences, or trends,  in both the subsystem to subsystem
velocity  clustering and  in the  overall kinematic  morphologies with
increasing equivalent width; we discuss how these may provide clues to
the   observed  differential   evolution  in   the   equivalent  width
distribution of {\MgII} absorbers.

\end{abstract}


\section{Introduction}
\label{sec:intro}

Ultimately, the  goal of studying  absorption lines in the  spectra of
quasars is to chart the details  of cosmic evolution and to develop an
understanding of  the role  of gas in  the formation and  evolution of
galaxies.   Strong   metal--line  absorption  systems   are  excellent
candidates  for pursuing  this  line of  research  because metals  are
produced in  stars, and stars are produced  in galaxies.  Furthermore,
an $\alpha$--group element is best suited  since it is a tracer of the
earliest stages  of stellar evolution (i.e.\ Type  II supernovae), and
is  not significantly  produced  in secondary  or  later stages.   The
resonant {\MgIIdblt}  doublet is ideally suited for  these reasons and
for the reason that it is a low ionization species that traces neutral
hydrogen column densities commonly associated with galaxy environments
[i.e.\ Lyman limit systems; $\log N({\HI}) \geq 17.3$~{\cmsq}].

Ground--based, low resolution  ($\sim 1$--$2$~{\AA}) surveys searching
for {\MgII} absorption systems in  the spectra of quasars have yielded
insights into the evolution  of their number density, absorption cross
section, and cosmological clustering over the redshift range $0.2 \leq
z  \leq  2.2$   (e.g.\  \cite{ltw87};  \cite{ssb88};  \cite{boisse92};
\cite{ss92}).  In  particular, Steidel \&  Sargent (1992\nocite{ss92},
hereafter   SS92)  have   presented  a   comprehensive  look   at  the
distributions of  the equivalent widths, doublet  ratios, and velocity
clustering  (on scales  greater than  500~{\kms}).   {\MgII} absorbers
having  rest--frame  equivalent  widths, $W_{r}(2796)$,  greater  than
$0.3$~{\AA}  are  consistent   with  a  population  of  cosmologically
distributed  objects, i.e.\  galaxies.  The  co--moving  redshift path
density\footnote{The redshift path density is defined as the number of
objects per  unit redshift.  It  constrains the product of  the number
density  and  physical  cross  section  projected  on  the  sky.}   is
consistent with  a non--evolving  population of objects.   However, as
the sample is progressively  restricted to stronger systems, evolution
becomes progressively more pronounced  with a high significance level;
the strongest  systems are evolving  away with cosmic time.   In other
words, the mean $W_{r}(2796)$ increases with redshift.

In principle,  this evolution  could be attributed  to changes  in the
mean quantity of gas, or in the chemical, ionization, and/or kinematic
conditions in  galaxies.  It is unlikely that  the chemical enrichment
of the gas is decreasing  with cosmic time, since the mean metallicity
in    the   universe    continues    to   increase.     Bergeron~\etal
(1994\nocite{bergeron94}) established  that the mean  ionization level
in  metal--line absorbers decreases  with cosmic  time over  this same
redshift  interval.  Bergeron \etal  also reported  finding sub--Lyman
limit {\MgII}  absorbers at  the lower end  of the  redshift interval,
indicating that the {\HI} gas  mass may be decreasing with cosmic time
in these systems.  In tandem with the ionization evolution, this would
indicate  that the  {\it total},  not just  the neutral,  gas  mass is
decreasing with cosmic time, and  this could be linked to the observed
{\MgII} evolution.

On  the other  hand, perhaps  the velocity  spreads of  the individual
systems   are   decreasing   with   time.   Petitjean   and   Bergeron
(1990\nocite{pb90}) studied $\sim 30$ {\MgII} absorbing systems with a
moderate  resolution  of  30~{\kms}.  They  resolved  multi--component
substructure in complex absorption profiles, measured the cloud--cloud
velocity clustering on the  scale of galaxy halo velocity dispersions,
and  established a correlation  between the  equivalent width  and the
number of  subcomponents\footnote{This relationship had  been proposed
by York \etal (1986) and Wolfe (1986).}.  Since the doublet ratio does
not evolve with  redshift (SS92), it has been  inferred that {\it each
of  the  individual  subcomponents\/}  within  an  overall  system  is
saturated,  or  nearly   saturated.   Indeed,  Petitjean  \&  Bergeron
(1990\nocite{pb90})  suggested  that the  typical  subcomponent has  a
strength  of $W_{r}(2796)  \simeq 0.1$~{\AA},  which is  saturated for
small Doppler parameters.  The  total equivalent width, then, would be
proportional to  the number and  velocity spread of  these individual,
relatively narrow, saturated subcomponents.  As another example of
what could be governing the evolution-- if the velocity spread is
decreasing with cosmic time, the evolution in $W_{r}(2796)$ may not be
due to a decrease in the  gas mass, but due to the kinematics becoming
more systematic or settled with time.

Now that the general statistics and evolution of {\MgII} absorbers are
established  and that  several important,  but  limited, observational
clues  are available for  inferring what  physically is  evolving, the
next  logical  step is  to  undertake  a  comprehensive study  of  the
kinematics.  The first  goal is to simply establish  the nature of the
gas  kinematics  and  the  second  is  to  investigate  what  physical
processes are  governing the  redshift evolution in  the $W_{r}(2796)$
distribution.

High--resolution  spectra are  required for  resolving  the absorption
profiles into  {\it individual\/} components for  determining both the
line--of--sight velocity  distributions (kinematics) and  the relative
component--to--component  absorption  strengths.   In  addition,  high
signal--to--noise ratios are required for measuring weaker transitions
with similar ionization potentials  (esp.\ {\MgI} and {\FeII}).  These
transitions can  provide clues to  the ionization conditions,  and, in
the cases  where the {\MgII}  profiles are saturated, provide  the gas
kinematics.   With the  advent  of 8--10  meter  class telescopes  and
efficient, very high--resolution spectrographs, it has become possible
to acquire the requisite data in sufficient quantities.

In  this  paper, we  present  the  high  resolution ($\sim  6$~{\kms})
absorption profiles of 23  {\MgII} selected absorbing systems observed
with      the     HIRES     spectrograph      (\cite{vogt94}).      In
\S~\ref{sec:obs-anal},  we present  the observed  quasars  and briefly
address the reduction and analysis  of the data (i.e.\ how the varied,
complex  absorption profiles  are  parameterized into  one to  several
quantified ``properties'').  The  statistical properties of the sample
and    the    individual    absorbing    systems    are    given    in
\S~\ref{sec:systems}.  The  observed distributions of  the kinematics,
absorption   strengths,  and   column  densities   are   presented  in
\S~\ref{sec:results},  followed by  discussions  in \S~\ref{sec:open},
\S~\ref{sec:model}, and  \S~\ref{sec:evolving}.  Conclusions are given
in \S~\ref{sec:summary}.

\section{Observations, Data Reduction and Analysis}
\label{sec:obs-anal}

A  total of  25  quasars  were observed  with  the HIRES  spectrometer
(\cite{vogt94}) on the  Keck~I telescope over the nights  of 4--5 July
1994,  23--25  January 1995,  and  19--20  July  1996 UT.   HIRES  was
configured in  first--order using the $0.861{\arcsec}\times7{\arcsec}$
decker, C1.  Because of the first--order HIRES format, there are small
gaps in  the spectral  coverage redward of  5100~{\AA} where  the free
spectral range exceeds  the width of the $2048  \times 2048$ Tektronix
CCD.  The  $0.861{\arcsec}$ slit width  disperses the light with  $R =
\lambda / \Delta \lambda  = 45,000$ ($\simeq 6.6$~{\kms}) projected to
$\simeq 3$ pixels per resolution element, $\Delta \lambda$.

Three separate, consecutive integrations were obtained for each quasar
spectrum, bracketed  by Th--Ar calibration  lamps.  In a  given night,
almost  all quasars  were observed  with identical  HIRES  echelle and
cross disperser settings, so the instrument was rarely reconfigured in
a given  night.  At the time  of the observations, there  was no image
rotator for HIRES, so the slit  was not aligned with nor held constant
with respect to the parallactic angle throughout the observations.

July runs  were adversely affected  by high, patchy cirrus  clouds, so
these  spectra  typically have  lower  signal--to--noise ratios.   The
January   1995   spectra   were   obtained   in   clear   and   stable
$\sim0.5${\arcsec}  seeing  conditions.   In  the continuum  near  the
observed {\MgII} profiles,  the rest--frame equivalent width detection
limit (5~$\sigma$) ranges from 0.008--0.016~{\AA}.

\begingroup 
\small
\begin{deluxetable}{lcclrc}
\tablewidth{0pc}
\tablecaption{Journal of Observations}
\tablehead
{
\colhead{Quasar} &
\colhead{V [mag]} & 
\colhead{$z_{\rm em}$} &  
\colhead{Date [UT]} & 
\colhead{Exp [s]} & 
\colhead{$\lambda $ Range [\AA]} 
} 
\startdata
$0002+051$ & 16.2 & 1.90 & 1994 Jul 5 &     2700 & 3655.7--6079.0 \nl
$0117+213$ & 16.1 & 1.49 & 1995 Jan 23 &    5400 & 4317.7--6775.1  \nl
$0420-014$ & 17.0 & 0.92 & 1995 Jan 23 &    3600 & 3810.5--6304.9  \nl
$0454+039$ & 16.5 & 1.34 & 1995 Jan 22 &    4500 & 3765.8--6198.9  \nl
$0454-220$ & 15.5 & 0.53 & 1995 Jan 22 &    5400 & 3765.8--6198.9  \nl
$0823-223$ & 15.7 & $>0.92$ & 1995 Jan 24 & 3600 & 3977.8--6411.8  \nl
$1101-264$\tablenotemark{a} & 16.1 & 2.14 & \nodata     & 100,200 & 3791.8--3847.8 \nl
$1148+384$ & 17.0 & 1.30 & 1995 Jan 24 &    5400 & 3986.5--6424.5  \nl
$1206+459$ & 16.1 & 1.16 & 1995 Jan 23 &    3600 & 3810.5--6304.9  \nl
$1222+228$ & 15.5 & 2.04 & 1995 Jan 23 &    3600 & 3810.5--6304.9  \nl
$1241+176$ & 15.4 & 1.28 & 1995 Jan 22 &    2400 & 3765.8--6189.9  \nl
$1248+401$ & 16.3 & 1.03 & 1995 Jan 22 &    4200 & 3765.8--6189.9  \nl
$1254+044$ & 16.0 & 1.02 & 1995 Jan 22 &    2400 & 3765.8--6189.9  \nl
$1317+274$ & 16.0 & 1.01 & 1995 Jan 23 &    3600 & 3810.5--6304.9  \nl
$1354+193$ & 16.0 & 0.72 & 1995 Jan 22 &    3600 & 3765.8--6189.9  \nl
$1421+331$ & 16.7 & 1.91 & 1995 Jan 23 &    3600 & 3818.6--6316.9  \nl
$1634+706$ & 14.9 & 1.34 & 1994 Jul 4,5 &   2700 & 3723.3--6185.7  \nl
$2128-123$ & 15.9 & 0.50 & 1996 Jul 19 &    3900 & 3766.2--5791.3  \nl
$2145+064$ & 16.3 & 1.00 & 1996 Jul 18 &    4500 & 3766.2--5791.3  \nl
\enddata
\tablenotetext{a}{Observed with the IPCS on the AAT by M.~Pettini and
R.~Hunstead.}
\label{tab:obsjournal}
\end{deluxetable}

\endgroup

\subsection{Sample Selection} 
\label{sec:sampsele}

The absorbing systems were  selected from the spectroscopic surveys of
SS92 and Sargent,  Steidel, \& Boksenberg (1988\nocite{ssb88}).  These
surveys  were complete  to a  5$~\sigma$ rest--frame  equivalent width
threshold of  0.3~{\AA}.  The spectra  presented here are  complete to
0.02~{\AA}  within  $\pm 500$~{\kms}  of  the  {\MgII} $\lambda  2796$
optical depth centroids.

The quasars  were selected based upon  the criteria that  each (1) was
observable during  the alloted telescope  time, (2) had $V  \leq 17.5$
(in order to  obtain a high signal--to--noise ratio  in roughly a 3600
second integration), and (3) preferably had multiple systems along the
line of  sight that  could be  observed with a  single setting  of the
HIRES echelle.

The journal  of observations is  listed in Table~\ref{tab:obsjournal}.
Tabulated  are  the $V$  magnitudes,  the  emission  redshifts of  the
quasars, the  UT date  of observation, the  total integration  time in
seconds,  and  the wavelength  coverage  (not  including breaks  above
5100~{\AA}).   Included  is the  spectrum  of  Q$1101-264$, which  was
observed with the IPCS on the AAT and generously donated to this study
by M.~Pettini.

\subsection{Data Reduction}
\label{sec:datared}

The HIRES data were reduced with the IRAF\footnote{IRAF is distributed
by the National Optical Astronomy Observatories, which are operated by
AURA,  Inc., under  contract to  the NSF.}  {\sc  Apextract\/} package
(V2.10.3) for echelle data.  Further details of the data reduction can
be   found    in   Churchill   (1995\nocite{lotr}),    and   Churchill
(1997\nocite{thesis}).

Each  observed   data  frame  was  overscan   subtracted,  bias  frame
corrected, scattered light corrected,  and flatfielded in the standard
fashion.  Cosmic  rays were removed  by averaging the frames  with the
IRAF  task {\sc Imcombine}  using the  {\it crreject\/}  option.  This
technique  was  possible  only  because  the  HIRES  spectrograph  was
extremely stable between exposures; there was no quantifiable shifting
of  the  cross  disperser  or  echelle  settings.   The  spectra  were
extracted using  the optimal routines  of Horne (1986\nocite{horne86})
and  Marsh  (1989\nocite{marsh89}) as  implemented  by  the IRAF  {\sc
Apextract} package.   Also extracted were the  uncertainty spectra and
the sky spectra.

Due  to the  instrument stability,  the Th--Ar  lamps  bracketing each
series  of quasar  observations  were simply  added  together for  the
wavelength  calibration.  The  wavelengths were  calibrated  to vacuum
using  the  ``vacthar.dat'' line  list  provided  with  the IRAF  {\sc
Ecidentify}   task.   For  the   dispersion  correction,   using  {\sc
Dispcorr}, the spectra  were {\it not\/} linearized, in  that the data
were not interpolated to enforce equal wavelength intervals per pixel.
This  preserves  uncorrelated counts  in  each  pixel  and leaves  the
profile   shapes  unaltered.    Finally,  the   heliocentric  velocity
correction  was manually  applied  to each  echelle  order, where  the
heliocentric velocity was calculated using the IRAF {\sc Rvcorr} task.

The  continuum flux  was  normalized using  the  methods described  by
Sembach  \& Savage  (1992\nocite{semsav92};  also see  \cite{thesis}),
where  the fitting  was performed  interactively using  the  IRAF {\sc
Sfit} task embedded in an iterative script.  

\subsection{Data Analysis}
\label{sec:dataanal}

\begingroup
\small

\begin{deluxetable}{rccc}
\tablewidth{0pc}
\tablecolumns{4}
\tablecaption{Transitions Covered}
\tablehead
{
\colhead{Tran} &
\colhead{$\lambda_{\rm vac}$} &
\colhead{$f$} &
\colhead{$\Gamma \times 10^8$} \\
 &
\colhead{[{\AA}]} &
 &
 }
\startdata
{\hbox{{\rm Fe}\kern 0.1em{\sc ii~}}2344}   & 2344.214 & 0.1097 & 2.6800 \nl
{\hbox{{\rm Fe}\kern 0.1em{\sc ii~}}2374}   & 2374.4612 & 0.0282 & 2.9900 \nl
{\hbox{{\rm Fe}\kern 0.1em{\sc ii~}}2383}   & 2382.765 & 0.3006 & 3.1000 \nl
{\hbox{{\rm Fe}\kern 0.1em{\sc ii~}}2587}   & 2586.650 & 0.0646 & 2.7200 \nl
{\hbox{{\rm Fe}\kern 0.1em{\sc ii~}}2600}   & 2600.1729 & 0.2239 & 2.7000 \nl
{\hbox{{\rm Mg}\kern 0.1em{\sc ii~}}2796}   & 2796.352 & 0.6123 & 2.6120 \nl
{\hbox{{\rm Mg}\kern 0.1em{\sc ii~}}2803}   & 2803.531 & 0.3054 & 2.5920 \nl
{\hbox{{\rm Mg}\kern 0.1em{\sc i~}}2853}    & 2852.964 & 1.8100 & 4.9500 \nl
\enddata
\label{tab:ions}
\end{deluxetable}

\endgroup

\subsubsection{Feature Finding}
\label{sec:featfind}

For each  quasar spectrum,  an objective line  list of  all 5~$\sigma$
absorption features was constructed using a method very similar to the
one  described in  Schneider  \etal (1993\nocite{schneider93}).   This
method is optimal for  finding unresolved lines (see \cite{archiveI}).
The HIRES instrumental spread function (ISF) is well approximated as a
Gaussian        with        $\sigma_{\rm        ISF}(\lambda)        =
0.0453(\lambda/5500)$~{\AA}.  In  terms of the  formalism of Schneider
\etal,  we set  $J_{0} =  6$.   The HIRES  spectra have  3 pixels  per
resolution  element  so that  this  value  samples  the ISF  over  4.3
resolution elements (13 pixels).

\subsubsection{Feature Identification}
\label{sec:lineid}

Line and  line blending identifications were  accomplished as follows.
A  list   of  roughly   100  transitions  with   accurate  rest--frame
wavelengths  (\cite{morton91})  was   compiled.   The  literature  was
exhaustively searched for known  absorption systems in each quasar and
a list of their redshifts  was constructed (even if the strong {\MgII}
and {\CIV} doublets were not  covered in our spectra).  This list also
included  the redshifts  of  the discovered  ``weak'' {\MgII}  systems
(\cite{weakI}).  The  point of including all known  redshifts from the
literature was to document line blending with transitions from other
redshifts with those in our sample.

We  then  scanned  the  spectra  at the  expected  location  for  each
transition  in  our list.   Output  was  a  list identifying  probable
blends, i.e.\ instances when the expected wavelength locations covered
by two or more different transitions from different redshifted systems
overlapped.   The  final   adopted  identifications  and  limits  were
obtained by examining the spectra by hand.  Very few blends were found
and when  they were, the details  are given in the  discussions of the
individual systems.

In  Table~\ref{tab:ions} we  present the  transitions included  in our
study.  From left to right, the columns are the ionization species and
transition identity, the vacuum laboratory wavelengths, the oscillator
strengths, and the natural  broadening constants.  The atomic data are
taken from  the compilation of  Morton (1991\nocite{morton91}), except
for the  {\FeII}~$\lambda 2587$ and $\lambda  2374$ transitions, which
are taken from Cardelli \& Savage (1995\nocite{carsav95}). Tripp \etal
(1996\nocite{tripp96}) review the reasons  why these $f$ values likely
represent an improvement over the Morton values.

For  this work,  we  include  only the  {\MgII},  {\MgI}, and  {\FeII}
transitions  that  have been  redshifted  into  our observed  spectral
range.   In some  systems, {\MnII},  {\CaII}, and  {\TiII} transitions
were  detected  (see \cite{archiveI}).   Since  this  work is  focused
primarily on the kinematics of {\MgII} selected absorption systems, we
present only  the transition  listed in Table~\ref{tab:ions}.   In the
velocity  ranges where  {\MgII}  profiles are  saturated, i.e.\  where
kinematic information  is lost, the {\FeII} and  {\MgI} kinematics can
be used assuming  that they arise in the same  ionization phase as the
{\MgII}.  Note  that there is now  evidence for both a  ``warm'' and a
``cold''  {\HI} component in  damped Lyman--$\alpha$  absorbers (DLAs)
(\cite{lane}); some of the {\MgI}  may arise in this separate ``cold''
phase  (\cite{rao-privcomm}).   It is  possible  that such  multiphase
structure is  also present in Lyman--limit {\MgII}  absorbers and that
{\MgI} is partially arising in the ``cold'' phase.

\subsubsection{Defining Kinematic Subsystems} 
\label{sec:subsystems}

Many  of the  overall  absorption  profiles are  comprised  of two  to
several  absorption features  that are  separated by  more  than three
pixels (i.e.\  a resolution element)  of continuum flux.   We quantify
the properties for each of these separate absorption regions, which we
call ``kinematic subsystems''.  Kinematic subsystems are defined using
only  the {\MgII}  $\lambda 2796$  transition.  The  wavelength region
defining  each  kinematic   subsystem  is  determined  objectively  by
searching to either  side of subsystem centroid for  the wavelength at
which  the   per  pixel  equivalent   width  (see  \cite{schneider93};
\cite{archiveI})   becomes  insignificant,  i.e.\   $W(\lambda_{j})  <
\sigma_{W}(\lambda_{j})$  where $j$  denotes  the pixel  index (for  a
graphical example, see Figure~3 of \cite{weakI}).

\begin{figure*}[ht]
\plotfiddle{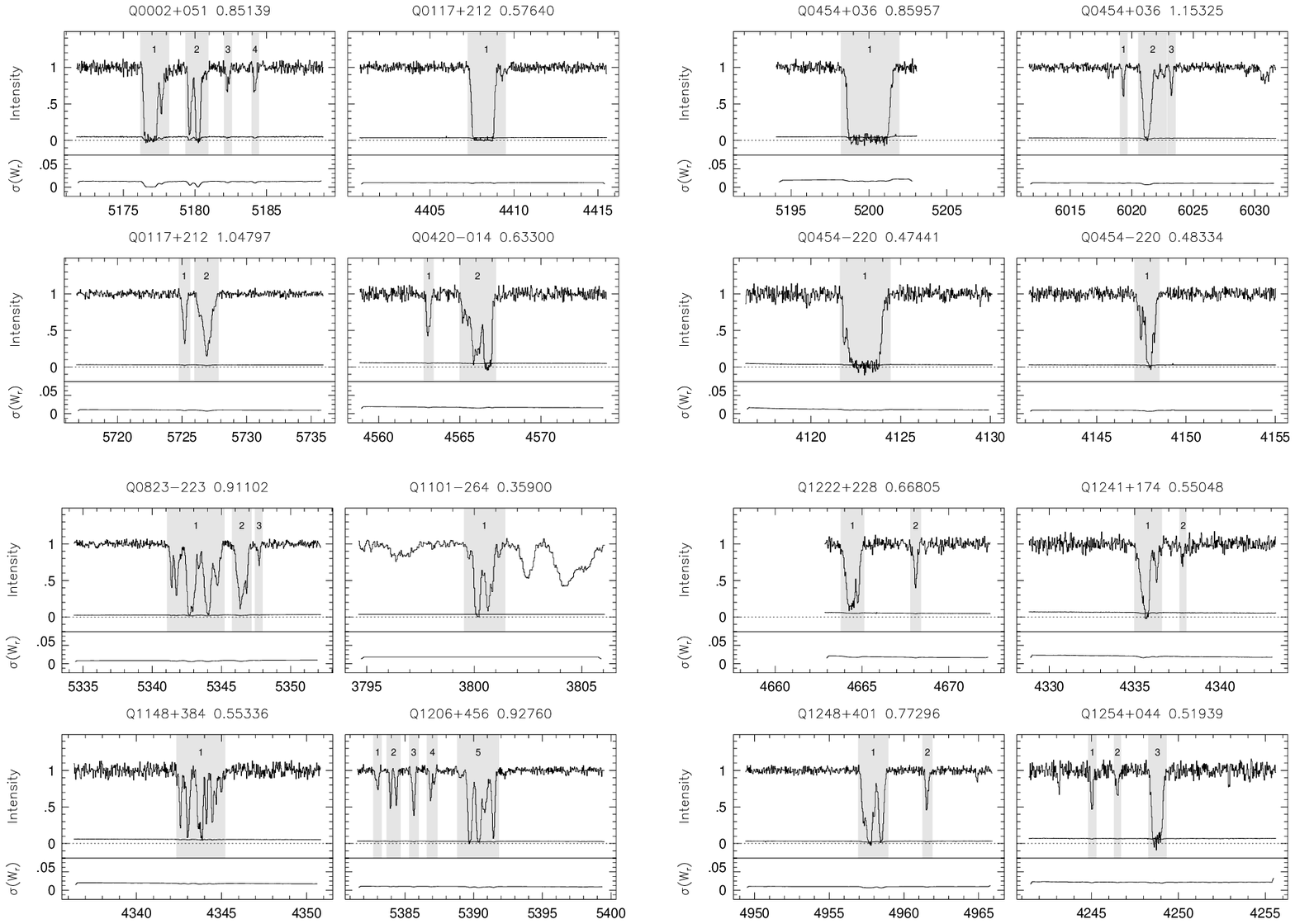}{5.6in}{0.}{75}{75}{-230}{101}
\figurenum{1}
\plotfiddle{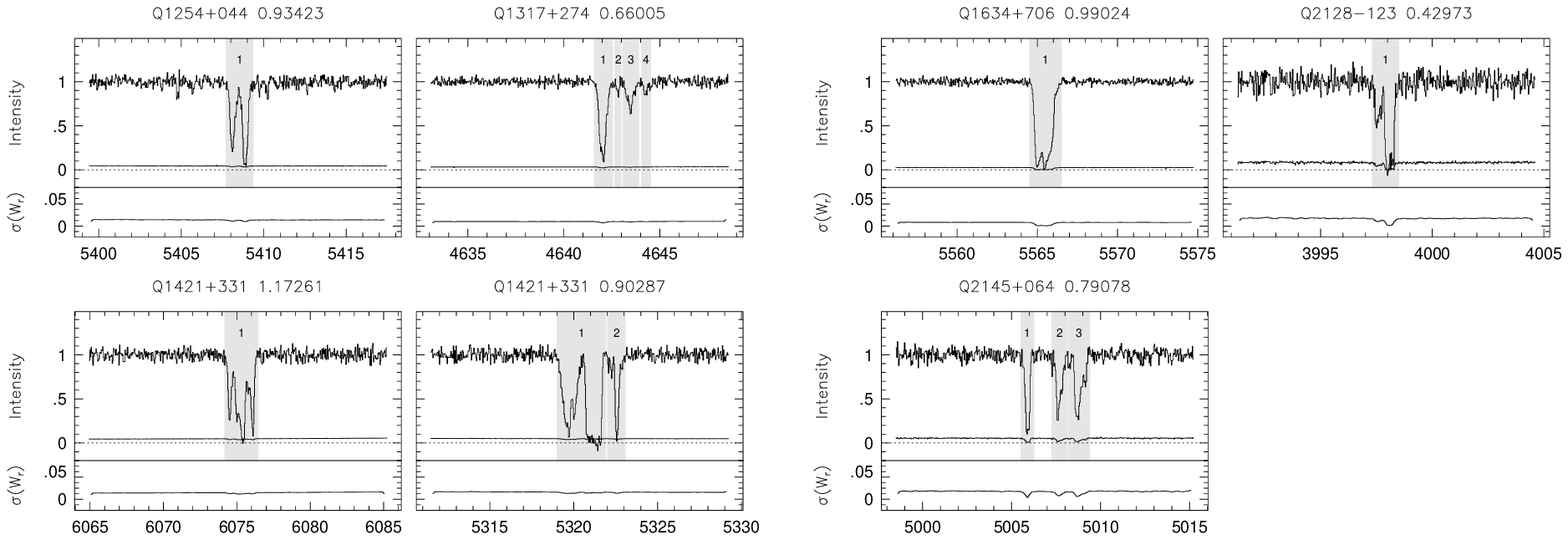}{0.in}{0.}{75}{75}{-230}{-25}
\caption[Fig1a.eps,fig1b.eps]  {The   normalized  {\MgII}  $\lambda
2796$  HIRES/Keck   profiles  and  the   5~$\sigma$  equivalent  width
detection thresholds  presented as a function  of observed wavelength.
Each  spectrum, however  is plotted  over  a 1000~{\kms}  span in  its
rest--frame.   The shaded  areas show  where the  kinematic subsystems
have  been defined.   Throughout  this work,  they  are identified  by
number from blue to red.
\label{fig:ewlims}}
\end{figure*}

The {\MgII}  $\lambda 2796$  profiles for each  of the systems  in our
sample are presented  in Figure~\ref{fig:ewlims}.  The upper subpanels
show  a rest--frame  $1000$~{\kms}  range about  the {\MgII}  $\lambda
2796$  transition  at its  observed  (redshifted) wavelength.   Shaded
regions indicate the wavelengths  regions of each kinematic subsystem,
which are  numbered from  blue to red.   The lower subpanels  show the
5~$\sigma$   rest--frame   equivalent   width  threshold,   given   by
$5~\sigma^{2}_{W(\lambda_{i})}$, as a function of wavelength.

\subsection{Calculating Absorption Properties}
\label{sec:calc}

We compute most absorption properties directly from the flux values of
the data.  For the overall  systemss, we compute the systemic redshift
from  the optical  depth  mean  of the  profile.   For each  kinematic
subsystem,  we  compute the  equivalent  widths, velocities,  velocity
widths,  and  apparent column  densities.   We  also parameterize  the
profiles using  Voigt profile decomposition  to measure the  number of
``clouds'',   their   column   densities,  velocities,   and   Doppler
parameters.     The    mathematical     formalism    is    given    in
Appendix~\ref{app:A}.

Analysis  of the VP  results will  be presented  in a  companion paper
(Churchill,  Vogt, \&  Charlton 2000\nocite{cvc00}).   Here  we simply
describe the  formalism and present  the VP decomposition  results for
two reasons:  (1) the model fits  need to be presented  with the data,
and (2) inferences  based upon VP results, whether  statistical or for
individual   systems,   absolutely   require   extensive   simulations
(e.g. \cite{hu95}; \cite{lu_lya}) in order to establish the systematic
effects  of the  fitting process.   We will  calibrate our  VP results
using simulations in the companion paper.
 
\section{The Systems}
\label{sec:systems}

The  HIRES   data  of  the  detected  transitions   are  presented  in
Figure~\ref{fig:data}  (at the end  of the  paper).  The  profiles are
aligned   in  rest--frame   velocity,  where   $v=0$   corresponds  to
$\lambda_{sys}$  computed  from  Equation~\ref{eq:zabs}.  Systems  are
defined to  be all  absorption that lies  with $\pm 500${\kms}  of the
systemic velocity zero  point.  In no case is there  a system, weak or
strong,  that lies anywhere  near this  velocity window;  in practice,
this velocity window could be increased to an arbitrarily large size.

\begin{figure*}[th]
\figurenum{3}
\plotfiddle{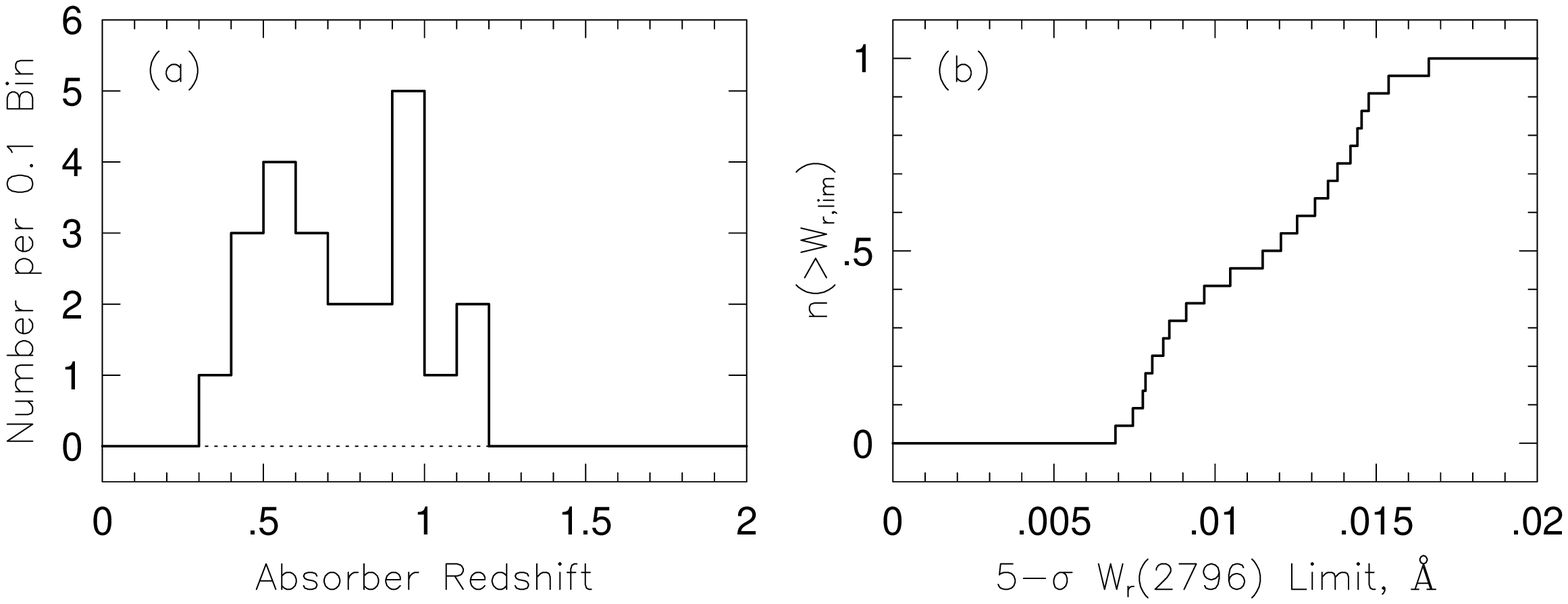}{2.3in}{0.}{70}{70}{-280}{-210}
\caption[fig3.eps] {  (a) The  binned redshift distribution  of our
sample.   ---  (b) The  cumulative  distribution  of the  5~$\sigma$,
rest--frame equivalent width  detection threshold within a 1000~{\kms}
range about {\MgII} $\lambda 2796$. \label{fig:z_ewlim} }
\end{figure*}

\begin{figure*}[bht]
\figurenum{4}
\plotfiddle{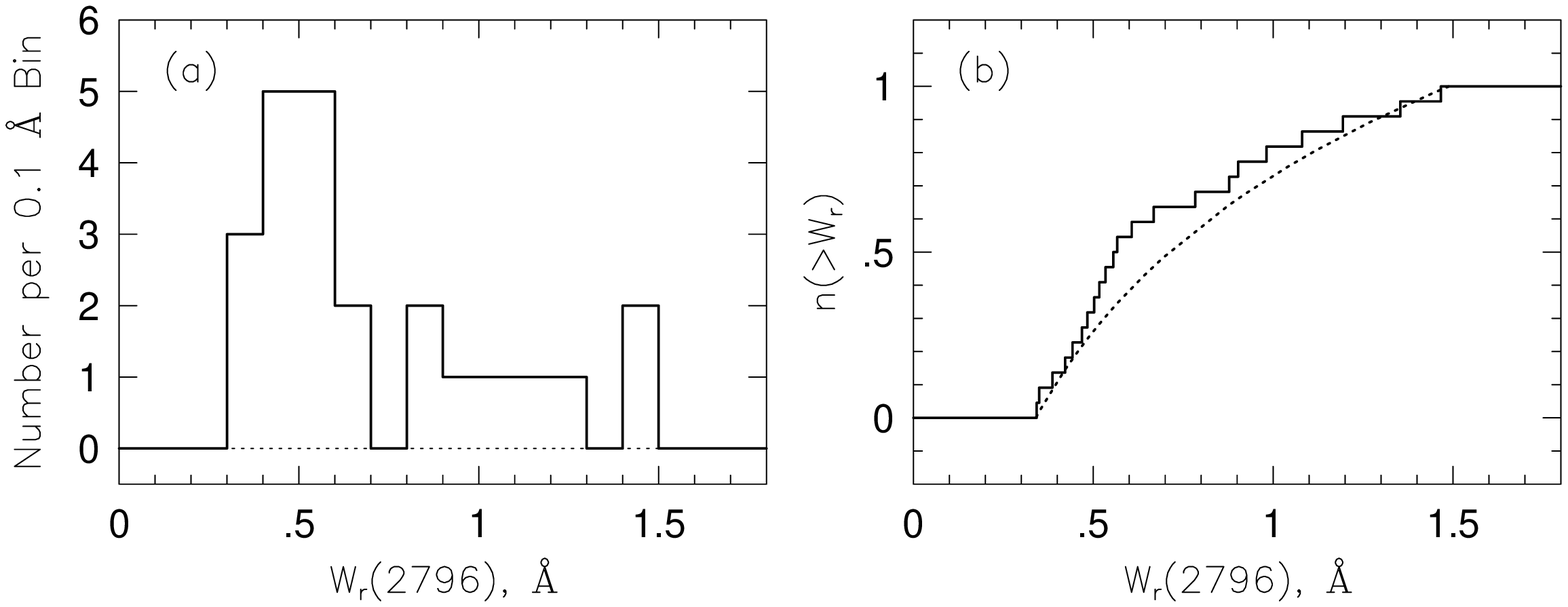}{2.3in}{0.}{70}{70}{-280}{-210}
\caption[fig4.eps] {  (a) The binned,  rest--frame equivalent width
distribution for  {\MgII} $\lambda 2796$  of our sample.  ---  (b) The
cumulative  distribution of  our  sample (histogram)  compared to  the
distribution  for an unbiased  survey with  the same  equivalent width
range as observed (dot--dot curve). \label{fig:ewdist} }
\end{figure*}

Ticks  above  the normalized  spectra  are  the  centroids of  the  VP
components.  The VP model spectra (the fits) are shown as solid curves
superimposed upon the data.  Because the component velocities are tied
together for all ions, ticks appear over spectral regions even where a
VP column density is an upper limit.

In Table~\ref{tab:absprops}, we list the absorption redshifts, $z_{\rm
abs}$, the kinematic spreads, $\omega_{v}$, the rest--frame equivalent
widths, $W_{r}(2796)$,  and the {\MgII} doublet ratios.   We also list
these  quantities  and  the  velocity,  $\left< v  \right>$,  for  all
kinematic   subsystem,   which   are   numbered  as   illustrated   in
Figure~\ref{fig:ewlims}.

\subsection{Sample Properties}

In  Figure~\ref{fig:z_ewlim}$a$,   we  present  the   binned  redshift
distribution of our  sample.  The redshift range is  $0.36 \leq z \leq
1.17$.   The mean  is $\left<  z  \right> =  0.77$, with  43\% of  the
systems between $0.4  \leq z\leq 0.7$ and 39\%  of the systems between
$0.7 \leq  z\leq 1.0$.  In  panel $b$ of  Figure~\ref{fig:z_ewlim}, we
show  the  cumulative  distribution  of the  5~$\sigma$,  rest--frame,
{\MgII}  $\lambda  2796$  equivalent  width detection  limit  for  the
sample.   More precisely,  this  illustrates the  distribution of  the
detection sensitivity to very  weak {\MgII} absorption averaged over a
1000~{\kms} region  ($\pm 500$~{\kms}) about each  absorber.  The data
are 100\% complete to  $0.017$~{\AA} and 80\% to $0.014$~{\AA}.  There
is no sensitivity below  $0.007$~{\AA}.  That is, unresolved kinematic
subsystems  with  $\log   N({\MgII})  \simeq  11.6$~{\cmsq}  could  be
detected  in  {\it  all\/}   absorbers  in  the  sample;  below  $\log
N({\MgII}) \simeq 11.2$~{\cmsq} there is no sensitivity.

The  rest--frame  equivalent  width  distribution for  the  sample  is
presented  in  Figure~\ref{fig:ewdist}.  In  panel  $a$,  we show  the
binned  distribution   and  in  panel  $b$  we   show  the  cumulative
distribution.  The dot--dot curve  is the cumulative distribution of a
``fair'' sample (i.e.\ unbiased) over the same $W_{r}(2796)$ range for
a     power--law    distribution    function,     $n(W_{r})    \propto
W_{r}^{-\delta}$.  Over  the range  $0.02 \leq W_{r}(2796)  \leq 1.3$,
the  power--law  exponent,  $\delta$,  was  shown  to  be  $1.0\pm0.1$
(\cite{weakI}).   The steep rise  to $W_{r}(2796)  \simeq 0.6$  in the
observed  distribution,  as  compared  to the  ``fair''  distribution,
indicates that the the sample  is slightly ``bottom heavy'', having an
overabundance  of  smaller   equivalent  width  systems.   However,  a
Kolmogorov--Smirnov (KS) test (\cite{recipes}) applied to the observed
and ``fair''  cumulative distribution functions yields  a KS statistic
of $0.21$ with significance $0.25$.  That the observed distribution is
fair cannot  be ruled out by  the KS statistic;  the sample apparently
does not contained any significant bias with equivalent width.

\begin{figure*}[th]
\figurenum{5}
\plotfiddle{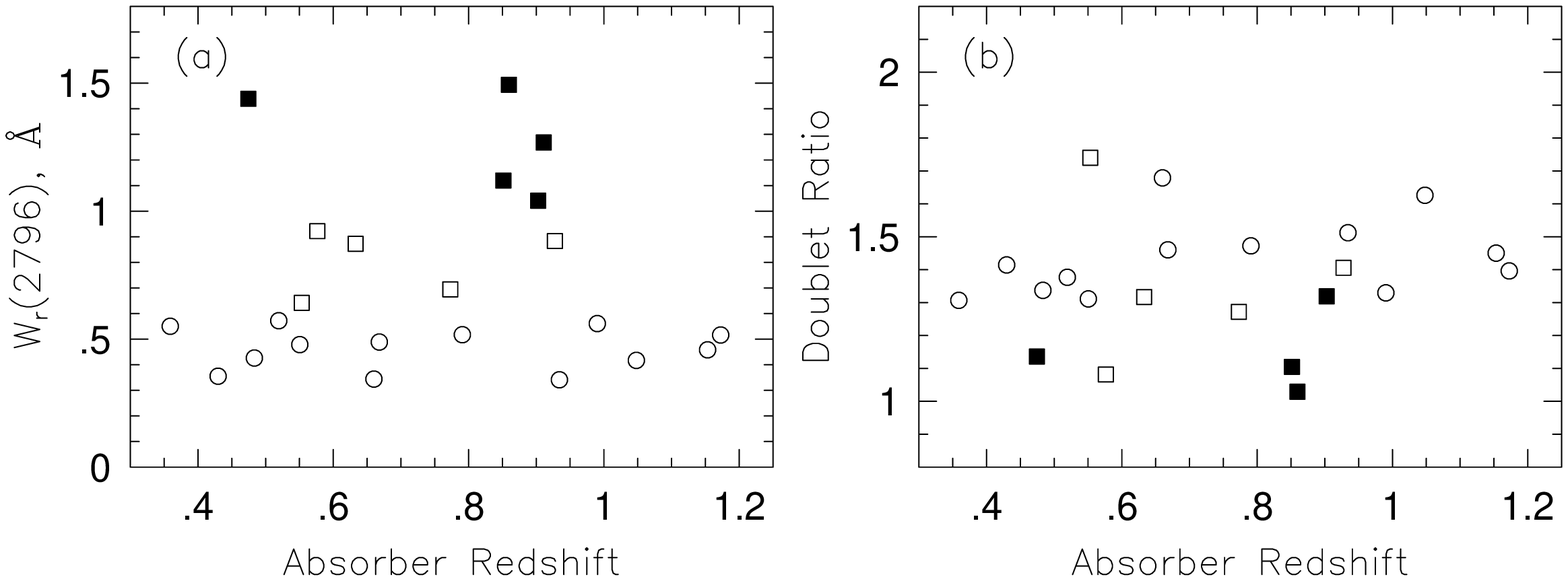}{2.3in}{0.}{70}{70}{-280}{-210}
\caption[fig5.eps]  { (a) The  rest--frame, {\MgII}  $\lambda 2796$
equivalent  width  vs.\  absorber  redshift.  Open  circles  represent
sample B ($0.3 \leq W_{r}(2796) < 0.6$~{\AA}) and open boxes represent
sample C ($0.6 \leq W_{r}(2796) < 1.0$~{\AA}).  sample E ($W_{r}(2796)
\geq  1.0$~{\AA})  is represented  with  filled  boxes.   --- (b)  The
{\MgII}  doublet ratio  vs.\ absorber  redshift.  Data  points  are as
defined in  panel $a$.  Measurement  errors are given, on  average, by
the   size   of    the   data   points   for   the    data   in   both
panels. \label{fig:ew_dr_z} }
\end{figure*}

\begin{figure*}[htb]
\figurenum{6}
\plotfiddle{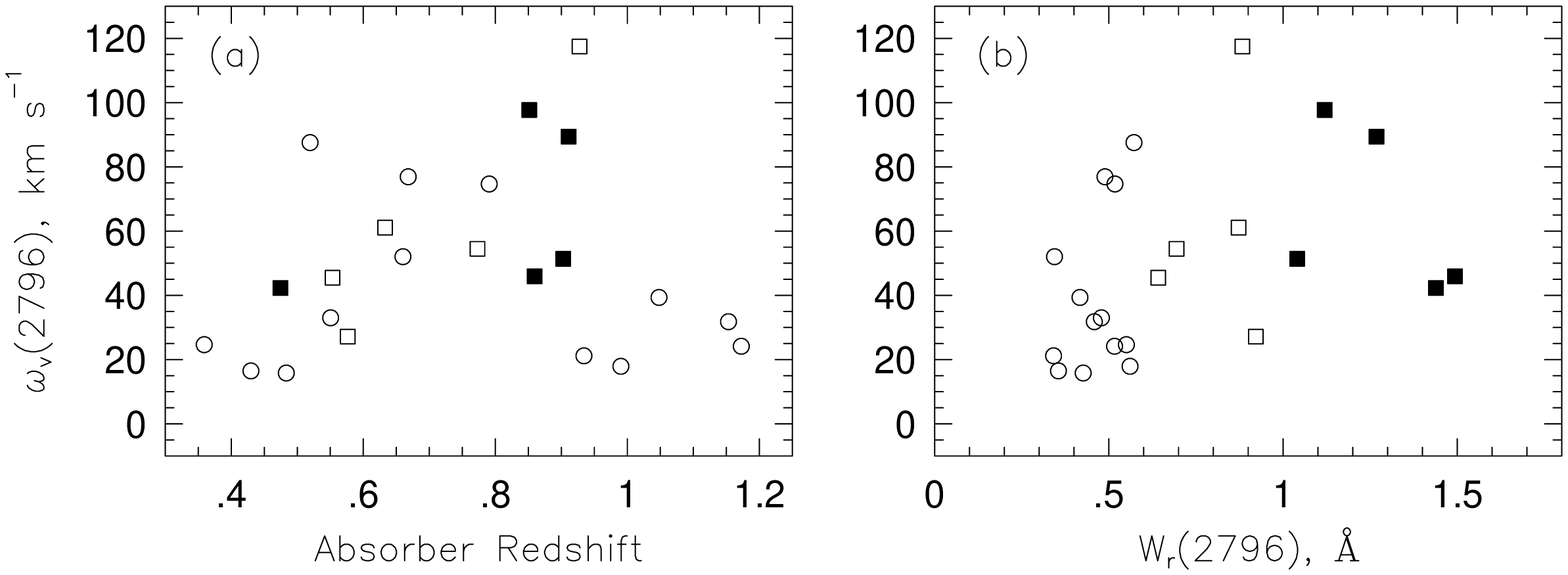}{2.3in}{0.}{70}{70}{-280}{-210}
\caption[fig6.eps] {  (a) The kinematic velocity  spread of {\MgII}
$\lambda 2796$ vs.\ absorber redshift.  Data points are as defined for
Figure~\ref{fig:ew_dr_z}.   --- (b) The  kinematic velocity  spread of
{\MgII}  $\lambda  2796$  vs.\  the rest--frame  equivalent  width  of
{\MgII} $\lambda 2796$. \label{fig:omega} }
\end{figure*}

Since we have a reasonably  fair equivalent width distribution, we can
contrast and compare inferences drawn from our moderate size sample to
those drawn  from the overall  population, including behavior  that is
differential  with $W_{r}(2796)$.  To  this end,  we have  divided our
full  sample  into  four  subsamples  based  upon  {\MgII}  absorption
strength.  In Table~\ref{tab:samples},  we present the mean properties
of  these samples  and the  sample members.   Sample A  is  the entire
sample, which has $W_{r}(2796)  \geq 0.3$~{\AA}; sample B systems have
$0.3 \leq  W_{r}(2796) < 0.6$~{\AA};  sample C systems have  $0.6 \leq
W_{r}(2796)  < 1.0$~{\AA};  sample  D systems  have $W_{r}(2796)  \geq
0.6$~{\AA};  and sample  E,  which is  a  subsample of  sample D,  has
$W_{r}(2796)  \geq  1.0$~{\AA}.   These  equivalent width  ranges  are
chosen based upon historical precedent.   SS92 have shown that each of
these  samples exhibits  a different  level of  cosmological evolution
(see \S~\ref{sec:intro}).

\subsection{Global Absorption Properties}

In Figures~\ref{fig:ew_dr_z}$a$--$b$, we present $W_{r}(2796)$ and the
{\MgII} doublet  ratio, DR, vs.\ absorber redshift.   Sample B systems
are plotted as open circles, sample C as open squares, and sample E as
filled  squares.  Sample  E is  highly concentrated  at $z  \sim 0.9$,
whereas sample  B is fairly  uniform over the covered  redshift range.
The mean DR  is 1.36 and shows no trend with  redshift.  The system DR
are  really the  flux weighted  averages of  the  individual kinematic
subsystems,  which  exhibit the  full  range  of  doublet ratios  (not
plotted); some  weak subsystems are unresolved  and somewhat saturated
while some are optically  thin.  As listed in Table~\ref{tab:samples},
there is a trend for DR to decrease from sample B to sample E.

In  panel $a$ of  Figure~\ref{fig:omega}, we  present the  full system
kinematic spread,  $\omega _{v}$, vs.\ absorption  redshift, where the
data point types denote samples B,  C, and E.  Note that the kinematic
spread of a system,  as defined in Equation~\ref{eq:omegavdef}, is not
the full  velocity spread.  For  example, the full velocity  spread of
the  $z=0.9276$  system  toward   Q~$1206+459$  is  $\Delta  v  \simeq
450$~{\kms},  whereas  its kinematic  spread  is  $\omega _{v}  \simeq
120$~{\kms}.  The kinematic spread  provides an optical depth weighted
velocity  spread; it  essentially  is  equivalent to  the  width of  a
Gaussian  fitted to  the complex,  full system  optical  depth profile
(i.e.\ it is column density weighted).

In  panel  $b$ of  Figure~\ref{fig:omega},  we  show  the full  system
kinematic spread vs.\  $W_{r}(2796)$.  Though a Spearman--Kendell test
reveals a significant correlation, what  is most telling is that there
is a large  scatter in the kinematic spread for  each sample, B, C, 
and  E;  it is  the  mean  of this  scatter  that  is increasing  with
equivalent width.

In  Table~\ref{tab:ewalltab}, we  present  the rest--frame  equivalent
widths  for  the  transitions  listed  in  Table~\ref{tab:ions}.   The
velocity  range  of each  kinematic  subsystems  is  given, where  the
subsystem  number  is  as  defined  in  Figure~\ref{fig:ewlims}.   The
velocities  are the  limits of  integration  for Equation~\ref{eq:ew}.
The system  totals are also listed.   No entry for  a given transition
means that it was not covered.

In Table~\ref{tab:aodcols}, we present the integrated column densities
using the apparent optical depth method (Equation~\ref{eq:na}).  Lower
limits are provided when all transitions of a given ion were saturated
within  the  subsystem.  For  {\MgI}  and  {\FeII},  upper limits  are
provided when the ion was not  detected in any of its transitions.  No
entry means that no transitions were covered for the given ion.

In  Table~\ref{tab:vptab},  we present  the  VP  parameters from  {\sc
Minfit}.  Following  each component's velocity, we  give the kinematic
subsystem number; this  allows the complexity of the  subsystems to be
examined. In the case of no detections, the column density is an upper
limit.  If there is no coverage of any transition for a given ion, the
entry is left blank.

\subsection{Individual Systems}

\subsubsection{\rm Q~$0002+051$, UM 18~~~~$z_{\rm abs}=0.851394$}

This system has a  kinematically intriguing absorption profile, with a
total  velocity spread  of roughly  475~{\kms}; it  was fitted  with a
total of twelve VP components.   Classified as a ``double'' systems by
Churchill \etal (2000b\nocite{archiveII}), the overall {\MgII} profile
was found to have four kinematic  subsystems.  This is one of only two
systems  in our  sample in  which two  subsystems (\#1  and  \#2) have
$W_{r}(2796)   >  0.3$~{\AA}.    Furthermore,  these   subsystems  are
separated by 180~{\kms}.  Subsystem \#1 is fully saturated, exhibiting
the kinematic  spread and morphology  characteristic of a  DLA {\MgII}
profile  [though it  is not  a DLA  (\cite{archiveII})].  But  for its
proximity  in  velocity  to  subsystem  \#1, subsystem  \#2  could  be
considered a system in its  own right.  Its large equivalent width and
kinematic  morphology   are  similar  to  those   of  Q~$1101-264$  at
$z=0.3590$,   Q~$1248+401$   at   $z=0.7730$,  and   Q~$1254+044$   at
$z=0.9342$.

\subsubsection{\rm PG~$0117+213$~~~~$z_{\rm abs}=0.576398$} 

This DLA (\cite{archiveII})  exhibits fully saturated {\MgII} profiles
and a {\MgI} profile with  definite variations in strength in velocity
space.  There is a slight blending in the blue wing of {\MgI} with the
red  wing of  {\FeII}  $\lambda 2600$  at  $z_{\rm abs}=0.7291$.   The
profiles  were fitted  with four  VP  components, though  this fit  is
constrained only  by the {\MgI} profile.  No  {\FeII} transitions were
captured by the CCD.

\subsubsection{\rm PG~$0117+213$~~~~$z_{\rm abs}=1.047971$} 

This system has two  kinematic subfeatures separated by roughly $v\sim
100$~{\kms}.  Both exhibit blueward  asymmetry and both have very weak
{\MgI}.   The overall  profile was  fitted with  seven  VP components.
This is a {\CIV}--deficient system (\cite{archiveII}).

\subsubsection{Q~$0420-014$~~~~$z_{\rm abs}=0.633004$}

The  overall  profile  is   comprised  of  two  kinematic  subfeatures
separated by  $v \sim 200$~{\kms}. One  of them is  broad and complex.
The red--most  saturated components in  subsystem \#2 have  {\MgI} and
{\FeII},  whereas the  bluer, more  complex components  do  not.  This
suggests  either  strong  variations   in  abundance  patterns  or  in
ionization/density  conditions.  A  total of  nine VP  components were
fitted.

\subsubsection{\rm PKS~$0454+039$~~~~$z_{\rm abs}=0.859565$} 

This DLA was studied by  Steidel \etal (1995\nocite{sbbd95}) and by Lu
\etal    (1996b\nocite{lu_dla96})   and   Churchill    \etal   (2000a,
b\nocite{archiveI}\nocite{archiveII}).  The profile  is a single fully
saturated feature  with a $\sim 200$~{\kms} spread.   Though eleven VP
components were  fitted, the column  densities and Doppler  $b$ values
for  components three  through  eight are  not  well constrained.   In
Table~\ref{tab:vptab},  formal  errors  are  not  provided  for  these
components, only  the approximate $N$  and $b$ values;  any inferences
regarding this system based upon the VP decomposition should be viewed
with  caution.  Based  upon the  profile fit  to the  {\FeII} $\lambda
2344$ transition,  there may be  an offset in the  continuum placement
across that profile.  There  is a spurious, unidentified, weak feature
at $v=+100$~{\kms} relative to {\MgII} $\lambda 2803$.

\subsubsection{\rm PKS~$0454+039$~~~~$z_{\rm abs}=1.153248$}

This ``classic'' system (\cite{archiveII}) has the only profile in the
sample  that  has  weak,  high--velocity absorption  distributed  {\it
symmetrically\/} about  the central main  absorption region.  However,
there is confusion near both  the {\MgII} doublet transitions, in that
weak, unidentified absorption  features are seen.  Additionally, there
appears  to be a  weak, unidentified  feature precisely  where {\FeII}
$\lambda 2374$  transition is predicted.  This is  apparent because of
the  unphysical strength  ratios  with respect  to  the other  {\FeII}
transitions  and  is  further  suggested by  another  possible  nearby
unidentified feature.  Very weak {\MgI} is detected in the core of the
strongest {\MgII} component, but, again, there is a weak, unidentified
feature within $-50$~{\kms}.

\subsubsection{\rm PKS~$0454-220$~~~~$z_{\rm abs}=0.474410$} 

Though  not a  DLA,  based  upon the  definition  $\log N({\HI})  \geq
20.3$~{\cmsq}, this system  has been classified as ``DLA/{\HI}--Rich''
by  Churchill \etal  (2000b\nocite{archiveII}).  A  total of  eight VP
components were fitted to this broad saturated profile.  Though formal
uncertainties are quoted in  Table~\ref{tab:vptab}, it should be noted
that the  column density uncertainties in most  central components are
greater than a dex.  Thus, inferences regarding this system based upon
the VP parameters should be  viewed with caution.  There is possibly a
weak  component missed  at  $v=+80$~{\kms} in  the  wings of  {\MgII};
however,   it   was   formally   rejected  by   {\sc   Minfit}.    The
{\FeII}~$\lambda 2344$,  2374, and 2383 transitions  were not captured
by the CCD.

\subsubsection{\rm PKS~$0454-220$~~~~$z_{\rm abs}=0.483340$}

This  system is  classified  as {\CIV}--deficient  (\cite{archiveII}).
There  are no  higher velocity  kinematic subsystems;  the  absence of
these  subsystems  is  a  common characteristic  of  {\CIV}--deficient
systems.  The profile was fitted  with a total of seven VP components,
which were  all well  constrained in that  they have  small fractional
errors.  The  strongest {\FeII} and  {\MgI} components are  aligned in
velocity with the maximum {\MgII} absorption.

\subsubsection{\rm PKS~$0823-223$~~~~$z_{\rm abs}=0.911017$} 

This systems is classified as a ``double'' (\cite{archiveII}).  Unlike
other  double systems,  this one  exhibits strong,  complex absorption
over most  of the kinematic  spread of $\sim$350~{\kms}.   Most double
systems are characterized by a single, large subsystem with a velocity
spread less than 20~{\kms}  and several weak, narrow, higher--velocity
subsystems.  The  {\MgII} $\lambda 2803$ transition  was captured only
for subsystem  \#1.  A total  of 18 VP  components were fitted  to the
system.  The system  is rich in {\FeII} absorption  in all subsystems.
{\MgI},  on  the  other  hand,  is unambiguously  strong  only  in  VP
components four, five,  and nine.  From this, it  can be inferred that
there is a  strong variation in the ionization  conditions between the
complex  at $-50  \leq v  \leq  -90$~{\kms} and  at $-10  \leq v  \leq
+30$~{\kms}, even  though the {\FeII} might  suggest otherwise (unless
there are abundance differences in the [$\alpha$/Fe]).

\subsubsection{\rm Q~$1101-264$~~~~$z_{\rm abs}=0.359153$}

The  fully reduced and  continuum normalized  spectra studied  in this
work  were kindly  provided  by Max  Pettini.   The observations  were
obtained with the  UCL echelle spectrograph on the  AAT, with the IPCS
as the  detector.  The resolution  is 6~{\kms} FWHM  (or approximately
0.07--0.08~{\AA} at the wavelengths covered here).

The  {\MgII} transitions lie  in the  {\Lya} forest  of this  line of
sight, but  there is no apparent  contamination from {\HI}.   A fit to
the profiles yields six  VP components.  Strong {\FeII} $\lambda 2600$
and    {\MgI}     have    been    reported     by    Carswell    \etal
(1991\nocite{carswell91}), but are not presented here.

\subsubsection{\rm Q~$1148+384$~~~~$z_{\rm abs}=0.553362$}

This system has a very  interesting profile that is comprised of eight
VP  components in a  single kinematic  system spread  over 200~{\kms}.
The  absorption  strengths   symmetrically  decrease  with  increasing
velocity,   giving  the  overall   profile  a   Gaussian  ``envelope''
appearance.  However, it is apparent that the {\FeII} to {\MgII} ratio
is  not  symmetric;  {\FeII}   is  relatively  stronger  in  the  blue
components of the system.   The {\FeII}~$\lambda 2344$, 2374, and 2383
transitions were not captured.

\subsubsection{\rm PG~$1206+459$~~~~$z_{\rm abs}=0.927602$} 

The kinematics and ionization  conditions of this rich, complex system
have   been    studied   in   detail   by    Churchill   \&   Charlton
(1999\nocite{q1206}).   It has  been  classified as  a double  systems
(\cite{archiveII}).   The overall  system  was fitted  with twelve  VP
components over a  total of five subsystems.  {\FeII}  and {\MgII} are
present only in VP components eight, ten, and twelve in subsystem \#5.

\subsubsection{\rm Q~$1222+228$~~~~$z_{\rm abs}=0.668052$}

The  profile is  characterized by  a central  strong absorption  and a
``high--velocity''  subsystem  at  $v  \sim  230$~{\kms}.   This  weak
subsystem, \#2,  has relatively strong {\FeII}, and  perhaps very weak
{\MgI}.  Velocities less than  $-100$~{\kms} were not observed for the
{\MgII} $\lambda  2796$ transitions, but  were for the  $\lambda 2803$
transition.  There is no detectable absorption at those velocities.  A
total  of  five  VP  components  were fitted  to  the  main,  stronger
subsystem  and a  single component  was fitted  to  the high--velocity
subsystem.  {\FeII} exhibits fine velocity structure in subsystem \#1.

\vglue 0.5in
\subsubsection{\rm PG~$1241+176$~~~~$z_{\rm abs}=0.550482$} 

The main {\MgII} profile of this ``classic'' system (\cite{archiveII})
exhibits a  redward subcomponent and a smooth  blueward asymmetry that
is  suggestive of  bulk motions  with varying  line--of--sight density
rather  than a  blend  of two  VP  components.  The  asymmetry is  not
significant  in {\FeII}  and {\MgI},  but  may be  present just  below
detection levels.  There is a subsystem at $v\simeq +140${\AA} with no
detected {\FeII} or {\MgI}.  A  total of four VP components was fitted
to the system.  The {\FeII}~$\lambda$ 2344, 2374, and 2383 transitions
were not captured by the CCD for the given HIRES configuration.

\subsubsection{\rm PG~$1248+401$~~~~$z_{\rm abs}=0.772953$} 

This system  is classified  as a ``classic''  (\cite{archiveII}).  The
{\MgII} profiles  exhibit the ``double  horn'' shape that  Lanzetta \&
Bowen (1992)  derived for a  smooth density, constant  infall velocity
kinematic  halo  model.   The  velocity  separation of  the  two  main
absorbing regions is $\sim 50$~{\kms}.  The blue ``horn'' is comprised
of three  VP components due to  the asymmetry in  the overall profile.
This central  profile is quite  similar to the Q~$1101-264$  system at
$z_{\rm abs} = 0.3590$.  At $v=+225$~{\kms}, there is an outlying high
velocity  optically thin  subsystem.   This outlier  exhibits a  small
redward asymmetry.  Several {\FeII} transitions ($\lambda 2344$, 2374,
2383, 2587, and 2600) have been detected.  The high velocity subsystem
is detected in  the $\lambda 2600$ transition, but  is not detected to
5~$\sigma$ in  the other {\FeII}  transitions.  A clear  detection of
{\MgI} is also  present in the high velocity  system.  Normally, these
optically  thin  high velocity  subsystems  are  not  seen to  exhibit
neutral absorption (this is a rare occurrence).

\subsubsection{\rm Q~$1254+044$~~~~$z_{\rm abs}=0.519389$}  

The main subsystem  in this absorber exhibits an  asymmetry, like that
in Q~$1241+176$  at $z=0.5505$, suggestive of bulk  motion and varying
line--of--sight  density.  The absorption  is stronger,  however, such
that the asymmetry is seen most  strongly in the neutral gas as traced
by {\MgI}.  This main feature has been fitted with three VP components
and shows a typical rotating disk  shape as can especially can be seen
in  the {\MgI} transitions.   There are  two high  velocity, optically
thin single  VP component subsystems at  negative velocities, $v=-160$
and $-360$~{\kms}.  Neither {\MgI}  nor {\FeII} were detected in these
two   subsystems.   The   {\FeII}~$\lambda  2344$,   2374,   and  2383
transitions were not captured by the CCD.

\subsubsection{\rm Q~$1254+044$~~~~$z_{\rm abs}=0.934232$}  

The  kinematics  of  this  systems   are  very  similar  to  those  of
Q~$1248+401$ at  $z=0.7730$ and also subsystem \#2  of Q~$0002+051$ at
$z=0.8514$.   Again, this is  the classic  ``double horn''  shape that
Lanzetta \& Bowen (1992) derived for a smooth density, constant infall
velocity  kinematic halo model.   The velocity  separation of  the two
main  absorbing  regions is  $\sim  40$~{\kms}.   A  total of  two  VP
components, one  modeling each  of the ``horns'',  were fitted  to the
profiles.  The  {\FeII} is  prominent only in  one of  the components.
{\MgI}  is also  present  in  this particular  component,  but is  not
detected in the other.

\subsubsection{\rm Q~$1317+277$, TON 153~~~~$z_{\rm em}=0.660051$}  

This {\CIV}--deficient  System (\cite{archiveII}) has  a full velocity
spread  of $\sim 150$~{\kms}  with a  main subsystem  that has  a full
width  of $\sim  50$~{\kms} and  an asymmetric  profile  shape.  Three
subsystems were found and a  total of eight VP components were fitted.
Subsystem \#3 is the most complex of any subsystem at high velocities.
Though fitted with three VP components, the profile is more suggestive
of  bulk motions.   {\FeII} has  been detected  in the  main  and next
strongest subsystems only.  Weak  {\MgI} absorption is detected in the
component with the highest {\MgII} column density.

\subsubsection{\rm Q~$1421+331$~~~~$z=0.902871$} 

This  complex  system  shows  rich variations  in  ionization,  and/or
chemical conditions  as can  be seen by  the variations in  {\FeII} to
{\MgII}  strengths across  subsystems.   The {\FeII}  profiles show  a
strong blueward  asymmetry, with a complexity  that suggest unresolved
fine  structure  in the  data.   {\MgI}  absorption  is seen  only  in
subsystem \#2  in the velocity  region where {\MgII} is  most strongly
saturated and exhibits a  slight blueward asymmetry.  The full profile
was  fitted  with  a  total  of  ten VP  components.   There  are  two
subsystems based  upon the objective algorithm, though  the profile is
more  suggestive of three.   In the  strongest subsystem,  the {\FeII}
transitions reveal a ``disk--like''  profile, or the characteristic of
organized motion with a density gradient along the line of sight.

\subsubsection{\rm Q~$1421+331$~~~~$z_{\rm  abs}=1.172609$} 

The  profile  is quite  symmetric,  though  it  exhibits three  strong
centers of higher column density  gas.  This systems has no associated
higher velocity  subsystems.  The complex {\MgII}  kinematics are seen
in  both  {\MgI}  and  {\FeII}  with remarkable  little  component  to
component variations.   The full  profile was fitted  with a  total of
seven VP components.

\subsubsection{\rm PG~$1634+706$~~~~$z_{\rm abs}=0.990239$} 

The  signal--to--noise ratio  for this  system is  the highest  in our
sample  for  the {\MgII}  and  {\MgI}  transitions.   It is  a  single
kinematic system  with no moderate  to high velocity subsystems  to an
equivalent width limit of $0.007$~{\AA}.  This system is classified as
{\CIV}--deficient  (\cite{archiveII}).  The  lack  of higher  velocity
subsystems is characteristic  of {\CIV}--deficient systems.  The total
profile velocity spread  is $\sim 80$~{\kms} and was  fitted with five
VP  components.  The  {\FeII}  profiles exhibit  the classic  ``double
horn'' shape, with an overall redward asymmetry.

\subsubsection{\rm PKS~$2128-123$, PHL~1598~~~~$z_{\rm em}=0.429735$}  

This system is  also {\CIV}--deficient (\cite{archiveII}), and typical
of this  class of {\MgII}  system, has no higher  velocity subsystems.
There  is  a ``double--like''  characteristic  to  the profiles.   The
system has been fitted with a  total of four VP components.  {\MgI} is
associated  with  all four  {\MgII}  components  and  shows a  redward
asymmetry in  the strongest two.   Unfortunately, none of  the {\FeII}
transitions  reported  by   Tytler  \etal  (1987\nocite{tbsyk87})  and
Petitjean \& Bergeron (1990\nocite{pb90}) were captured by the CCD,

\vglue 0.5in
\subsubsection{\rm PKS~$2145+067$~~~~$z_{\rm abs}=0.790777$}

This ``classic'' system  (\cite{archiveII}) has three subsystems, each
having  been fitted  with  two VP  components.   These subsystems  are
separated  by $\sim  90$~{\kms}.   Only subsystems  \#1  and \#3  have
detected {\FeII} and none  have detected {\MgI}.  The {\FeII} $\lambda
2587$  transition was  not included  in  the analysis  because it  was
positioned near the pen mark on the HIRES CCD.

\section{Discussion of Properties}
\label{sec:results}

In  order to  investigate  trends in  the  absorption properties  with
kinematics (i.e.\  velocities and kinematic spreads),  we have divided
the subsystems into three  classes, the ``low'', ``intermediate'', and
``high'' velocity  subsystems.  Recall  that the velocity  zero points
are  set  by  the  absorption  redshifts,  which  are  computed  using
Equation~\ref{eq:zabs}.   There is  a natural  break in  the subsystem
velocity  distribution at  $v=40$~{\kms}, so  we divided  the  low and
intermediate velocity classes at  $40$~{\kms} (note that there is only
one  subsystems with  $20 \leq  v \leq  40$~{\kms};  we conservatively
chose $40$~{\kms}).  The  median velocity of all subsystems  with $v >
40$~{\kms} is $165$~{\kms}, which is where we divided the intermediate
and high  velocity classes.  For  the following, we represent  the low
velocity class  with open diamond data points,  the intermediate class
with  open  triangles,  and   the  high  velocity  class  with  filled
triangles.

\subsection{Kinematics}
\label{sec:kinematics}

\begin{figure*}[thb]
\figurenum{7}
\plotfiddle{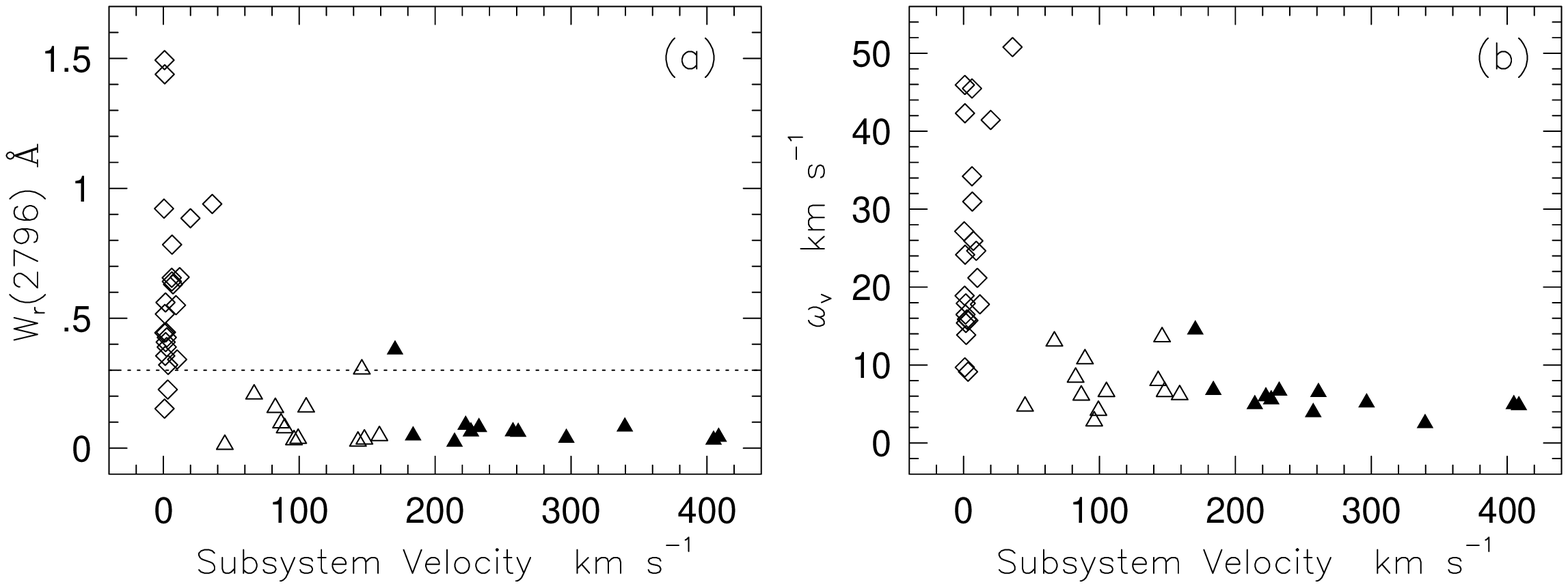}{2.3in}{0.}{70}{70}{-280}{-210}
\caption[fig7.eps]   {  (a)   {\MgII}   $\lambda  2796$   subsystem
equivalent  width   vs.\  subsystem  velocity,   $\left<  v  \right>$.
Diamonds represent subsystems with  $\left< v \right> \leq 40$~{\kms},
open triangles  $ 40  < \left< v  \right> \leq 165$~{\kms},  and solid
triangles  $\left<  v  \right>   >  165$~{\kms}.   ---  (b)  Subsystem
kinematic spread, $\omega _{v}$,  vs.\ subsystem velocity.  Data types
are the same as in panel (a). \label{fig:kin_ew_w-v} }
\end{figure*}

In Figure~\ref{fig:kin_ew_w-v}$a$  and $b$, we  show subsystem {\MgII}
$\lambda  2796$  equivalent widths  vs.\  subsystem  velocity and  the
subsystem kinematic spreads  vs.\ subsystem velocities, respectively.
A horizontal dot--dot line at $W_{r}(2796)$~{\AA} indicates the sample
minimum equivalent  width cutoff for the full  absorption systems.  In
general,  {\MgII}  systems  characteristically  have  a  ``dominant'',
optically  thick   absorbing  subsystem.   The   majority  of  {\MgII}
absorbers  have one to  a few  ``intermediate'' and  ``high'' velocity
subsystems.

The  equivalent widths of  the low  velocity, dominant  subsystems are
$W_{r}(2796) <  1.0$~{\AA}, with the  majority ranging from  $0.3 \leq
W_{r}(2796)  \leq  0.6$~{\AA}.   Their kinematic  spreads  typically
range between $10 \leq \omega _{v} \leq 30$~{\kms}.  The two systems in
Figure~\ref{fig:kin_ew_w-v}$a$ with $W_{r}(2796) \simeq 1.5$~{\AA} and
$\omega _{v} \simeq 50$~{\kms} are DLA/{\HI}--Rich systems; neither of
which have  intermediate or high  velocity subsystems [which may  be a
characteristic of lower redshift DLAs (\cite{archiveII})].

The  moderate and  high velocity  subsystems have  $W_{r}(2796) \simeq
0.1$~{\AA} and kinematic spreads less than 10~{\kms}.  They contribute
only 10--20\%  to the system total  equivalent width.  Out  of 24, two
have $W_{r}(2796)  > 0.3$~{\AA} (\#2  in the $z=0.9110$  system toward
Q$0823-223$  and  \#2 in  the  $z=0.8514$  system toward  Q$0002+051$,
respectively).   They  are  contributing  members  in  double  systems
(\cite{archiveII}).

There is a  break in the velocity distribution  at $v \sim 40$~{\kms},
which arises  because the dominant  subsystems, many of  which exhibit
complex kinematic structure, remove flux over a total velocity span of
$30$--$40$~{\kms}.  The  intermediate velocity subsystems  have a mean
kinematic  spread of $\simeq  6.8$~{\kms} and  those at  high velocity
have a  mean of $\simeq 5.5$~{\kms}.   The large majority  of the high
velocity  subsystems  are  unresolved,  whereas roughly  half  of  the
intermediate velocity subsystems  are resolved.  The anti--correlation
between  subsystem kinematic  spread  and velocity  is significant  at
$3~\sigma  $.  This  implies that  the high  velocity  subsystems have
little to  no bulk motions, whereas the  moderate velocity subsystems
exhibit a slightly broader range of bulk motions.

What  is gleaned  from  Figures~\ref{fig:kin_ew_w-v}$a$--$b$, is  that
moderate   redshift   {\MgII}   systems   have   {\it   characteristic
kinematics}.  They are {\it not\/} characterized by multiple ``large''
subsystems  of comparable kinematic  spreads and/or  equivalent widths
separated in  velocity.  They {\it are\/} characterized  by a dominant
subsystem with  a kinematic  spread of  a few tens  of {\kms}  that is
often  accompanied  by significantly  smaller  subsystems at  relative
velocities ranging  from a few  tens to a  few hundreds of  {\kms} and
kinematics spreads  of a few  to several {\kms}.  These  facts suggest
that the lines of  sight through intermediate redshift {\MgII} systems
with $W_{r}(2796) > 0.3$~{\AA}  are picking out a dominant, relatively
more massive  structure with somewhat  systematic kinematics of  a few
tens of  {\kms}, often surrounded by  smaller fragments of  gas over a
fairly large range of relative velocities.

\begin{figure*}[thb]
\figurenum{8}
\plotfiddle{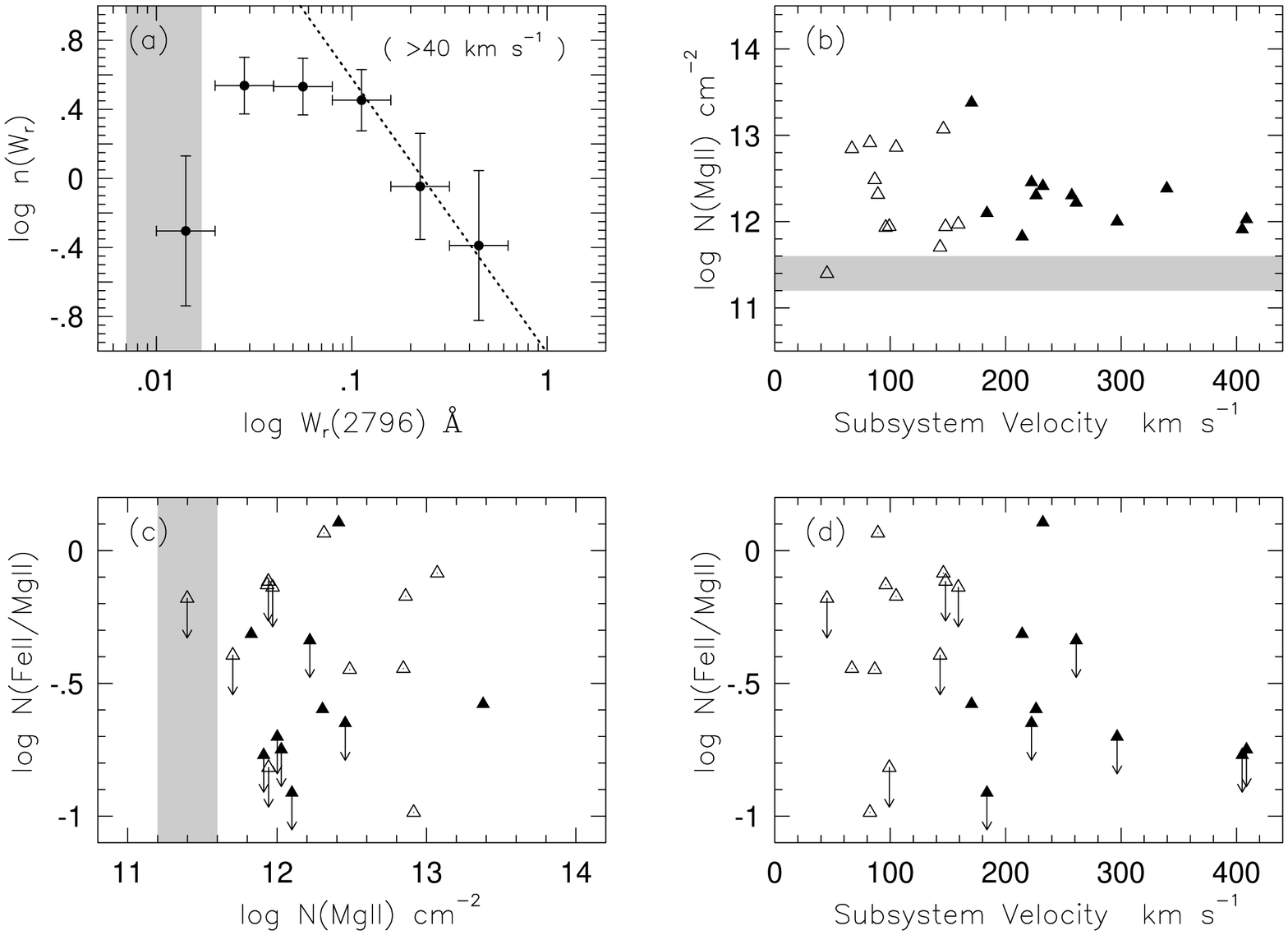}{4.6in}{0.}{70}{70}{-280}{-45}
\caption[fig8.eps]  {  (a)  The  equivalent width  distribution  of
moderate and high velocity  kinematic subsystems (those with $\left< v
\right> >  40$~{\kms}).  The shaded  region gives the range  where the
detection sensitivity drops (partial completeness); below $W_{r}(2796)
=    0.007$~{\AA}   there   is    no   detection    sensitivity   (see
Figure~\ref{fig:z_ewlim}).  The dot--dot line  is a power law fit with
slope of  $-1.6$ (see  text).  --- (b)  Apparent optical  depth column
densities, $\log  N({\MgII})$ vs.\ velocities, $\left<  v \right>$, of
the  kinematic  subsystems.   Data   points  types  are  same  as  for
Figure~\ref{fig:kin_ew_w-v}.  Typical  errors in $N({\MgII})$  are 0.2
dex.   --- (c)  Logarithmic ratio  $N({\FeII})/N({\MgII})$  vs.\ $\log
N({\MgII})$ for moderate and high velocity kinematic subsystems.  The
shaded  regions  is  the  same  as  in  (d) ---  Logarithmic  ratio
$N({\FeII})/N({\MgII})$ vs. subsystem velocities.
\label{fig:vgt40ewdist} \label{fig:NMgIIvsv} }
\end{figure*}

\begin{figure*}[bht]
\figurenum{9}
\plotfiddle{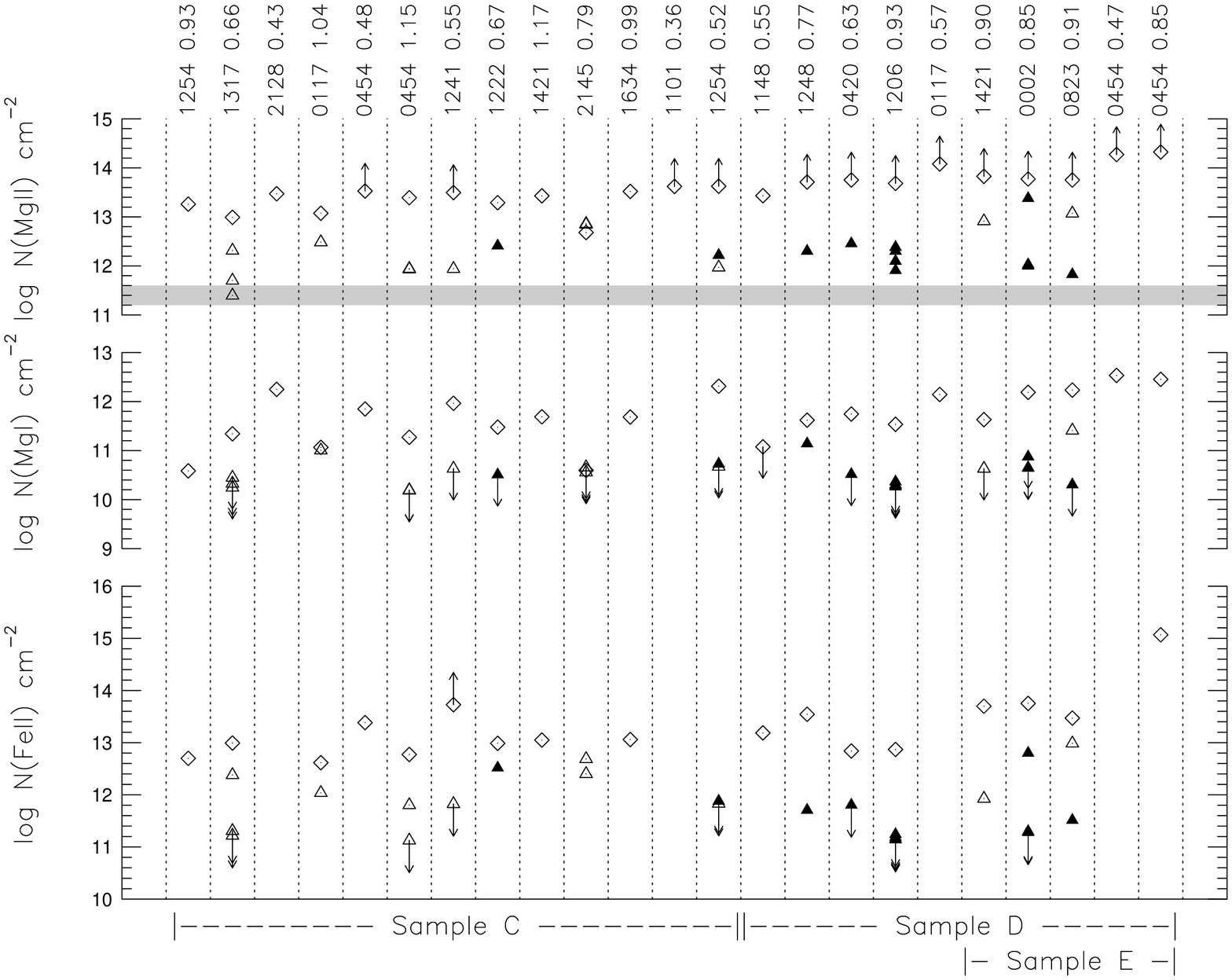}{5.2in}{0.}{70}{70}{-280}{-45}
\caption[fig9.eps]   { Apparent optical  depth column  densities of
$N({\MgII})$ [top panel], $N({\MgI})$ [center panel], and $N({\FeII})$
[bottom  panel]  for  each  system.   The  systems  are  presented  in
increasing  equivalent  width order.  Sample  memberships are  labeled
along   the   bottom.   Data    points   types   are   same   as   for
Figure~\ref{fig:kin_ew_w-v}.  Typical  errors in $N({\MgII})$  are 0.2
dex.   The shaded  area gives  the range  of {\MgII}  column densities
where  the detection  sensitivity  drops from  complete  to none  (for
unresolved features). \label{fig:Naod} }
\end{figure*}

\subsection{Equivalent Width Distribution}
\label{sec:ewdist}

The {\MgII}  equivalent widths of  the intermediate and  high velocity
subsystems  are similar  in  strength  to those  of  the weak  {\MgII}
systems (\cite{weakI}).   It is of interest, therefore,  to explore if
the physical  nature of isolated weak  systems is similar  to those of
the  moderate and  high velocity  subsystems clustered  within systems
with $W_{r}(2796) \geq  0.3$~{\AA}.  The equivalent width distribution
provides one means for comparison.

In Figure~\ref{fig:vgt40ewdist}$a$, we plot the rest--frame equivalent
width distribution for the  intermediate and high velocity subsystems,
those with $v > 40$~{\kms}.  The  data have been binned for purpose of
presentation.  The shaded region gives the equivalent width range over
which  the  detection  completeness  drops  from  100\%  to  zero  for
unresolved absorption.   Compared to a  power--law distribution, there
appears to  be a turnover below $W_{r}(2796)  \simeq 0.08$~{\AA}, well
above  the region  of  partial  completeness.  The  dotted  line is  a
maximum likelihood power--law fit to $n(W_{r}) = CW_{r}^{-\delta}$ for
$W_{r}> 0.08$~{\AA}, which yielded $\delta = 1.6\pm0.7$.

Churchill \etal (1999b\nocite{weakI}) showed that the equivalent width
distribution of an unbiased sample  of {\MgII} systems follows a power
law with $\delta = 1.0\pm0.1$ down to $W_{r}(2796) = 0.02$~{\AA}.  The
number density of very small equivalent width systems in the ``field''
continues   to  increase   as  $W_{r}(2796)$   decreases   well  below
0.08~{\AA}.  In contrast, the {\MgII} equivalent width distribution of
intermediate   and   high  velocity   subsystems   turns  over   below
$W_{r}(2796) \simeq  0.08$~{\AA}.  It may  be that both the  shape and
the  slope of the  equivalent width  distribution differs  between the
population  of weak  systems and  the intermediate  and  high velocity
subsystems of ``strong'' systems.

Because  of the  large  error in  the  slope fitted  to the  kinematic
subsystems,  however,  we cannot  discern  if  the  two slopes  differ
significantly.  The turnover below  $W_{r}(2796) < 0.08$~{\AA}, on the
other hand, is unambiguous  and represents a fundamental difference in
the two distributions.

\subsection{AOD Column Densities}
\label{sec:aod}

In Figure~\ref{fig:NMgIIvsv}$b$, we present the apparent optical depth
(AOD) {\MgII} column densities  for the intermediate and high velocity
subsystems vs.\  the subsystem velocity.  The shaded  region shows the
range of column densities  where the detection completeness drops from
100\% to zero for unresolved absorption.  The sample is 100\% complete
to  $\log  N({\MgII})  =  11.6$~{\AA}.   The  largest  columns,  $\log
N({\MgII})  \simeq  13$~{\cmsq} are  primarily  distributed among  the
intermediate   velocity  subsystems   of  the   double   systems  (see
\cite{archiveII}).  The more typical column density is less than $\log
N =  12.5$~{\cmsq}.  As  evident from the  turnover in  the equivalent
width  distribution   (Figure~\ref{fig:vgt40ewdist}$a$),  there  is  a
paucity of subsystems with $\log  N \leq 12$~{\cmsq}, which is 0.4 dex
above  where  the  column  density  detection  sensitivity  begins  to
decline.

It could be argued that  smaller column density subsystems are present
but that they have very  broad kinematic spreads and are therefore not
detectable  due  to  the noise  in  the  data  (i.e.\ they  have  been
``removed''  during  the continuum  fitting  process).  However,  this
argument is not supported by  the data.  The kinematic spreads of very
small column  density subsystems are  tightly clustered about  a small
value  (see   Figure~\ref{fig:kin_ew_w-v});  if  there   were  broader
subsystems  it would  be expected  that a  range of  kinematic spreads
would be seen in the data.  Otherwise, it would have to be argued that
a   bimodal  distribution   exists  and   that  one   mode   has  been
systematically missed.

In Figures~\ref{fig:NMgIIvsv}$c$ and $d$, we show the logarithmic
ratio of $N({\FeII})/N({\MgII})$ vs.\ $N({\MgII})$ and vs.\ subsystem
velocity, respectively.  In the moderate and high velocity subsystems,
the ratio $N({\FeII})/N({\MgII})$ varies in the log from $0$ to $-1$.
It is fair to say that, over the range $12.2 \leq N({\MgII}) \leq
13.4$~{\cmsq} observed in moderate and high velocity subsystems, the
likelihood of finding $\log N({\FeII})/N({\MgII}) = 0$ is as likely as
finding $\log N({\FeII})/N({\MgII}) = -0.5$.  It is not clear if
subsystems in this {\MgII} column density range commonly have $\log
N({\FeII})/N({\MgII}) = -1$, though one has been found.  It does
appear that subsystems with $N({\MgII}) \leq 12.2$~{\cmsq} may often
have $\log N({\FeII})/N({\MgII}) \sim -1$, though again, there is an
exception to this possibility.  If all this holds true, it may be that
the highest velocity clouds, which have some of the smallest
$N({\MgII})$, will have some of the smallest $N({\FeII})$ to
$N({\MgII})$ ratios.  The data in Figure~\ref{fig:NMgIIvsv}$d$ are
somewhat suggestive of this trend.  However, more systems would need
to be analysed in order to make difinitive statements.  It is
interesting to speculate, however, that a trend in
$N({\FeII})/N({\MgII})$ with subsystem velocity might indicate a very
real physical origin and evolution of the highest velocity
subsystems.

In Figure~\ref{fig:Naod}, we present the apparent optical depth column
densities of  {\MgII} [top panel], {\MgI} [center  panel], and {\FeII}
[bottom panel]  for all  subsystems in a  given system.   The systems,
which  are   labeled  across  the  top,  are   ordered  by  increasing
$W_{r}(2796)$ from  left to right.  Sample membership  is marked below
the bottom  panel.  This ordering  is to allow comparisons  within and
across samples.  The shaded region in the top panel gives the range of
detection thresholds of unresolved {\MgII} features.  Data point types
denote velocity class (low,  intermediate, and high).  Upper and lower
limits   are  shown   with  downward   and  upward   pointing  arrows,
respectively.

From Figure~\ref{fig:Naod},  it can  be seen how  often and  with what
strengths  {\MgI} and {\FeII}  are detected  in intermediate  and high
velocity subsystems.   An interesting  contrast, for example,  is seen
between the  high velocity subsystem  in the $z=0.7730$  system toward
Q~$1248+401$ and  that in  the $z=0.6330$ system  toward Q~$0420-014$.
Both  have comparable  {\MgII}  column densities,  yet the  $z=0.7730$
system has a fairly large  {\MgI} column density, while the $z=0.6330$
system has  a restrictive upper  limit.  The {\FeII} columns  may also
track this behavior.

\subsection{Kinematics and System Absorption Strength}
\label{sec:kinew}
 
\begin{figure*}[thb]
\figurenum{10}
\plotfiddle{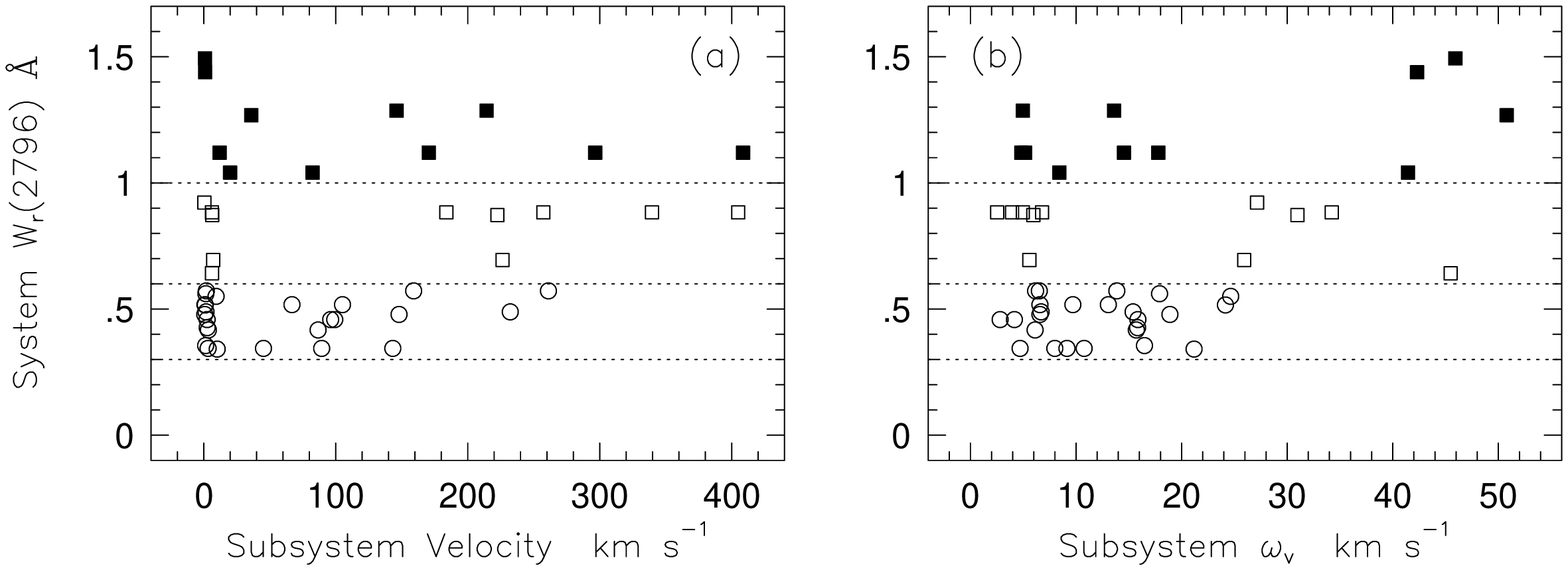}{2.3in}{0.}{70}{70}{-280}{-210}
\caption[fig10.eps] {  (a) {\MgII} $\lambda  2796$ equivalent width
of the full system  vs.\ kinematic subsystem velocities.  Open circles
represent Sample  B and  open boxes represent  Sample C.  Sample  E is
represented  with  filled boxes.   ---  (b)  The  same vs.\  kinematic
velocity  spread of the  subsystems.  Data  types are  the same  as in
panel (a). \label{fig:CDEkin} }
\end{figure*}

In  Figure~\ref{fig:CDEkin}$a$--$b$, we  show  {\MgII} $\lambda  2796$
equivalent width of the full system vs.\ subsystem velocities and vs.\
the  velocity  spreads  of  the  subsystems,  respectively.   As  with
Figures~\ref{fig:ew_dr_z} and  \ref{fig:omega}, open circles represent
sample B,  open boxes represent  sample C, and filled  boxes represent
sample E.  Horizontal dot--dot lines separate the samples.

The striking trend seen  in Figure~\ref{fig:CDEkin}$a$ is the relative
paucity  of high  velocity  ($v>165$~{\kms}) subsystems  in sample  B.
Though  the demarcation  between samples  B  and C  at $W_{r}(2796)  =
0.6$~{\AA}  is arbitrary  (being based  upon previous  studies), there
appears  to  be  a  trend  with decreasing  system  equivalent  width.
Smaller equivalent width systems  have fewer high velocity subsystems;
most of their subsystems  are at intermediate velocities.  

In  Figure~\ref{fig:CDEkin}$b$,  the  intermediate and  high  velocity
subsystems occupy the left  hand region ($\omega _{v} \leq 10$~{\kms})
of  the diagram,  whereas the  low velocity,  dominant  subsystems are
distributed over a broad range of kinematic spreads.  Sample B systems
have  no dominant  subsystems with  $\omega _{v}$  greater  than $\sim
25$~{\kms},  whereas  all  sample   C  and  E  systems  have  dominant
subsystems with $\omega _{v} > 25$~{\kms}.

To investigate  if these  apparent kinematic trends  are statistically
significant, we computed  the two--point velocity correlation function
(TPCF) of  the kinematic subsystems for  samples B, C, D,  and E.  The
TPCF  is  simply  the  frequency distribution  of  subsystem  velocity
splittings, $\Delta  v_{ij} = |v_{i}-v_{j}|$ for $i\neq  j$, where $i$
and $j$ represent any two subsystems in the sample.

\begin{figure*}[thb]
\figurenum{11}
\plotfiddle{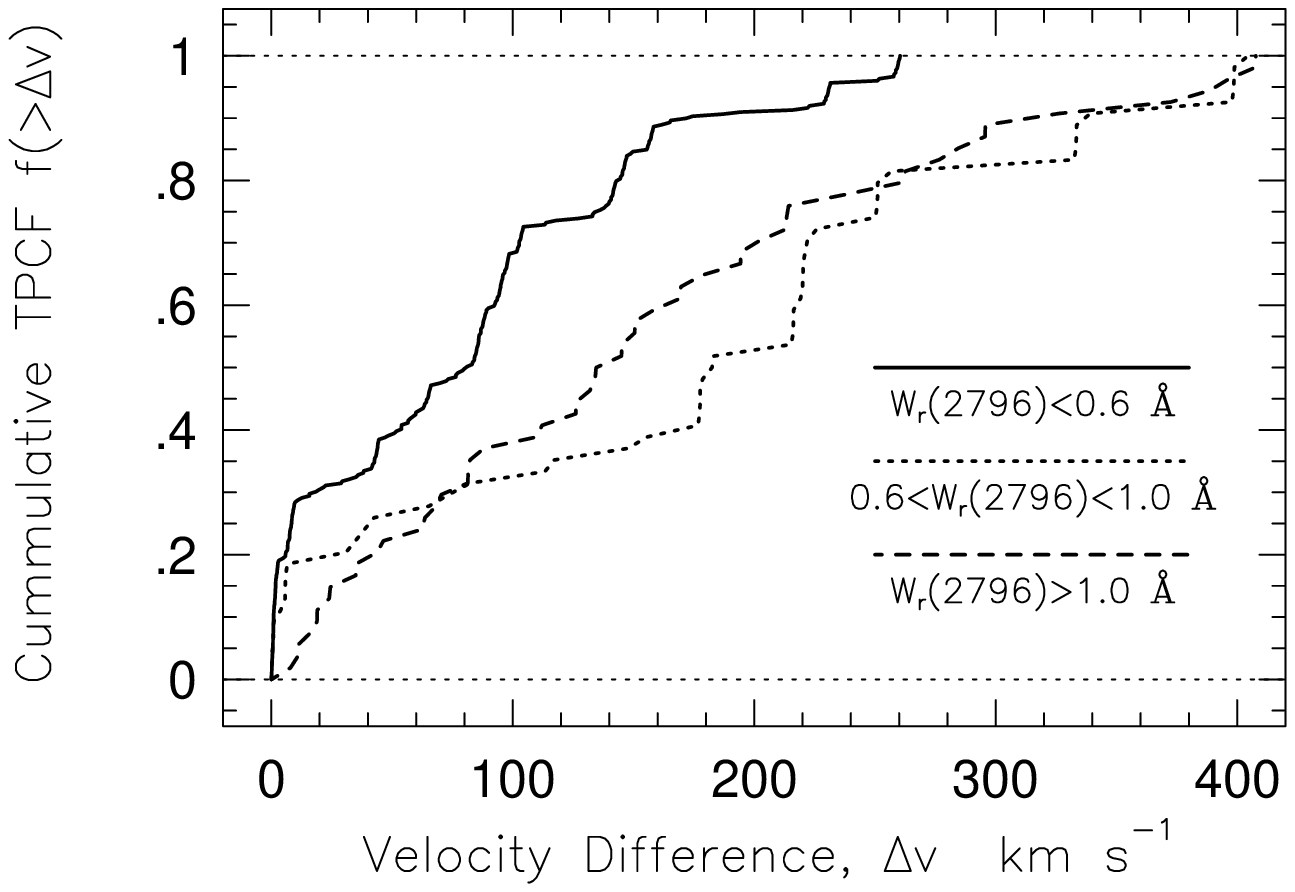}{2.4in}{0.}{70}{70}{-175}{-200}
\caption[fig11.eps]  {  (a)  The  cumulative  distribution  of  the
velocity  two--point correlation function  for all  subsystems.  Three
subsamples  are shown,  Sample  B (solid  curve),  Sample C  (dot--dot
curve), and Sample E (dash--dot curve). \label{fig:cdfCDE} }
\end{figure*}

In Figure~\ref{fig:cdfCDE}, we present the cumulative distribution of
the TPCFs for  sample B (solid curve), sample  C (dot--dot curve), and
sample E (dash--dash curve). Sample D, which is the union of samples C
and E,  is not plotted.  There is  a fast rise at  very small velocity
separations for  samples B  and C  that is not  present for  sample E.
This reflects the  fact that dominant, low velocity  subsystems in the
largest  equivalent  width  systems  tend  to  have  larger  kinematic
spreads,  which suppresses  small velocity  separations.   The smaller
equivalent  width  systems in  sample  B  do  not have  velocity  pair
separations  greater  than  $\sim  300$~{\kms}, whereas  {\it  both\/}
samples  C and  E have  velocity pair  separations as  large  as $\sim
400$~{\kms}.

Kolmogorov--Smirnov (KS) tests of sample  B vs.\ samples, C, D, and E,
yielded probabilities no greater  than $10^{-5}$ that either sample C,
D, or E exhibits  the same TPCF as sample B.  On  the other hand, a KS
test of  sample C  vs.\ sample  E yielded a  probability of  $0.1$; it
cannot be ruled out that samples C and E have the same distribution of
subsystem velocity splittings.

\section{On the Nature of the Kinematic Subsystems}
\label{sec:open}

\subsection{Dominant Kinematic Subsystems}

Virtually all {\MgII}  absorbers in our sample are  characterized by a
dominant,  often saturated,  kinematic subsystem.   For sample  B, the
kinematic spreads, $\omega _{v}$,  of these dominant subsystems are no
greater than 25~{\kms}.  It is rare for there to be a second kinematic
subsystem with  comparable kinematic spread and  equivalent width (the
$z=0.79$ system toward PKS~$2145+064$ is an exception).

As reported by Steidel, Dickinson, \& Persson (1994\nocite{sdp94}) and
by  Steidel (1995\nocite{steidel95}),  there  appears to  always be  a
normal, bright  galaxy within  $\sim 40h^{-1}$~kpc (sky  projected) of
the absorbing gas.  Furthermore,  the physical extent of the absorbing
gas was shown to have near--unity covering factor and to be related to
the  galaxy  $K$  luminosity,  such  that  $R(L_{K})  \simeq  40h^{-1}
(L_{K}/L_{K}^{\ast})^{0.15}$~kpc.   These results further  support the
scenario in which  absorbing gas is associated, in  some way, with the
total mass (number of stars) in  a ``host'' galaxy, or that there is a
strong  link between  the host  galaxy mass  and the  extended gaseous
environment in  which the galaxy is evolving  (also see \cite{csv96}).
Since  virtually  the  entire  {\MgII}  equivalent width  in  a  given
absorber  arises from the  dominant kinematic  subsystem, it  has been
these   subsystems  alone   that  are   governing  the   statistics of
absorber--galaxy associations.

There  is an  observed one--to--one  correspondence  between optically
thick {\HI} and  {\MgII} systems with $W_{r}(2796) >  0.3$~{\AA} and a
one--to--one  correspondence  between optically  thin  {\HI} and  weak
systems        with         $W_{r}(2796)        \leq        0.2$~{\AA}
(\cite{archiveI})\footnote{We caution  that the sample  size is small,
$\sim 15$.}.  This  would suggest that the moderate  and high velocity
kinematic subsystems  are optically thin in {\HI},  that the optically
thick  {\HI} arises  in  one or  a  few of  the  subcomponents in  the
dominant   kinematic  subsystem.    Photoionization   models  (Cloudy;
\cite{ferland})  are  consistent  with this  statement  (\cite{weakI};
\cite{q1206}).

The narrow kinematic spreads and  larger {\HI} column densities of the
dominant subsystems,  would suggest that dominant  subsystems arise in
higher density  structures with  some level of  systematic kinematics.
An unresolved  issue is whether the {\it  optically thick\/} absorbing
gas  is  distributed  quasi--uniformly  (giving  rise  to  the  unity
covering  factor)  in an  extended,  ``spherical'' halo  (\cite{lb90};
\cite{bb91}; \cite{steidel93};  \cite{steidel95}; \cite{3c336}), or in
a  more  flattened,   ``disk''  geometry  (\cite{bbp95};  \cite{cc96};
\cite{kinematicpaper}).  Absorption line  models using random lines of
sight through extended disk  structures, with flat rotation curves and
reasonable  vertical  velocity   dispersions,  naturally  explain  the
characteristic     properties    of     the     dominant    subsystems
(\cite{kinematicpaper}).

This by no means renders the ``disk model'' a unique interpretation of
the strongest  subsystems in  complex absorption line  data.  However,
any  simple  interpretation  must  explain  the  ubiquity  and  narrow
kinematic range of the dominant subsystems.  If galaxy disks are not a
significant contribution  to the absorption,  then we are  required to
suggest a  more complex scenario; spatial  and kinematic distributions
of  the absorbing gas,  from galaxy  to galaxy,  would be  required to
conspire such  as to  virtually always give  rise to a  {\it single\/}
optically  thick subsystems with  a kinematic  spread less  than $\sim
25$~{\kms} and still yield a unity covering factor, statistically.  As
such, the  presence and properties  of dominant subsystems  in {\MgII}
systems are important constraints for models of absorbers.

\subsection{Moderate and High Velocity Kinematic 
Subsystems}

There is  no direct evidence  whether or not the  small $W_{r}(2796)$,
moderate  to  high velocity  subsystems  are  components  in the  same
spatial distribution around galaxies as the dominant subsystems (i.e.\
reside   within  $R(L_{K})$   or  reside   at   larger  galactocentric
distances).  As such, inferring their nature and their relationship to
the  dominant kinematic  subsystem is  difficult.  However,  there are
some clues to be drawn from the data, which we discuss below.

\subsubsection{Comparison with Galactic HVCs}

The  physical  locations,  sizes,  metallicities,  and  kinematics  of
Galactic  high  velocity  clouds  (HVCs;  e.g.\  \cite{wakker97})  are
currently a matter of  debate (e.g.\ Blitz \etal 1999\nocite{blitz99};
Braun  \&  Burton  1999\nocite{bb99};  \cite{hvcpaper}; also  see  the
series   of    articles   in    Hibbard,   Rupen,   \&    van   Gorkom
2000\nocite{vla2000}).   It  is  clear  that  {\it  some\/}  HVCs  are
associated with the  Galaxy in that they are  within a few kiloparsecs
and/or are members  of large complexes that appear  to lie along great
circles on the  sky. What is not definitive,  however, is whether some
HVCs are  remnants of galaxy group  formation and whether  or not they
are spatially and kinematically distributed throughout the Local Group
consistent   with  material   inflowing  from   the   intersection  of
intergalactic   filaments,  where   multiple   galaxies  form   (e.g.\
\cite{zhang}).   Note, however, that  the results  of Zwaan  \& Briggs
(1999\nocite{zwaan99}) and Charlton \etal (2000\nocite{hvcpaper}) have
placed  strong  constraints  limiting  the parameter  space  of  group
formation models.

HVCs, including  the compact HVCs (\cite{bb99}), are  detected in {\HI}
21--cm  emission, for  which the  present sensitivity  is  typically a
${\rm few} \times 10^{18}$~atoms~{\cmsq};  they are optically thick in
{\HI}  and  would give  rise  to Lyman  limit  breaks  if detected  in
absorption.  Photoionization  modeling [using Cloudy (\cite{ferland})]
has revealed that {\MgII} absorption with $W_{r}(2796) \sim 0.1$~{\AA}
arises   in   optically  thin   {\HI}   gas,   i.e.\  $N({\HI})   \leq
10^{17.3}$~{\cmsq},     over     four     decades    of     ionization
parameter\footnote{The ionization parameter is the ratio of the number
density  of  hydrogen ionizing  photons  to  that  of hydrogen.}   for
metallicities greater  than 0.01~$Z_{\odot}$ (see Figures 11  \& 12 of
\cite{weakI}).   Detailed  modeling  of  the  $z=0.93$  system  toward
PKS~$1206+459$  has  revealed  that  the moderate  and  high  velocity
subsystems  in  that  absorber  are  tightly  constrained  to  be  low
ionization,  sub--Lyman limit,  pockets  of gas  with relatively  high
metallicities, i.e.\  $Z \geq 0.1$~$Z_{\odot}$  (\cite{q1206}).  These
``clouds'', which  have inferred masses  and sizes of $\sim  {\rm few}
\times  10$~M$_{\odot}$   and  $\sim  {\rm   few}  \times  10-100$~pc,
respectively,  are aligned  in velocity  with highly  ionized material
(i.e.\ {\CIV}, {\NV}, and {\OVI}) that arises in a separate gas phase.
This multiphase  ionization structure is  present in most  all {\MgII}
absorbers    having   moderate    and    high   velocity    subsystems
(\cite{archiveII}).

Though the ionization structure and metallicities of HVCs are not well
known, a  comparison of  the {\HI} column  densities and  the inferred
masses and sizes of moderate and high velocity subsystems reveals that
they very probably are not higher redshift analogues to Galactic HVCs.
Because of  the large  disparity in the  {\HI} column  densities, this
holds  independent  of  whether   HVCs  are  closely  associated  with
galaxies, remnants of group formation, or both.

\subsubsection{Comparison with Weak Systems}

The population of  weak {\MgII} systems (\cite{weakI}) has  a range of
equivalent widths, line widths, and doublet ratios similar to those of
the intermediate  and high  velocity subsystems.  As  mentioned above,
the  kinematic subsystems  are  optically thin  in  {\HI}, have  small
masses and sizes,  and are aligned in velocity  with higher ionization
gas in a separate phase.  Rigby \etal (2000\nocite{weakII}) have shown
that most single--cloud weak systems are also aligned in velocity with
a separate high  ionization gas phase and that  the {\HI} is optically
thin.  Furthermore, the masses,  sizes, and metallicities of many weak
systems are similar to the kinematic subsystems, being $\sim {\rm few}
\times  10$~M$_{\odot}$   and  $\sim  {\rm   few}  \times  10-100$~pc,
respectively (\cite{q1206}; \cite{weakII}).

At face  value, these considerations  would suggest that  the origins,
sizes,  lifetimes, and  chemical  and ionization  conditions of  small
$W_{r}(2796)$  subsystems  and   of  single--cloud  weak  systems  are
governed by  similar physical processes,  even if the  environments in
which   they  arise   are  quite   varied.   However,   as   shown  in
Figure~\ref{fig:vgt40ewdist}  and  discussed in  \S~\ref{sec:results},
there is  a clear  drop in  the number of  moderate and  high velocity
subsystems with  $W_{r}(2796) < 0.08$~{\AA}, whereas  no such decrease
is  seen for  the  weak  systems down  to  $W_{r}(2796) =  0.02$~{\AA}
(\cite{weakI}).  Therefore, for  the weakest subsystems, the governing
physical  processes  must  diverge  from  those of  the  weakest  weak
systems.

It  is  reasonable  to  explore  the  possibility  that  the  putative
different processes  are induced by  environment, such as the  mass of
the  potential well  (galactic  environment) or  the  distance of  the
absorbing gas from the potential center.  The kinematic subsystems are
known to  be associated  with bright, galaxies  with a range  of mass,
including  those  with $L_{K}  \sim  {\rm  few} \times  L_{K}^{\ast}$.
Though the  impact parameters  are $\sim 40$~kpc  or less,  the actual
galactocentric distances  of these weak subsystems are  not known.  On
the  other  hand,  the  majority  of  {\MgII}  absorbers  selected  by
$W_{r}(2796)   <   0.3$~{\AA}   do   {\it  not\/}   have   $L_K   \geq
0.05L_{K}^{\ast}$ galaxies within $\sim 40$~{kpc} (\cite{weakI}); they
are either  further out  in similar halos  or are associated  with the
small halos of dwarf galaxies with $L_{K} \leq 0.05 L_{K}^{\ast}$.

Consider  photoionized  absorbing,  cool--phase  gas clouds  that  are
pressure  confined  in hot,  virialized  gaseous  galactic halos  with
$T_{h} \sim V_{200}^{2} \times  10^{6}$~K, where $V_{200}$ is the halo
circular speed in units of  200~{\kms} (proportional to the total mass
within  the virial  radius).  The  dominant destruction  methods would
include   Rayleigh--Taylor   instability,   cloud--cloud   collisions,
Kelvin--Helmoltz instability,  and evaporation or  condensation (e.g.\
\cite{mo96}).  The  time scales for  these are dependent  upon various
combinations  of  the  ``cloud''  masses, $M_{cl}$,  sizes,  $r_{cl}$,
densities,  $n_{cl}$,   and  temperatures,  $T_{cl}$,   and  upon  the
virialized halo temperature,  $T_{h}$, density, $n_{h}$, and pressure,
$P_{h}$, the latter  two being functions of distance  out in the halo,
$R$.  For the following, we  will assume that $W_{r}(2796)$ is a rough
indicator of cloud size, $r_{cl}$.

Rayleigh--Taylor  instability  sets  in   when  the  drag  force  (ram
pressure)  on a  cloud exceeds  the surface  self--gravitational force
(surface tension).  For  a cloud of fixed mass,  the velocity at which
instability  sets in  varies  as $(n_{cl}/n_{h})^{1/2}  V_{200}^{-3/2}
(M_{cl}/r_{cl})  $,  where $n_{h}$  scales  as  $R^{-1}$ and  $n_{cl}$
scales  with  slightly less  sensitivity,  resulting  in  a very  weak
anti--proportionality with  $R$.  Smaller  clouds are much  favored in
lower mass halos with little dependence upon galactocentric distance.

The  time scale  for cloud--cloud  collisions depends  upon  the cloud
covering  factor, $f_{c}$,  and  scales as  $f_{c}^{-1}RV_{200}^{-1}$.
For  fixed  $f_{c}$, the  destruction  rate  of  clouds by  collisions
decreases at  large galactocentric distances  and is smaller  in lower
mass halos.  If $f_{c}$ decreases  with increasing $R$, the time scale
increases with galactocentric distance  more rapidly than to the first
power.

Pressure  confined clouds  moving at  the sound  speed in  an external
medium are  destroyed by Kelvin--Helmholtz  (hydrodynamic) instability
on  the   order  of  the   cloud  dynamical  time,  which   scales  as
$r_{cl}T^{1/2}_{cl}$  (\cite{murray93}).  Clouds with  lower densities
and smaller sizes  are most likely to be  destroyed by heat conduction
(evaporation)  on a  time  scale very  sensitive  to the  mass of  the
virialized halo.   The evaporation time scales  as $M_{cl} r_{cl}^{-1}
V_{200}^{-5}$   (\cite{cowie-mckee77}).   Mo   \&  Miralda--Escud\'{e}
(1996\nocite{mo96})  have suggested  that,  in cases  where the  cloud
covering factor approaches unity,  cloud creation should balance cloud
disruption  and evaporation.  In  such a  scenario, the  minimum cloud
mass in this equilibrium process will be smaller in environments where
the destruction time scales are  longer for small clouds, which occurs
in smaller mass halos.

The  upshot is  that each cloud  destruction  mechanism favors  longer
lifetimes  for  small  clouds  in  smaller mass  halos  and  at  large
galactocentric distances.   This by no means  explains the differences
in  the equivalent width  distribution of  moderate and  high velocity
subsystems and  weak systems for small $W_{r}(2796)$.   However, it is
suggestive that small $W_{r}(2796)$  clouds are less favored in larger
mass halos.  Of course,  this assumes the ionization conditions and/or
metallicities do not systematically vary between weak systems and weak
subsystems.  As stated above, current evidence is that they do not.

\section{Profile Asymmetries and Kinematics}
\label{sec:model}

One clue  to the nature of  the moderate and  high velocity subsystems
may be the kinematic asymmetries of the overall {\MgII} profiles; {\it
in  virtually  every  system  with  multiple  subsystems,  those  with
moderate  and  high  velocity   are  either  all  blueshifted  or  all
redshifted with  respect to the dominant  subsystem}.  This phenomenon
would  naturally arise if  moderate and  high velocity  subsystems are
small  components in a  {\it systematic  distribution\/} of  gas (both
spatially and  kinematically) that is offset relative  to the dominant
subsystem,  even if  this  structure extends  well beyond  $R(L_{K})$.
Examples are  small clouds within a  Magellanic Stream--like structure
(which  does not  have a  unity  covering factor  around the  Galactic
halo),  or between  the large  column density,  {\HI}  bridges flowing
between M81 and its satellite galaxies (\cite{yun95}).  The systematic
asymmetries would not {\it typically\/} arise, however, from absorbing
gas with both unity covering factor and a random spatial and kinematic
distribution with respect to the dominant kinematic subsystem.

\begin{figure*}[thb]
\figurenum{12}
\plotfiddle{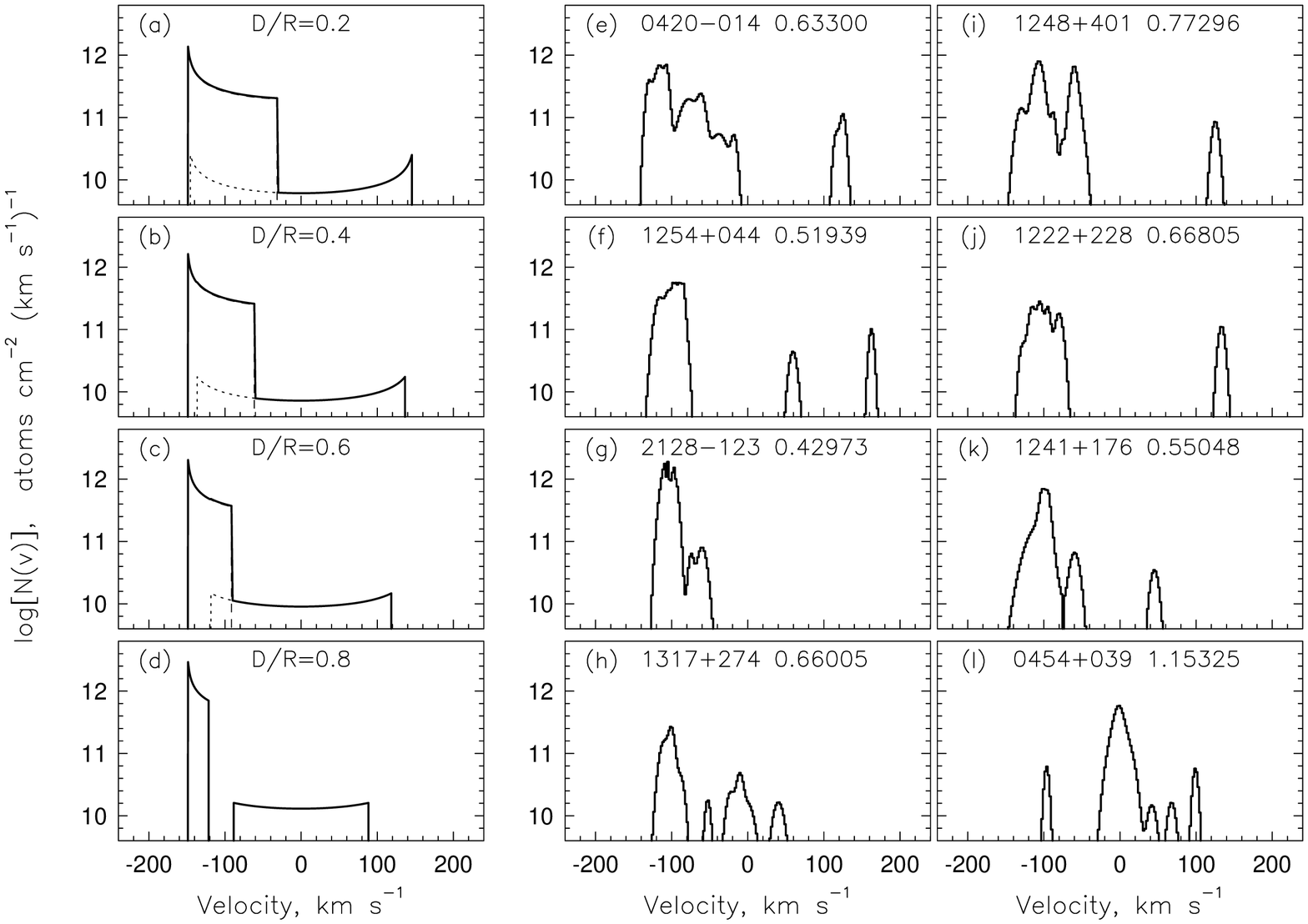}{4.6in}{0.}{70}{70}{-280}{-45}
\caption[fig12.eps] {  (a--d) The  logarithm of the  column density
probability distribution  per unit velocity, $N(v)$  based upon simple
disk/halo models.  The impact parameter, $D$, is given in units of the
absorber radius,  $R$, as labeled  in each panel.  The  dot--dot curve
represents absorption from the halo  and the dash--dash curve from the
disk.  The solid curve is the total column density probability at each
velocity.   The $N(v)$  have  been normalized  to  $\log N({\MgII})  =
13.6$~{\cmsq}  for   the  disk   component  and  $\log   N({\MgII})  =
12.3$~{\cmsq}  for  the  halo  component.  ---  (e--l)  The  logarithm
apparent  column densities  (Equation~\ref{eq:nov}) for  eight  of the
absorbers in the sample.  The  velocities zero points and the signs of
the  velocities have  been adjusted  arbitrarily to  better  match the
probability distributions for display purposes. \label{fig:nov} }
\end{figure*}

To illustrate  a possible spatial--kinematic  relationship between the
dominant subsystem and weak,  moderate to high velocity subsystems, we
examine  a  simplistic  model  of  an edge--on  rotating  disk  and  a
spherical  infalling  halo.    We  use  the  probability  distribution
functions,  $\Phi  (v)$,   for  intercepting  an  absorbing  component
(``cloud'') at  a given  line of sight  velocity through the  disk and
halo  derived  by  Lanzetta  \& Bowen  (1992\nocite{lb92};  see  their
Equations  1--5 and Figures  9--11).  This  model assumes  an $r^{-2}$
spatial  density  profile  for  the  absorbing  components.   

We  have normalized  these probability  distributions such  that $\int
_{-\infty}^{\infty}  \Phi (v) dv  = N_{tot}$,  where $N_{tot}$  is the
integrated {\MgII} column density of the line of sight passing through
the  model structure.   Using  this normalization,  one can  interpret
$\Phi  (v)dv$  as  $N(v)dv$,  the  probable column  density  per  unit
velocity in  the velocity  interval $v$ to  $v+dv$.  We  normalize the
disk   component    to   the   average    {\MgII}   column   densities
(Equation~\ref{eq:na})  of  the   dominant  subsystems  and  the  halo
component to that of the moderate and high velocity subsystems (sample
B).   These  are  $\log  N_{tot}  = 13.6$~{\cmsq}  for  the  disk  and
$12.3$~{\cmsq} for the halo, respectively.

In Figure~\ref{fig:nov}$a$--$d$, we show the logarithm probable column
density velocity distribution for  a combined halo/disk structure with
$v=0$~{\kms}  set  to  the  disk/halo  systemic  velocity.   The  disk
rotation and  halo infall  velocities are set  to $w_0  = 150$~{\kms},
typical of a  sub--$L^{\ast}$ galaxy.  The sense of  the disk rotation
is  chosen toward  the line  of sight.   Each panel  illustrates $\log
N(v)$ for the  ratio of the impact parameter, $D$, to  the size of the
absorbing  structure,  $R$  (the  model  geometry  is  illustrated  in
Figure~9  of  Lanzetta \&  Bowen).  The  dot--dot  curve is  the  halo
absorption and the dash--dash curve is the disk absorption.  The solid
curve is the combined absorption.

The most salient feature of this  model is that the disk absorption is
systematically  offset  from  the   halo  absorption  for  all  impact
parameters; the model  consistently predicts asymmetric profiles.  The
peak  predicted column  density per  unit  velocity from  the disk  is
always larger than that predicted  from the halo.  This is governed by
the normalization, which is tuned to the data.  As impact parameter is
increased, the kinematic spreads of  both the disk and halo components
decrease.  For $D/R > 0.7$, the disk absorption and halo absorption no
longer overlap in velocity space.

\begin{figure*}[thb]
\figurenum{13}
\plotfiddle{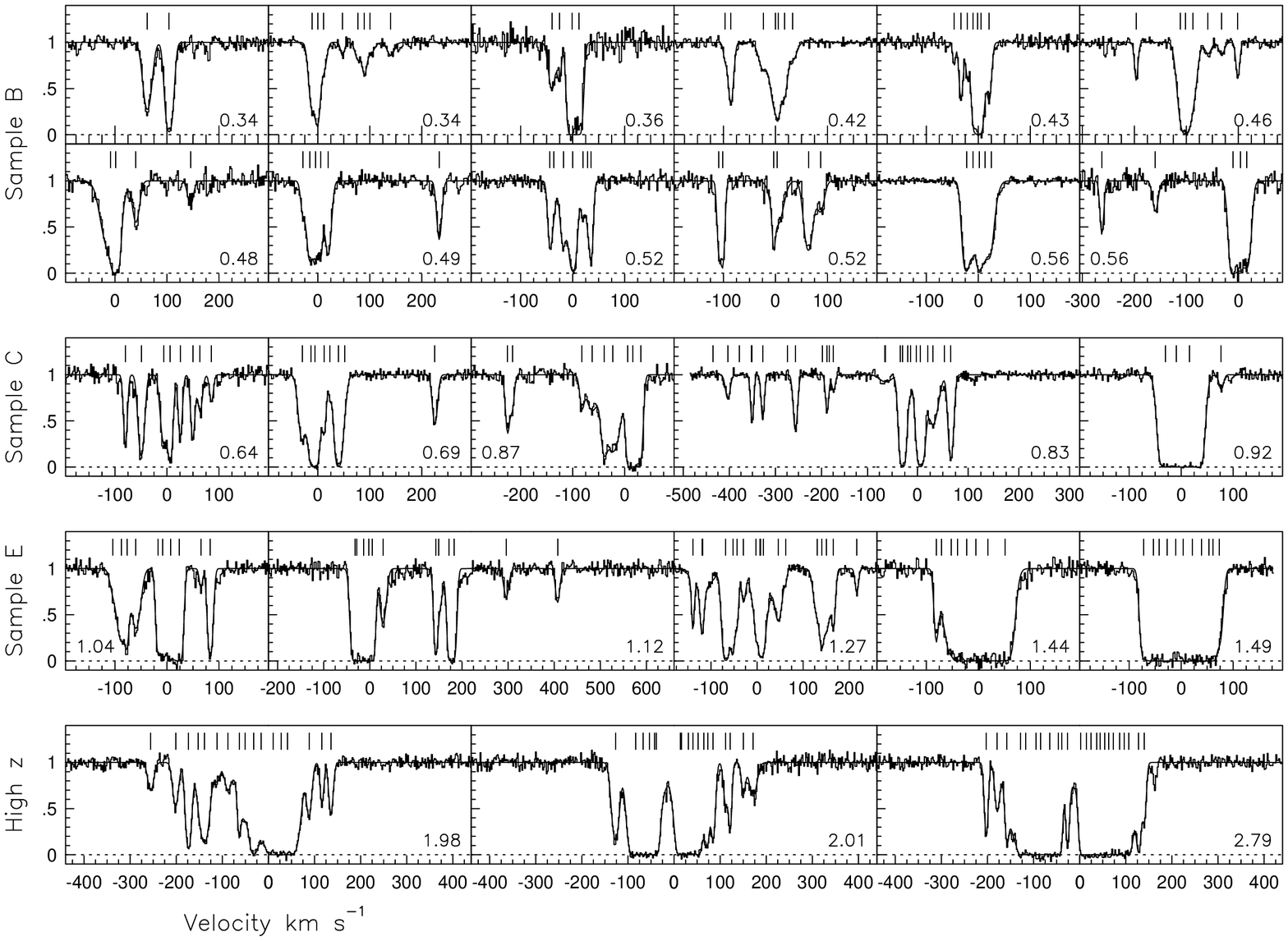}{4.6in}{0.}{70}{70}{-280}{-45}
\caption[fig13.eps] { A gallery  of {\MgII} $\lambda 2796$ profiles
as  a function of  rest--frame velocity.   The profiles  are separated
into subsamples  (B, C,  and E) and  ordered by  increasing equivalent
width.   The rest--frame equivalent  widths are  labeled in  the lower
corners of each subpanel. \label{fig:profiles} }
\end{figure*}

In  Figure~\ref{fig:nov}$e$--$l$,  we  show {\MgII}  apparent  optical
depth  column density  profiles,  $\log N_{a}(v)$,  for  eight of  the
systems  in our  sample that  show strong  asymmetries  (computed from
Equation~\ref{eq:nov}).  These profiles are generated by inverting the
smooth  Voigt  profile  models  for  display  purposes.   The  {\MgII}
$\lambda 2803$ transition is used because it is often not as saturated
in the strong line  cores\footnote{Saturation results in a lower limit
for in the $N_{a}(v)$ profiles.  The exact limit is dependent upon the
uncertainty  in  the flux  values  in the  line  core.   At very  high
signal--to--noise ratio, the limit  is determined by the read noise.}.
The  velocity zero points  and the  sign of  the velocities  have been
adjusted to ease visual comparison with the model distributions.

Qualitatively, the  observed profiles are remarkably  similar to those
predicted by the  simple models of discreet absorbing  components in a
disk/halo   absorbing  structure.   The   disk  absorption   could  be
interpreted to represent the  probability distribution of the dominant
kinematic  subsystem   and  the  halo  absorption   to  represent  the
probability distribution  of the  weak, intermediate to  high velocity
subsystems.

Such  an  interpretation is  consistent  with  the observed  kinematic
asymmetries  {\it  and\/}  provides  a  prediction of  what  parts  of
galaxies  give rise to  the observed  kinematic subsystems.   That is,
many of the  observed profile velocity asymmetries may  be governed by
the  passage of  the  line of  sight  through a  rotating disk.   This
structure would give rise to a dominant kinematic subsystems offset in
velocity from the systemic velocity.  Its covering factor must be near
unity in order to account for the ubiquity of absorption features with
velocity spreads  on the  order of the  disk component of  the models.
The  weak subsystems  at intermediate  to high  velocities  would then
arise in the ``halo'' structure comprised of smaller absorbing clouds.
This structure  would have a somewhat  broader, symmetric distribution
of velocities.   Statistically, this  ``halo'' structure would  have a
covering factor  near unity, based  upon 24 moderate to  high velocity
subsystems observed in 23  systems.  However, only 13 systems actually
have these  subsystems.  

If disks play a role in governing the kinematics of {\MgII} absorbers,
the disk orientation  is going to give rise  to variations in location
of  the disk  component of  the model  velocity  distribution (recall,
those presented are for edge--on disks only).  In the case of face--on
orientations,  the  velocity  width  of  the disk  component  will  be
proportional to  the velocity dispersion  of the gas  perpendicular to
the disk (see Charlton \& Churchill 1998\nocite{kinematicpaper}).  The
disk systemic velocity will be  centered at $\sim v=0$~{\kms} with the
velocities  of any ``halo''  structure distributed  symmetrically.  In
Figure~\ref{fig:nov}$l$,  we   present  a  possible   example  of  the
kinematics expected from a face--on disk orientation.

This absorber  is the  only one with  weak subsystems to  {\it both\/}
sides  of   the  the   dominant  subsystem.   At   $z=1.15325$  toward
PKS~$0454+039$  (Figure~\ref{fig:data}$f$), the candidate  absorber is
difficult  to resolve  in an  WFPC2/{\it  HST\/} image  from the  {\it
HST\/}  archive  (also  see  Le~Brun  \etal  1997\nocite{lebrun_dla}).
However, there is an example  of a $D=13.2$~kpc, face--on, disk galaxy
(see  Figure~1 in  Churchill 2000\nocite{vla2K})  associated  with the
$z=0.55336$  absorber  toward Q~$1148+384$  (Figure~\ref{fig:data}$k$)
that has a central  dominant absorption component.  Though this system
technically  is  a  single  kinematic subsystem,  several  clouds  are
present  whose   column  densities  decrease   as  subsystem  velocity
separations  from the central  component increase.   Overall, however,
profiles  in  which  the  weaker,  and/or moderate  to  high  velocity
subsystems are distributed  symmetrically about the dominant subsystem
are rare in our sample.

\section{What is Evolving?}
\label{sec:evolving}

The   differential   evolution  in   the   {\MgII}  equivalent   width
distribution  has  been  determined  by measuring  the  redshift  path
density, $dN/dz$, for samples with different minimum equivalent widths
(see  SS92).   Absorbers with  $W_{r}(2796)  >  0.3$~{\AA} exhibit  no
cosmological  evolution.   However, as  the  minimum $W_{r}(2796)$  is
increased,  both the  mean  redshift of  the  sample and  the rate  of
evolution (slope  of the  redshift path density)  increase, indicating
that the strongest  systems evolve away with cosmic  time and that the
largest systems evolve the most rapidly.

Though we  have quantified the {\MgII} absorption  properties from the
profiles, we point out that  qualitative visual inspection can be more
revealing  than  parameterized  representations,  especially  for  the
profile ``shapes'',  or kinematic  morphologies.  We have  divided our
data into samples B, C, and  E using the same minimum equivalent width
cut  offs employed  by  SS92,  i.e.\ greater  than  $0.3$, $0.6$,  and
$1.0$~{\AA},   respectively.    In   the   upper   three   panels   of
Figure~\ref{fig:profiles},  we illustrate  the {\MgII}  $\lambda 2796$
profiles in increasing equivalent width order, with the data separated
into these  three samples.  The profiles are  presented in rest--frame
velocity  with  the  flux   normalized  and  VP  models  superimposed,
including ticks  that indicate the  number of VP components  and their
velocities.   In the  lower corner  of each  subpanel,  the equivalent
width is given in angstroms.

The mean kinematic spread of the {\it dominant subsystems\/} increases
along the sequence  from sample B to C to E.   This is not surprising,
given that the dominant  subsystems of larger equivalent width systems
are  often saturated  over a  fair fraction  of their  velocity range.
This is reflected  in the trend for the  largest $W_{r}(2796)$ systems
to have only lower limits  on the integrated apparent column densities
of their  dominant subsystems (see  Figure~\ref{fig:Naod}).  Thus, the
equivalent width is directly proportional to the velocity width of the
dominant subsystems,  which would  not hold true  if the  moderate and
high  velocity subsystems  were  often saturated.   This could  imply,
then,  that  the  differential   evolution  in  the  equivalent  width
distribution  is primarily  due  to kinematic  evolution (decrease  in
velocity spread) of  the dominant subsystems.  However, this  is not a
unique interpretation.

In the bottom panel  of Figure~\ref{fig:profiles}, we present examples
of  higher redshift  systems as  representatives of  the  very largest
$W_{r}(2796)$   absorbers\footnote{These  systems   are   part  of   a
complementary, targeted  survey of  the higher redshift,  very largest
equivalent width  {\MgII} absorbers.}.   Each has an  equivalent width
consistent  with or greater  than $2.0$~{\AA}  and a  redshift greater
than  $z=1.3$.   If  these   profiles  are  representative  at  higher
redshift, then  they indicate  that large, ``double''  optically thick
subsystems  are  characteristic  of  systems  that  evolve  away  most
rapidly.  It has  been shown, as one interpretation,  that the {\MgII}
evolution  rate  is consistent  with  the  evolution  of galaxy  pairs
(\cite{archiveII}), assuming that each saturated region of the {\MgII}
profiles arise from a separate galaxy,

However, and  this cannot be addressed  by the current  data, there is
the  question  of  whether   the  weak,  moderate  and  high  velocity
subsystems are  vestiges of what is  evolving.  If they  are, then the
disappearance  of  multiple, optically  thick  subsystems, like  those
shown  in  Figure~\ref{fig:profiles},  could  be explained  as  either
ionization  or structure evolution  into the  weak, moderate  and high
velocity   subsystems.   Incorporating   knowledge   of  the   profile
asymmetries discussed in \S~\ref{sec:model}, this would imply that the
{\MgII}  evolution  is  kinematically  asymmetric.  In  the  disk/halo
scenario outlined above for connecting galaxy structure to the profile
asymmetries, such evolution  would imply a reduction in  the number of
``halo'' absorbing components, or clouds. 

\section{Conclusion}
\label{sec:summary}

We  have presented  the {\MgIIdblt}  absorption profiles,  resolved at
$\simeq  6$~{\kms} with  HIRES/Keck~I, for  23 quasar  absorption line
systems  having $W_{r}(2796)  >  0.3$~{\AA}.  We  have also  presented
{\FeII}, and {\MgI} profiles  when the transitions were detected.  The
overall  absorption properties  of  many of  these  systems have  been
previously  investigated  (\cite{archiveI}).   

The  complex  kinematics were  parameterized  by defining  ``kinematic
subsystems''  and by  Voigt profile  fitting.  We  have  discussed the
absorber properties, including the distributions of equivalent widths,
subsystem to subsystem column densities, and velocity separations (see
\S~\ref{sec:results}).  A companion  paper (\cite{cvc00}) will present
analysis of  the Voigt  profile results.  The  data were  divided into
three samples  by equivalent width  range.  We found that  there are
systematic differences,  or trends, in the  kinematics with increasing
equivalent  width, and  that this  may provide  clues to  the observed
differential evolution in the equivalent width distribution of {\MgII}
absorbers (e.g.\ SS92).

The  properties of  the ``moderate''  and ``high''  velocity kinematic
subsystems have been  compared to those of the  Galactic high velocity
clouds (HVCs) and to those  of the population of single ``cloud'' weak
systems.   We  have  inferred  that  the moderate  and  high  velocity
subsystems are not higher redshift analogues to HVCs primarily because
of  the  low  {\HI}  column  densities  in  the  former.   Though  the
subsystems and  weak systems are  comparable in their  inferred sizes,
metallicities,  and   ionization  conditions,  the   turnover  in  the
subsystem   equivalent  width   distribution   at  $W_{r}(2796)   \sim
0.08$~{\AA} would indicate that they reside in different environments.
We discussed the time scales  for cloud destruction and point out that
the smallest clouds  are most favored to survive  in the smallest mass
halos and at the largest galactocentric distances.

Direct knowledge of the  kinematic evolution of {\MgII} absorbers will
require  an  unbiased,  higher   redshift  data  set  with  HIRES/Keck
resolution.   It is  likely that  a clear  understanding  will require
charting the  kinematics of the  higher ionization phases (as  seen in
{\CIV} absorption, for example)  in order to understand the multiphase
structure  of the  gas.  It  is in  the relative  kinematics  of these
ionization phases that  the energetics and gas phase  structure can be
employed to  infer the actual  mechanisms (i.e.\ star  formation, pair
evolution,  intergalactic medium  infall,  chemical enrichment,  etc.)
that drive the differential  evolution in the {\MgII} equivalent width
distribution.

\acknowledgements  

Support for this work was provided by the NSF (AST--9617185), and NASA
(NAG 5--6399),  California Space Institute (CS--194),  and a Sigma--Xi
Grants in Aid of Research.  We thank M.~Keane for assistance with July
1994 observations and discussions very  early in the project.  We have
enjoyed and  learned from countless  discussions during the  course of
this work,  most notably  with J.~Bergeron, M.~Bolte,  J.~Charlton, M.
Dickinson,    F.~Drake,    S.~Faber,   R.~Guhathakurta,    R.~Ganguly,
J.~Hibbard,  B.~Jannuzi,  E.~Jenkins,  M.~Keane,  K.~Lanzetta,  L.~Lu,
P.~Petitjean,  M.~Pettini, J.~Rigby, J.~Rachen,  M.~Rauch, W.~Sargent,
B.~Savage, D.~Schneider,  K.~Sembach, G.~Smith, C.~Steidel, F.~Timmes,
D.~York,  and  D.~Zaritsky.   We  thank J.~Charlton  for  reading  and
commenting on the draft of this manuscript.

\
\appendix
\section{Formalism of Measured Quantities}
\label{app:A}

Here, we  outline the mathematical formalism  for measuring absorption
properties  directly  from the  data  themselves.   We computed  these
quantities and  their measurement errors (uncertainties)  for both the
overall  systems   and  each  of  their   kinematic  subsystems.   The
computational   formalism,   including   careful  treatment   of   the
uncertainties,  has been  presented  in detail  in  the appendices  of
Sembach   \&   Savage   (1992\nocite{semsav92})   and   in   Churchill
(1997\nocite{thesis}),  where  a few  corrections  to  the Sembach  \&
Savage derivations have been noted.

\subsection{Systemic Redshifts}
\label{sec:zabs}
 
The systemic redshift  of each absorbing system is  used to define the
velocity zero point.  We determined each absorber's systemic redshift,
$z_{\rm abs}$,  from the  optical depth mean  of the  {\MgII} $\lambda
2796$ profiles.  We solved the implicit equation,
\begin{equation}
\int _{\lambda_{b}}^{\lambda_{\rm sys}} \tau_{a}(\lambda)d\lambda =
\int _{\lambda_{\rm sys}}^{\lambda_{r}} \tau_{a}(\lambda)d\lambda ,
\label{eq:zabs}
\end{equation}
for $\lambda _{\rm sys}$, the observed wavelength at the optical depth
median, where $\lambda_{b}$ and $\lambda_{r}$ are the extreme blueward
and redward wavelengths of the overall system, respectively, and $\tau
_{a}(\lambda)$ is the apparent optical depth (\cite{savage_aod}),
\begin{equation}
\tau_{a}(\lambda)   =  \ln   \left[  \frac{I_{c}(\lambda)}{I(\lambda)}
\right] .
\label{eq:aod}
\end{equation}

The  optical depth  median, as  opposed to  the flux  median [i.e.~$\{
1-I(\lambda)/I_{c}(\lambda)\}  $],  is  used  because it  weights  the
absorption by  the {\it quantity  of gas}, providing a  column density
weighted median.  Where the profiles are saturation, i.e.\ $I(\lambda)
\le | \sigma  _{I}(\lambda) |$, the apparent optical  depth is a lower
limit, given by  $\tau_{a}(\lambda) = \ln [ I_{c}(\lambda)  / | \sigma
_{I}(\lambda)  | ]$.  Between  kinematic subsystems,  $\tau_{a} \equiv
0$, in order to eliminate continuum noise.

The  absorber   redshift  is  then   obtained  from  $z_{\rm   abs}  =
\lambda_{\rm  sys}/2796.352   -  1$.   The   {\it  formal\/}  redshift
uncertainities are  typically $(1-5) \times 10^{-6}$,  less than $\sim
1$~{\kms} (co--moving).

\subsection{Equivalent Widths}
\label{sec:ews}
 
The standard expression for the observed equivalent width is 
\begin{equation}
W = 
            \int _{\lambda_{b}}^{\lambda_{r}}
            \left[
              1 - \frac{I(\lambda)}{I_{c}(\lambda)}
            \right] d\lambda .
\label{eq:ew}
\end{equation}
The integration is performed for each kinematic subsystem and then the
total system equivalent  width is obtained by the  sum.  The limits of
integration,  $\lambda _{b}$  and  $\lambda _{r}$,  are determined  as
described in \S~\ref{sec:subsystems}.

\subsection{Subsystem Velocities}
\label{sec:meanv}
 
Each kinematic subsystem  has a non--zero velocity.   As with the
absorber systemic  redshifts, we compute the velocities from the
apparent optical depth across the {\MgII} $\lambda 2796$ profile.  The
velocity, $\left< v \right>$, is the first velocity moment of the
optical depth,
\begin{equation}
\left< v \right>   =    \frac{\tau_{a}^{(1)}}{\tau_{a}} 
  =  \frac {\int _{v_{b}}^{v_{r}} 
            \tau_{a}(v) v dv }
           {\int _{v_{b}}^{v_{r}} 
            \tau_{a}(v) dv} 
\label{eq:vbardef}
\end{equation}
where  $v=  (c/\lambda _{\rm  sys})  \cdot  (\lambda  - \lambda  _{\rm
sys})$,  and where the  integration is  performed over  the wavelength
region defining each kinematic subsystem.

\subsection{Velocity Widths}
\label{sec:vwidths}
 
The velocity width of a kinematic subsystem, $\omega _{v}$, is defined
by the second moment of the apparent optical depth,
\begin{equation}
\omega _{v}^{2}   =
    \frac{\tau_{a}^{(2)}}{\tau_{a}} 
=   \frac { \int _{v_{b}}^{v_{r}} 
          \tau _{a}(v) \left[ v - \left< v \right> \right] ^{2}  dv }
    { \int _{v_{b}}^{v_{r}} 
          \tau _{a}(v) dv} .
\label{eq:omegavdef}
\end{equation}
This  quantity is  the  equivalent Gaussian  width  of the  optical
depth profile.

\subsection{Apparent Column Densities}
\label{sec:aoddef}

The  ``apparent  optical  depth''   technique  of  Savage  \&  Sembach
(1991\nocite{savage_aod})  was  employed  to  compute  column  density
profiles, $N(v)$~[atoms~{\cmsq}], from
\begin{equation}
N_{a}(v) =  \frac{m_{e}c}{\pi e^{2}}
           \frac{\tau_{a}(v)}{f\lambda_{\rm o}} ,
\label{eq:nov}
\end{equation}
where fundamental constants have  the usual meanings, and the apparent
optical  depth  is  given  by Equation~\ref{eq:aod}.   The  $N_{a}(v)$
profiles   of  doublets,   or  other   same--ion   transitions  having
$f\lambda_{\rm  o}$ differing  by  a factor  of  two, can  be used  to
discern the  velocity location of unresolved  saturation in absorption
profiles.

Savage \&  Sembach \nocite{savage_aod} have shown  that this technique
yields accurate  column densities when  the width of  the instrumental
spread function does  not exceed the {\it intrinsic\/}  line widths by
more  than  a factor  of  a few.   No  corrections  [also see  Jenkins
(1997\nocite{jenkins})] have been applied to the measured $N_{a}(v)$.

The integrated  apparent column  density of a  system or  subsystem is
given by
\begin{equation}
N_{a} = \int _{v_{b}}^{v_{r}} N_{a}(v) dv , 
\label{eq:na} 
\end{equation}
where $v_{b}$  and $v_{r}$ are the  velocity extrema of  the system or
kinematic subsystem being measured.

For adopting the ``best value'' of $N_{a}$ for an ion, there are three
possibilities: (1)~All  transitions of an ion  exhibit some saturation
between  $v_{b}$ and  $v_{r}$, so  that $N_{a}(v)$  is a  lower limit.
Often, the  transition with the smallest $f\lambda$  provides the best
constraints on the  lower limit; (2)~All but one  transition of an ion
exhibits  saturation, in  which case,  the adopted  column  density is
taken from  the unsaturated transition;  and (3)~All or more  than one
transition of  an ion are unsaturated,  providing multiple independent
measurements.   The  adopted  value  is computed  from  the  optimally
weighted mean [i.e.\ weighted by $1/\sigma ^{2}(N_{a})$].

\subsection{Voigt Profile Decomposition}
\label{sec:vpfit}

Voigt  profile  (VP) decomposition  is  an  established technique  for
modeling   absorption  profiles   because  it   is  well   suited  for
parameterizing the absorbing gas properties into physically meaningful
quantities,   i.e.\   the  number   of   components  (clouds),   their
line--of--sight   velocities,  column   densities,  and   Doppler  $b$
parameters.

When  an absorption  line  having optical  depth  $\tau (\lambda)$  is
recorded by  an instrument having  a finite resolution defined  by its
instrumental  spread  function   (ISF),  $\Phi(\delta  \lambda)$,  the
measured spectrum is the convolution of the intrinsic spectrum and the
ISF,
\begin{equation}
I_{\rm obs}(\lambda) = \Phi(\delta \lambda) \otimes
\left\{ I_{\rm o}(\lambda)
\exp \left[ - \tau(\lambda) \right] \right\} ,
\label{eq:Iobs}
\end{equation}
where $I_{\rm o}(\lambda)$ is the emitted intensity, and $I_{\rm
obs}(\lambda)$ is the observed intensity.

The  optical depth  is the  product of  the line  of  sight integrated
column density,  $N= \int  n(l)dl$~{\cmsq}, and the  atomic absorption
coefficient, $\alpha(\lambda)$, and is given by
\begin{equation}
\tau(\lambda) = N\alpha(\lambda)
              = \frac{\sqrt{\pi}e^{2}}{m_{e}c^{2}}
                \frac{Nf\lambda_{\rm o}^{2}}{\Delta _{\lambda}}
                u(x,y) ,
\label{eq:taulambda}
\end{equation}
where  $u(x,y)$ is  the Voigt  function evaluated  at $  x  = (\lambda
-\lambda   _{\rm  o})/\Delta   _{\lambda}  $   and   $y=  \lambda_{\rm
o}^{2}\Gamma/4\pi  c \Delta  _{\lambda}$,  and where  $\Gamma$ is  the
transition  damping constant  (see  Table~\ref{tab:ions}) and  $\Delta
_{\lambda}$  is the Doppler  width.  This  formalism assumes  that the
dominant line  broadening mechanism is  a Lorentzian convolved  with a
Gaussian (the  former being  the intrinsic line  width and  the latter
being due  to line--of--sight thermal  and/or bulk motions).   For the
computation  of the  Voigt  function, we  used  the real  part of  the
complex probability function,
\begin{eqnarray}
w(z) & = & e^{-z^{2}} \left[ 1 + \frac{2i}{\sqrt{\pi}} \int _{0}^{z}
          e^{-t^{2}} dt \right] \\ 
     & = & e^{-z^{2}} {\rm erfc} (-iz) \nonumber \\ 
    & = & u(x,y) + iv(x,y),  \nonumber
\end{eqnarray}
which  is accurately computed  using the  very efficient  routine {\sc
Cpf12} (\cite{humlicek79}).

Thus, for each VP, there are three free parameters, $N$, $\lambda_{\rm
o}$, and $\Delta _{\lambda}$. When a systems is a complex blend of
multiple VP components, the optical depth used in
Equation~\ref{eq:Iobs} is the sum of the individual components,
\begin{equation}
\tau(\lambda) = \sum _{i=1}^{N_{\rm vp}} \tau_{i}(\lambda),
\end{equation}
where $\tau_{i}(\lambda)$ is given by Equation~\ref{eq:taulambda}, and
the  number of free parameters becomes $3N_{\rm vp}$.

The component velocities of  all transitions are ``tied'' together, as
are  the column  densities of  all transitions  of a  given  ion.  The
column density  of each ion is allowed  to vary freely as  are the $b$
parameters.   For  the  statistical  modeling,  we  use  a  $\chi^{2}$
minimization  with   a  code  of   our  own  design,   {\sc  Minfit\/}
(\cite{thesis}),  which uses  the least  squares  minimizing algorithm
{\sc Dnlse1} (\cite{more78}) and  supporting {\sc Slatec} library made
publicly available by {\it netlib.org}.





\begingroup
\small

\begin{deluxetable}{llrrrcc}
\tablenum{3}
\tablewidth{387pt}
\tablecolumns{7}
\tablecaption{Kinematic Subsystem Properties and System Totals}
\tablehead
{
\colhead{QSO} &
\colhead{$z_{\rm abs}$} &
\colhead{System} &
\colhead{$\left< v \right>$} &
\colhead{$\omega$} &
\colhead{$W_{\rm r}(2796)$} &
\colhead{DR} \\
 &
 &
 &
\colhead{[{\kms}]} &
\colhead{[{\kms}]} &
\colhead{[{\AA}]} &
}
\startdata
Q0002+051 & 0.85139 & 1 & $-12.0\pm  0.4$ & $ 17.8\pm  0.2$ & $0.658\pm0.005$ & $1.06\pm0.01$ \nl
 & & 2 & $170.5\pm  0.3$ & $ 14.5\pm  0.3$ & $0.379\pm0.006$ & $1.13\pm0.03$ \nl
 & & 3 & $296.6\pm  1.0$ & $  5.2\pm  1.0$ & $0.039\pm0.004$ & $1.63\pm0.33$ \nl
 & & 4 & $408.6\pm  0.9$ & $  4.8\pm  0.9$ & $0.043\pm0.004$ & $1.32\pm0.22$ \nl
 & & Total & \nodata & $ 97.7\pm  0.9$ & $1.120\pm0.010$ & $1.10\pm0.01$ \nl
\multicolumn{7}{c}{ } \nl
Q0117+213 & 0.57640 & Total & \nodata & $ 27.1\pm  0.5$ & $0.922\pm0.006$ & $1.08\pm0.01$ \nl
\multicolumn{7}{c}{ } \nl
Q0117+213 & 1.04797 & 1 & $-86.6\pm  0.3$ & $  6.1\pm  0.4$ & $0.097\pm0.003$ & $1.80\pm0.09$ \nl
 & & 2 & $  3.5\pm  0.3$ & $ 15.7\pm  0.3$ & $0.321\pm0.004$ & $1.58\pm0.03$ \nl
 & & Total & \nodata & $ 39.4\pm  0.3$ & $0.417\pm0.005$ & $1.63\pm0.03$ \nl
\multicolumn{7}{c}{ } \nl
Q0420-014 & 0.63300 & 1 & $-222.4\pm  0.6$ & $  6.0\pm  0.7$ & $0.089\pm0.005$ & $1.50\pm0.14$ \nl
 & & 2 & $ -6.4\pm  1.2$ & $ 31.0\pm  0.4$ & $0.783\pm0.009$ & $1.30\pm0.02$ \nl
 & & Total & \nodata & $ 61.1\pm  1.4$ & $0.873\pm0.010$ & $1.32\pm0.02$ \nl
\multicolumn{7}{c}{ } \nl
Q0454+039 & 0.85957 & Total & \nodata & $ 45.9\pm  0.6$ & $1.494\pm0.009$ & $1.03\pm0.01$ \nl
\multicolumn{7}{c}{ } \nl
Q0454+039 & 1.15325 & 1 & $-96.1\pm  0.6$ & $  2.8\pm  1.0$ & $0.032\pm0.003$ & $1.72\pm0.25$ \nl
 & & 2 & $  2.5\pm  0.3$ & $ 15.9\pm  0.5$ & $0.389\pm0.005$ & $1.39\pm0.03$ \nl
 & & 3 & $ 99.2\pm  0.6$ & $  4.2\pm  0.8$ & $0.037\pm0.003$ & $2.20\pm0.33$ \nl
 & & Total & \nodata & $ 31.8\pm  0.8$ & $0.458\pm0.006$ & $1.45\pm0.03$ \nl
\multicolumn{7}{c}{ } \nl
Q0454-220 & 0.47441 & Total & \nodata & $ 42.3\pm  0.7$ & $1.439\pm0.013$ & $1.14\pm0.01$ \nl
\multicolumn{7}{c}{ } \nl
Q0454-220 & 0.48334 & Total & \nodata & $ 15.8\pm  0.5$ & $0.426\pm0.006$ & $1.34\pm0.03$ \nl
\multicolumn{7}{c}{ } \nl
Q0823-223 & 0.91102\tablenotemark{a} & 1 & $-36.0\pm  0.8$ & $ 50.8\pm  0.3$ & $0.940\pm0.005$ & $1.40\pm0.02$ \nl
 & & 2 & $146.1\pm  0.2$ & $ 13.6\pm  0.2$ & $0.304\pm0.004$ & \nodata \nl
 & & 3 & $214.3\pm  0.9$ & $  5.0\pm  1.0$ & $0.024\pm0.003$ & \nodata \nl
 & & Total & \nodata & $ 89.4\pm  0.6$ & $1.268\pm0.007$ & \nodata \nl
\multicolumn{7}{c}{ } \nl
Q1101-264 & 0.35900 & Total & \nodata & $ 24.6\pm  0.5$ & $0.550\pm0.006$ & $1.31\pm0.02$ \nl
\multicolumn{7}{c}{ } \nl
Q1148+384 & 0.55336 & Total & \nodata & $ 45.5\pm  0.8$ & $0.642\pm0.011$ & $1.74\pm0.06$ \nl
\multicolumn{7}{c}{ } \nl
Q1206+459 & 0.92760 & 1 & $-404.9\pm  0.7$ & $  5.0\pm  0.7$ & $0.032\pm0.002$ & $1.82\pm0.27$ \nl
 & & 2 & $-339.6\pm  0.6$ & $ 12.5\pm  0.4$ & $0.083\pm0.003$ & $1.66\pm0.11$ \nl
 & & 3 & $-257.3\pm  0.3$ & $  3.9\pm  0.4$ & $0.064\pm0.002$ & $1.71\pm0.12$ \nl
 & & 4 & $-183.8\pm  0.5$ & $  6.8\pm  0.4$ & $0.049\pm0.003$ & $2.15\pm0.24$ \nl
 & & 5 & $  6.2\pm  0.9$ & $ 34.2\pm  0.5$ & $0.656\pm0.005$ & $1.31\pm0.01$ \nl
 & & Total & \nodata & $117.5\pm  1.8$ & $0.883\pm0.007$ & $1.41\pm0.02$ \nl
\tablebreak
\multicolumn{7}{c}{ } \nl
Q1222+228 & 0.66805 & 1 & $ -1.7\pm  0.5$ & $ 15.4\pm  0.3$ & $0.408\pm0.007$ & $1.46\pm0.03$ \nl
 & & 2 & $232.2\pm  0.5$ & $  6.7\pm  0.5$ & $0.081\pm0.004$ & $1.48\pm0.12$ \nl
 & & Total & \nodata & $ 76.9\pm  1.8$ & $0.489\pm0.008$ & $1.46\pm0.03$ \nl
\multicolumn{7}{c}{ } \nl
Q1241+176 & 0.55048 & 1 & $  0.7\pm  0.6$ & $ 18.9\pm  0.7$ & $0.444\pm0.009$ & $1.30\pm0.04$ \nl
 & & 2 & $147.9\pm  1.1$ & $  6.6\pm  0.7$ & $0.035\pm0.004$ & $1.41\pm0.29$ \nl
 & & Total & \nodata & $ 33.0\pm  1.7$ & $0.479\pm0.010$ & $1.31\pm0.04$ \nl
\multicolumn{7}{c}{ } \nl
Q1248+401 & 0.77296 & 1 & $  7.1\pm  0.7$ & $ 25.9\pm  0.3$ & $0.631\pm0.004$ & $1.26\pm0.01$ \nl
 & & 2 & $226.3\pm  0.6$ & $  5.6\pm  0.8$ & $0.064\pm0.003$ & $1.42\pm0.12$ \nl
 & & Total & \nodata & $ 54.5\pm  1.1$ & $0.694\pm0.005$ & $1.27\pm0.02$ \nl
\multicolumn{7}{c}{ } \nl
Q1254+044 & 0.51939\tablenotemark{b} &  1 & $-261.1\pm  1.0$ & $  6.5\pm  0.9$ & $0.062\pm0.006$ & $2.63\pm0.64$ \nl
 & & 2 & $-159.0\pm  1.1$ & $  6.2\pm  0.8$ & $0.048\pm0.006$ & \nodata \nl
 & & 3 & $   1.9\pm  0.8$ & $ 13.9\pm  0.4$ & $0.446\pm0.009$ & $1.17\pm0.04$ \nl
 & & Total & \nodata & $ 87.5\pm  4.7$ & $0.556\pm0.013$ & $1.38\pm0.05$ \nl
\multicolumn{7}{c}{ } \nl
Q1254+044 & 0.93423 & Total & \nodata & $ 21.2\pm  0.3$ & $0.341\pm0.005$ & $1.51\pm0.04$ \nl
\multicolumn{7}{c}{ } \nl
Q1317+274 & 0.66005\tablenotemark{b} & 1 & $ -3.2\pm  0.2$ & $  9.1\pm  0.2$ & $0.226\pm0.003$ & $1.49\pm0.04$ \nl
 & & 2 & $ 45.2\pm  1.2$ & $  4.7\pm  1.1$ & $0.014\pm0.002$ & \nodata \nl
 & & 3 & $ 89.3\pm  0.7$ & $ 10.8\pm  0.7$ & $0.078\pm0.003$ & $1.67\pm0.16$ \nl
 & & 4 & $143.2\pm  1.1$ & $  8.0\pm  0.7$ & $0.026\pm0.003$ & \nodata \nl
 & & Total & \nodata & $ 52.0\pm  0.6$ & $0.344\pm0.006$ & $1.68\pm0.06$ \nl
\multicolumn{7}{c}{ } \nl
Q1421+331 & 0.90287 & 1 & $-19.9\pm  1.4$ & $ 41.4\pm  0.6$ & $0.885\pm0.008$ & $1.32\pm0.02$ \nl
 & & 2 & $ 82.4\pm  0.4$ & $  8.4\pm  0.7$ & $0.156\pm0.005$ & $1.29\pm0.07$ \nl
 & & Total & \nodata & $ 51.4\pm  1.0$ & $1.041\pm0.010$ & $1.32\pm0.02$ \nl
\multicolumn{7}{c}{ } \nl
Q1421+331 & 1.17261 & Total & \nodata & $ 24.1\pm  0.5$ & $0.516\pm0.006$ & $1.40\pm0.03$ \nl
\multicolumn{7}{c}{ } \nl
Q1634+706 & 0.99024 & Total & \nodata & $ 17.9\pm  0.1$ & $0.561\pm0.003$ & $1.33\pm0.01$ \nl
\multicolumn{7}{c}{ } \nl
Q2128-123 & 0.42973 & Total & \nodata & $ 16.5\pm  0.4$ & $0.355\pm0.008$ & $1.41\pm0.06$ \nl
\multicolumn{7}{c}{ } \nl
Q2145+064 & 0.79078 & 1 & $-105.1\pm  0.3$ & $  6.6\pm  0.3$ & $0.158\pm0.004$ & $1.33\pm0.05$ \nl
 & & 2 & $  0.9\pm  0.5$ & $  9.7\pm  0.5$ & $0.152\pm0.005$ & $1.60\pm0.10$ \nl
 & & 3 & $ 66.8\pm  0.5$ & $ 13.1\pm  0.4$ & $0.208\pm0.005$ & $1.51\pm0.07$ \nl
 & & Total & \nodata & $ 74.7\pm  0.4$ & $0.517\pm0.008$ & $1.47\pm0.04$ \nl
\enddata
\tablenotetext{a}{The $\lambda 2803$ transition was only partially
covered by the CCD.}
\tablenotetext{b}{The $\lambda 2803$ transition was not formally
detected in some subsystems.  However, the subsystems are deemed real based
upon VP decomposition.}
\label{tab:absprops}
\end{deluxetable}


\endgroup

\begingroup
\small

\begin{deluxetable}{lccccc}
\tablenum{4}
\tablewidth{0pc}
\tablecolumns{6}
\tablecaption{Samples and Subsamples}
\tablehead
{
\colhead{ } &
\colhead{Sample A} &
\colhead{Sample B} &
\colhead{Sample C} &
\colhead{Sample D} &
\colhead{Sample E} \\
\colhead{ } &
\colhead{$W_{r} \geq 0.3$~{\AA}} &
\colhead{$0.3 \leq W_{r} < 0.6$~{\AA}} & 
\colhead{$0.6 \leq  W_{r} < 1.0$~{\AA}} &
\colhead{$W_{r} \geq 0.6$~{\AA}} &
\colhead{$W_{r} \geq 1.0$~{\AA}} 
 }
\startdata
$N$ & $23$ & $13$ & $5$ & $10$ & $5$ \nl
$\left< W_{r}(2796) \right>$, {\AA} & $0.56$ & $0.48$ & $0.80$ & $0.98$ & $1.04$\nl
$\left< z \right>$ & $0.77$ & $0.66$ & $0.69$ & $0.81$ & $0.86$\nl
$\left< DR \right>$ & $1.4$ & $1.4$ & $1.4$ & $1.3$ & $1.1$\nl
\cutinhead{Membership Chart}
$0002+051~~~0.85139$ & X &   &   & X & X \nl
$0117+213~~~0.57640$ & X &   & X & X &   \nl
$0117+213~~~1.04797$ & X & X &   &   &   \nl
$0420-014~~~0.63300$ & X &   & X & X &   \nl
$0454+039~~~0.85957$ & X &   &   & X & X \nl
$0454+039~~~1.15325$ & X & X &   &   &   \nl
$0454-220~~~0.47441$ & X &   &   & X & X \nl
$0454-220~~~0.48334$ & X & X &   &   &   \nl
$0823-223~~~0.91102$ & X &   &   & X & X \nl
$1101-264~~~0.35900$ & X & X &   &   &   \nl
$1148+384~~~0.55336$ & X &   & X & X &   \nl
$1206+459~~~0.92760$ & X &   & X & X &   \nl
$1222+228~~~0.66805$ & X & X &   &   &   \nl
$1241+176~~~0.55048$ & X & X &   &   &   \nl
$1248+401~~~0.77296$ & X &   & X & X &   \nl
$1254+044~~~0.51939$ & X & X &   &   &   \nl
$1254+044~~~0.93423$ & X & X &   &   &   \nl
$1317+274~~~0.66005$ & X & X &   &   &   \nl
$1421+331~~~0.90287$ & X &   &   & X & X \nl
$1421+331~~~1.17261$ & X & X &   &   &   \nl
$1634+706~~~0.99024$ & X & X &   &   &   \nl
$2128-123~~~0.42973$ & X & X &   &   &   \nl
$2148+064~~~0.79078$ & X & X &   &   &   \nl
\enddata 
\label{tab:samples}
\end{deluxetable}

\endgroup

\begingroup
\small
 

%
%

\footnotesize
\begin{deluxetable}{rccccccc}
\tablenum{5}
\tablecolumns{8}
\tablecaption{Subfeature Equivalent Widths for Target Transitions}
\tablehead{
\colhead{} &
\colhead{} &
\colhead{\MgI} &
\colhead{\FeII} &
\colhead{\FeII} &
\colhead{\FeII} &
\colhead{\FeII} &
\colhead{\FeII} \\
\colhead{System} &
\colhead{($v^{-}, v^{+})$} &
\colhead{$W_r(2853)$} & 
\colhead{$W_r(2344)$} & 
\colhead{$W_r(2374)$} &
\colhead{$W_r(2383)$} &
\colhead{$W_r(2587)$} &
\colhead{$W_r(2600)$} \\
}
\startdata
\cutinhead{Q$0002+051~~z_{\rm abs}=0.851394$}
1 &   $ ( -55,  60)$      &   $ 0.166\pm0.027$    &   $ 0.225\pm0.022$    &   $ 0.077\pm0.025$    &   $ 0.347\pm0.016$    &   $ 0.170\pm0.025$    &   $ 0.349\pm0.014$    \nl
2 &   $ ( 125, 219)$      &   $<0.027        $    &   $ 0.047\pm0.025$    &   $<0.023        $    &   $ 0.082\pm0.019$    &   $<0.025        $    &   $ 0.070\pm0.017$    \nl
3 &   $ ( 282, 314)$      &   $<0.012        $    &   $ 0.025\pm0.010$    &   $<0.010        $    &   $<0.009        $    &   $<0.011        $    &   $<0.007        $    \nl
4 &   $ ( 393, 423)$      &   $<0.011        $    &   $<0.010        $    &   $<0.009        $    &   $<0.008        $    &   $<0.011        $    &   $<0.008        $    \nl
Total &   $ ( -55, 423)$     &   $ 0.161\pm0.027$   &   $ 0.297\pm0.035$   &   $ 0.077\pm0.025$   &   $ 0.429\pm0.025$   &   $ 0.170\pm0.025$   &   $ 0.419\pm0.022$   \nl
\cutinhead{Q$0058+019~~z_{\rm abs}=0.612672$}
1 &  $(-114,  97)$   &  $ 0.300\pm0.062$  &  $ 1.230\pm0.416$  &  $ 0.460\pm0.592$  &  $<5.357        $  &  $ 1.043\pm0.057$  &  $ 1.274\pm0.037$  \nl
Total &   $ (-114,  97)$     &   $ 0.300\pm0.062$   &   $ 1.230\pm0.416$   &   $ 0.460\pm0.592$   &   $<5.357        $   &   $ 1.043\pm0.057$   &   $ 1.274\pm0.037$   \nl
\cutinhead{Q$0117+212~~z_{\rm abs}=0.576398$}
Total &  $ ( -62,  91)$  &   $ 0.155\pm0.023$   & \nodata & \nodata & \nodata & \nodata & \nodata \nl
\cutinhead{Q$0117+212~~z_{\rm abs}=1.047971$ }
1 &   $ (-109, -63)$        &   $ 0.012\pm0.005$    &   $<0.006        $    &   $<0.006        $    &   $ 0.018\pm0.005$    &   $<0.006        $    &   $ 0.014\pm0.005$    \nl
2 &   $ ( -46,  51)$   &   $ 0.015\pm0.010$  &   $ 0.016\pm0.012$  &   $<0.012        $  &   $ 0.061\pm0.009$  &   $<0.012        $  &   $ 0.051\pm0.009$  \nl
Total &   $ (-109, 51)$       &   $ 0.027\pm0.011$   &   $ 0.022\pm0.012$   &   $<0.018        $   &   $ 0.080\pm0.011$   &   $<0.019        $   &   $ 0.065\pm0.010$   \nl
\cutinhead{Q$0420-014~~z_{\rm abs}=0.633004$}
1 &   $ (-241, -201)$     &   $<0.009        $    &   $<0.073        $    &   $<0.049        $    &   $<0.051        $    &   $<0.027        $    &   $<0.018        $    \nl
2 &   $ ( -96,  51)$     &   $ 0.062\pm0.031$   &   $<0.229        $   &   $<0.153        $   &   $<0.174        $   &   $<0.097        $   &   $ 0.076\pm0.054$   \nl
Total &   $ (-241,  51)$     &   $ 0.062\pm0.031$   &   $<0.302        $   &   $<0.202        $   &   $<0.225        $   &   $<0.124        $   &   $ 0.076\pm0.054$   \nl
\cutinhead{Q$0450-132~~z_{\rm abs}=0.493937$}
Total &   $ ( -91,  80)$     &   $ 0.078\pm0.068$   & \nodata & \nodata & \nodata & \nodata & \nodata \nl
\cutinhead{Q$0450-132~~z_{\rm abs}=1.174615$}
1 &   $ (-119, 135)$      & \nodata &   $ 0.765\pm0.020$    &   $ 0.485\pm0.033$    &   $ 1.072\pm0.019$    & \nodata &   $ 1.131\pm0.019$    \nl
2 &   $ ( 143, 176)$     & \nodata &   $<0.005        $   &   $<0.007        $   &   $<0.006        $   & \nodata &   $ 0.014\pm0.006$   \nl
Total &   $ (-119, 176)$       & \nodata &   $ 0.765\pm0.020$   &   $ 0.485\pm0.033$   &   $ 1.072\pm0.019$   & \nodata &   $ 1.146\pm0.020$   \nl
\cutinhead{Q$0454+039~~z_{\rm abs}=0.859565$}
Total &   $ ( -95, 121)$     &   $ 0.306\pm0.022$   &   $ 0.978\pm0.015$   &   $ 0.719\pm0.021$   &   $ 1.144\pm0.018$   &   $ 1.007\pm0.016$   &   $ 1.232\pm0.014$   \nl
\cutinhead{Q$0454+039~~z_{\rm abs}=1.153254$}
1 &   $ (-109, -79)$      &   $<0.004        $    &   $<0.005        $    &   $<0.004        $    &   \nodata   &   $<0.004        $    &   $ 0.009\pm0.005$    \nl
2 &   $ ( -35,  81)$      &   $ 0.026\pm0.012$    &   $ 0.048\pm0.017$    &   $ 0.019\pm0.014$    &   $ 0.056\pm0.020$    &   $ 0.014\pm0.013$    &   $ 0.075\pm0.014$    \nl
3 &   $ (  83, 115)$    &   $<0.004        $  &   $<0.006        $  &   $<0.005        $  &   $<0.007 $  &   $<0.004        $  &   $<0.005        $  \nl
Total &   $ (-109, 115)$       &   $ 0.026\pm0.012$   &   $ 0.048\pm0.017$   &   $ 0.019\pm0.014$   &   $ 0.056\pm0.020$   &   $ 0.014\pm0.013$   &   $ 0.084\pm0.015$   \nl
\cutinhead{Q$0454-220~~z_{\rm abs}=0.474410$}
Total &   $ ( -98, 106)$     &   $ 0.333\pm0.014$   & \nodata & \nodata & \nodata &   $ 0.677\pm0.067$   &   $ 0.975\pm0.029$   \nl
\cutinhead{Q$0454-220~~z_{\rm abs}=0.483340$}
Total &   $ ( -60,  41)$     &   $ 0.070\pm0.010$   & \nodata & \nodata & \nodata &   $ 0.069\pm0.028$   &   $ 0.162\pm0.043$   \nl
\cutinhead{Q$0823-223~~z_{\rm abs}=0.911017$}
1 &   $ (-157,  75)$      &   $ 0.189\pm0.029$    &   $ 0.123\pm0.050$    &   $ 0.006\pm0.054$    &   $ 0.371\pm0.039$    &   $ 0.118\pm0.026$    &   $ 0.298\pm0.024$    \nl
2 &   $ ( 105, 185)$      &   $ 0.034\pm0.011$    &   $ 0.048\pm0.020$    &   $<0.022        $    &   $ 0.123\pm0.014$    &   $ 0.037\pm0.009$    &   $ 0.111\pm0.009$    \nl
3 &   $ ( 199, 230)$    &   $<0.005 $   &   $<0.010        $  &   $<0.011        $  &   $<0.007        $  &   $<0.005        $  &   $ 0.007\pm0.005$  \nl
Total &   $ (-157, 230)$     &   $ 0.224\pm0.031$   &   $ 0.171\pm0.054$   &   $ 0.006\pm0.054$   &   $ 0.494\pm0.041$   &   $ 0.155\pm0.028$   &   $ 0.416\pm0.026$   \nl
\tablebreak
\cutinhead{Q$1101-264~~z_{\rm abs}=0.359002$}
Total &   $ ( -57,  94)$     & \nodata & \nodata & \nodata & \nodata & \nodata & \nodata \nl
\cutinhead{Q$1148+384~~z_{\rm abs}=0.553362$}
Total &   $ ( -96, 101)$     &   $<0.063        $   & \nodata & \nodata & \nodata &   $ 0.065\pm0.081$   &   $ 0.151\pm0.071$   \nl
\cutinhead{Q$1206+459~~z_{\rm abs}=0.927602$~A\tablenotemark{E}}
1 &   $ (-421, -386)$     &   $<0.005        $    &   $<0.006        $    &   $<0.006        $    &   $<0.006        $    &   $<0.006        $    &   $<0.006        $    \nl
2 &   $ (-366, -310)$     &   $<0.007        $    &   $<0.009        $    &   $<0.009        $    &   $<0.009        $    &   $<0.010        $    &   $<0.008        $    \nl
3 &  $(-275, -236)   $  &   $<0.005        $  &   $<0.007        $  &   $<0.007        $  &   $<0.007        $  &   $<0.008        $  &   $<0.006        $  \nl
4 &    $ (-204, -160)$   &    $<0.006        $  &    $<0.007        $  &    $<0.007        $  &    $<0.008        $  &    $<0.008        $  &    $<0.006        $  \nl
5 &   $ ( -80,  89)$     &   $ 0.042\pm0.020$   &   $ 0.033\pm0.024$   &   $ 0.018\pm0.024$   &   $ 0.098\pm0.026$  &   $ 0.030\pm0.030$   &   $ 0.077\pm0.020$   \nl
Total &   $ (-421, 89)$       &   $ 0.042\pm0.020$   &   $ 0.033\pm0.024$   &   $ 0.018\pm0.024$   &   $ 0.098\pm0.026$   &   $ 0.030\pm0.030$   &   $ 0.077\pm0.020$   \nl
\cutinhead{Q$1213-003~~z_{\rm abs}=1.320059$}
Total &   $ (-221, 175)$     &   $ 0.686\pm0.054$   &   $ 1.073\pm0.091$   &   $ 0.417\pm0.067$   &   $ 1.668\pm0.057$   &   $ 0.875\pm0.066$   &   $ 1.585\pm0.031$   \nl
\cutinhead{Q$1213-003~~z_{\rm abs}=1.554148$ }
1 &   $ (-273, -235)$     &   $<0.010        $    &   $<0.008        $    &   $<0.007        $    &   $<0.008        $    &   $<0.010        $    &   $<0.008        $    \nl
2 &   $ (-218, 155)$    &   $ 0.215\pm0.075$  &   $ 0.499\pm0.060$  &   $ 0.278\pm0.049$  &   $ 0.849\pm0.046$  &   $ 0.464\pm0.064$  &   $ 0.853\pm0.047$  \nl
Total &   $ (-273, 155)$     &   $ 0.215\pm0.075$   &   $ 0.499\pm0.060$   &   $ 0.278\pm0.049$   &   $ 0.849\pm0.046$   &   $ 0.464\pm0.064$   &   $ 0.853\pm0.047$   \nl
\cutinhead{Q$1222+228~~z_{\rm abs}=0.668052$ }
1 &   $ ( -43,  42)$      &   $ 0.038\pm0.016$    &   $ 0.063\pm0.045$    &   $<0.035        $    &   $ 0.113\pm0.035$    &   $<0.027        $    &   $ 0.119\pm0.022$    \nl
2 &   $ ( 213, 253)$     &   $<0.009        $   &   $<0.024        $   &   $<0.018        $   &   $ 0.028\pm0.021$   &   $<0.015        $   &   $ 0.047\pm0.011$   \nl
Total &   $ ( -43, 253)$     &   $ 0.038\pm0.016$   &   $ 0.063\pm0.045$   &   $<0.052        $   &   $ 0.140\pm0.041$   &   $<0.042        $   &   $ 0.166\pm0.025$   \nl
\cutinhead{Q$1225+317~~z_{\rm abs}=1.794833$}
1 &   $ (-147, 136)$      &   $ 0.275\pm0.046$    &   $ 0.300\pm0.043$    &   $ 0.086\pm0.035$    &   $ 0.661\pm0.025$    & \nodata &   $ 0.620\pm0.050$    \nl
2 &   $ ( 136, 195)$     &   $<0.013        $   &   $<0.013        $   &   $<0.009        $   &   $<0.009        $   & \nodata &   $<0.016        $   \nl
Total &   $ (-147, 195)$     &   $ 0.275\pm0.046$   &   $ 0.300\pm0.043$   &   $ 0.086\pm0.035$   &   $ 0.661\pm0.025$   & \nodata &   $ 0.620\pm0.050$   \nl
\cutinhead{Q$1241+176~~z_{\rm abs}=0.550482$}
1 &   $ ( -49,  62)$      &   $ 0.098\pm0.030$    & \nodata & \nodata & \nodata &   $ 0.096\pm0.047$    &   $ 0.236\pm0.048$    \nl
2 &   $ ( 133, 161)$     &   $<0.011        $   & \nodata & \nodata & \nodata &   $<0.017        $   &   $<0.018        $   \nl
Total &   $ ( -49, 161)$     &   $ 0.098\pm0.030$   & \nodata & \nodata & \nodata &   $ 0.096\pm0.047$   &   $ 0.236\pm0.048$   \nl
\cutinhead{Q$1248+401~~z_{\rm abs}=0.772957$}
1 &   $ ( -52,  69)$      &   $ 0.049\pm0.021$    &   $ 0.105\pm0.026$    &   $ 0.045\pm0.026$    &   $ 0.241\pm0.021$    &   $ 0.116\pm0.019$    &   $ 0.233\pm0.018$    \nl
2 &   $ ( 207, 248)$     &   $ 0.016\pm0.007$   &   $<0.011        $   &   $<0.011        $   &   $<0.010        $   &   $<0.008        $   &   $ 0.014\pm0.008$   \nl
Total &   $ ( -52, 248)$       &   $ 0.065\pm0.022$   &   $ 0.105\pm0.026$   &   $ 0.045\pm0.026$   &   $ 0.241\pm0.021$   &   $ 0.116\pm0.019$   &   $ 0.247\pm0.020$   \nl
\cutinhead{Q$1254+044~~z_{\rm abs}=0.519389$}
1 &   $ (-277, -244)$     &   $<0.013        $    & \nodata & \nodata & \nodata &   $<0.018        $    &   $<0.021        $    \nl
2 &   $ (-173, -143)$     &   $<0.012        $    & \nodata & \nodata & \nodata &   $<0.016        $    &   $<0.018        $    \nl
3 &   $ ( -33,  40)$      &   $ 0.169\pm0.019$    & \nodata & \nodata & \nodata &   $ 0.185\pm0.029$    &   $ 0.298\pm0.024$    \nl
Total &   $ (-407,  40)$       &   $ 0.169\pm0.019$   & \nodata & \nodata & \nodata &   $ 0.185\pm0.029$   &   $ 0.298\pm0.024$   \nl
\cutinhead{Q$1254+044~~z_{\rm abs}=0.934232$}
Total &   $ ( -61,  32)$     &   $ 0.005\pm0.019$   &   $ 0.021\pm0.022$   &   $<0.023        $   &   $ 0.056\pm0.022$   &   $ 0.040\pm0.020$   &   $ 0.051\pm0.015$   \nl
\cutinhead{Q$1317+274~~z_{\rm abs}=0.660051$ }
1 &   $ ( -33,  30)$      &   $ 0.026\pm0.009$    &   $ 0.060\pm0.021$    &   $<0.028        $    &   $ 0.104\pm0.014$    &   $ 0.028\pm0.010$    &   $ 0.099\pm0.011$    \nl
2 &   $ (  36,  58)$      &   $<0.004        $    &   $<0.010        $    &   $<0.013        $    &   $<0.007        $    &   $<0.005        $    &   $<0.006        $    \nl
3 &   $ (  62, 118)$      &   $<0.008        $    &   $<0.021        $    &   $<0.028        $    &   $ 0.049\pm0.013$    &   $<0.009        $    &   $ 0.027\pm0.012$    \nl
4 &   $ ( 126, 158)$      &   $<0.006        $    &   $<0.015        $    &   $<0.018        $    &   $<0.009        $    &   $<0.006        $    &   $<0.008        $    \nl
Total &   $ ( -33, 158)$     &   $ 0.026\pm0.009$   &   $ 0.060\pm0.021$   &   $<0.087        $   &   $ 0.153\pm0.020$   &   $ 0.028\pm0.010$   &   $ 0.126\pm0.016$   \nl
\tablebreak
\cutinhead{Q$1329+412~~z_{\rm abs}=0.893337$}
1 &   $ ( -45,  25)$      &   $<0.033        $    &   $<0.037        $    &   $<0.030        $    &   $ 0.069\pm0.021$    &   $<0.034        $    &   $ 0.051\pm0.031$    \nl
2 &   $ ( 102, 139)$     &   $<0.022        $   &   $<0.022        $   &   $<0.016        $   &   $ 0.032\pm0.013$   &   $<0.021        $   &   $ 0.029\pm0.018$   \nl
Total &   $ ( -45, 139)$       &   $<0.055        $   &   $<0.058        $   &   $<0.046        $   &   $ 0.101\pm0.025$   &   $<0.054        $   &   $ 0.080\pm0.035$   \nl
\cutinhead{Q$1354+193~~z_{\rm abs}=0.456588$}
1 &   $ ( -90, -61)$      &   $<0.009        $    & \nodata & \nodata & \nodata & \nodata &   $<0.033        $    \nl
2 &   $ ( -51,  56)$      &   $ 0.038\pm0.028$    & \nodata & \nodata & \nodata & \nodata &   $ 0.149\pm0.088$    \nl
Total &   $ ( -90,  56)$       &   $ 0.038\pm0.028$   & \nodata & \nodata & \nodata & \nodata &   $ 0.149\pm0.088$   \nl
\cutinhead{Q$1421+331~~z_{\rm abs}=0.902871$ }
1 &   $ (-118,  46)$      &   $ 0.048\pm0.032$    &   $ 0.163\pm0.030$    &   $ 0.063\pm0.028$    &   $ 0.309\pm0.027$    &   $ 0.121\pm0.045$    &   $ 0.223\pm0.034$    \nl
2 &   $ (  50, 112)$     &   $<0.014        $   &   $<0.014        $   &   $<0.012        $   &   $ 0.022\pm0.015$   &   $<0.022        $   &   $<0.016        $   \nl
Total &   $ (-118, 112)$       &   $ 0.048\pm0.032$   &   $ 0.163\pm0.030$   &   $ 0.063\pm0.028$   &   $ 0.331\pm0.031$   &   $ 0.121\pm0.045$   &   $ 0.223\pm0.034$   \nl
\cutinhead{Q$1421+331~~z_{\rm abs}=1.172609$}
Total &   $ ( -60,  53)$     &   $ 0.057\pm0.026$   &   $ 0.058\pm0.020$   &   $<0.022        $   &   $ 0.140\pm0.018$   &   $<0.031        $   &   $ 0.131\pm0.018$   \nl
\cutinhead{Q$1622+235~~z_{\rm abs}=0.656114$}
Total &   $ ( -91,  89)$     & \nodata &   $<0.130        $   &   $ 0.447\pm0.152$   &   $<0.155        $   &   $ 0.700\pm0.090$   &   $ 1.015\pm0.050$   \nl
\cutinhead{Q$1622+235~~z_{\rm abs}=0.797107$}
1 &   $ ( -83, -46)$      &   $<0.021        $    &   $<0.055        $    &   $<0.049        $    &   \nodata           &   $<0.024        $    &   $<0.027        $    \nl
2 &   $ ( -40,  26)$      &   $<0.035        $    &   $<0.068        $    &   $<0.070        $    &   \nodata         &   $<0.038        $    &   $<0.040        $    \nl
3 &   $ (  47,  80)$     &   $<0.198        $   &   $<0.042        $   &   $<0.038        $   &    \nodata  &   $<0.021 $    &   $<0.024        $   \nl
Total &   $ ( -83,  80)$     &   $<0.254        $   &   $<0.166        $   &   $<0.157        $   &   \nodata   &   $<0.083        $   &   $<0.092        $   \nl
\cutinhead{Q$1622+235~~z_{\rm abs}=0.891253$}
Total &   $ (-114,  95)$     &   $ 0.307\pm0.033$   &   $ 0.659\pm0.102$   &   $ 0.431\pm0.089$   &   $ 1.517\pm1.587$   &   $ 0.754\pm0.344$   &   $ 1.080\pm0.421$   \nl
\cutinhead{Q$1634+704~~z_{\rm abs}=0.990240$}
Total &   $ ( -48,  59)$      &   $ 0.058\pm0.010$    &   $ 0.069\pm0.037$    &   $<0.033        $    &   $ 0.135\pm0.018$    &   $ 0.042\pm0.015$    &   $ 0.127\pm0.011$    \nl
\cutinhead{Q$2128-123~~z_{\rm abs}=0.429735$}
Total &   $ ( -54,  35)$     &   $ 0.162\pm0.026$   & \nodata & \nodata & \nodata & \nodata & \nodata \nl
\cutinhead{Q$2145+064~~z_{\rm abs}=0.790777$}
1 &   $ (-127, -82)$      &   $<0.011        $    &   $ 0.008\pm0.016$    &   $<0.017        $    &   $ 0.053\pm0.019$    &   \nodata    &   $ 0.041\pm0.014$    \nl
2 &   $ ( -23,  30)$      &   $<0.012        $    &   $<0.019        $    &   $<0.019        $    &   $<0.026        $    &    \nodata    &   $<0.019        $    \nl
3 &   $ (  32, 105)$     &   $<0.015        $   &   $<0.026        $   &   $<0.026        $   &   $ 0.034\pm0.032$   &   \nodata   &   $<0.023        $   \nl
Total &   $ (-127, 105)$     &   $<0.038        $   &   $ 0.008\pm0.016$   &   $<0.063        $   &   $ 0.087\pm0.038$   &    \nodata   &   $ 0.041\pm0.014$   \nl
\enddata
\label{tab:ewalltab}
\end{deluxetable}


\endgroup

\begingroup
\small
 

%
%

\begin{deluxetable}{rcccc}
\tablenum{6}
\tablecolumns{5}
\tablecaption{Subsystem AOD Column Densities}
\tablehead{
\colhead{System} &
\colhead{($v^{-}, v^{+})$} &
\colhead{\MgII} &
\colhead{\MgI} &
\colhead{\FeII} \\
}
\startdata
\cutinhead{Q$0002+051~~z_{\rm abs}=0.851394$}
1 &   $ ( -55,  60)$      & $ 13.773\pm 0.001$ & $ 12.188\pm 0.024$ & $ 13.753\pm 0.003$ \nl  
2 &   $ ( 125, 219)$      & $ 13.379\pm 0.003$ & $<10.874 $         & $ 12.801\pm 0.024$ \nl
3 &   $ ( 282, 314)$      & $ 12.001\pm 0.035$ & $<10.655 $         & $<11.300$ \nl
4 &   $ ( 393, 423)$      & $ 12.028\pm 0.031$ & $<10.642 $         & $<11.279$ \nl
Total &   $ ( -55, 423)$  & $ 13.931\pm 0.001$ & $ 12.232\pm 0.022$ &$ 13.802\pm 0.004$ \nl
\cutinhead{Q$0117+212~~z_{\rm abs}=0.5764$}
Total & $ ( -62, 91)$     & $>14.080$ & $12.144\pm0.019$ & \nl
\cutinhead{Q$0117+212~~z_{\rm abs}=1.047971$ }
1 &   $ (-109, -63)$ & $ 12.486\pm 0.010$ & $ 11.006\pm 0.077$ & $ 12.038\pm 0.045$ \nl
2 &   $ ( -46,  51)$ & $ 13.075\pm 0.005$ & $ 11.065\pm 0.097$  & $ 12.616\pm 0.018$ \nl
Total &   $ (-109, 51)$ & $ 13.174\pm 0.004$ & $ 11.337\pm 0.063$ & $12.718\pm 0.017$ \nl
\cutinhead{Q$0420-014~~z_{\rm abs}=0.633004$}
1 &   $ (-241, -201)$ & $ 12.456\pm 0.019$ &  $<10.517$ & $<11.806$ \nl
2 &   $ ( -96,  51)$ & $>13.753$ &$ 11.745\pm 0.057$  & $ 12.840\pm0.081$ \nl
Total &   $ (-241,  51)$ & $ 13.774\pm 0.001$ & $ 11.770\pm 0.054$ & $12.879\pm 0.075$  \nl
\cutinhead{Q$0454+039~~z_{\rm abs}=0.859565$}
Total &   $ ( -95, 121)$ & $>14.323$ &  $ 12.453\pm 0.009$  &  $ 15.068\pm 0.028$ \nl
\cutinhead{Q$0454+039~~z_{\rm abs}=1.153254$}
1 &   $ (-109, -79)$      & $ 11.933\pm 0.028$  &  $<10.186$ &  $11.804\pm 0.108$ \nl
2 &   $ ( -35,  81)$      & $ 13.391\pm 0.008$ &  $ 11.271\pm 0.072$& $ 12.773\pm 0.020$ \nl
3 &   $ (  83, 115)$   &  $ 11.943\pm 0.028$ & $<10.195$ & $<11.126$ \nl
Total &   $ (-109, 115)$ & $ 13.421\pm 0.008$  &  $ 11.338\pm 0.061$ & $ 12.826\pm 0.020$ \nl
\cutinhead{Q$0454-220~~z_{\rm abs}=0.474410$}
Total &   $ ( -98, 106)$     & $>14.272$ & $ 12.532\pm 0.005$ &$>14.515$ \nl
\cutinhead{Q$0454-220~~z_{\rm abs}=0.483340$}
Total &   $ ( -60,  41)$     & $>13.527$ & $ 11.849\pm 0.018$ & $ 13.384\pm 0.051$ \nl
\cutinhead{Q$0823-223~~z_{\rm abs}=0.911017$} 
1 &   $ (-157,  75)$      &  $>13.752$ & $ 12.234\pm 0.016$ &   $ 13.470\pm 0.007$ \nl 
2 &   $ ( 105, 185)$      & $13.072\pm 0.007$ & $ 11.411\pm 0.057$ & $ 12.987\pm 0.012$ \nl
3 &   $ ( 199, 230)$    & $ 11.827\pm 0.038$ & $<10.305$ & $ 11.513\pm 0.198$ \nl
Total &   $ (-157, 230)$ & $ 13.839\pm 0.001$ &  $ 12.299\pm 0.016$ & $ 13.597\pm 0.006$  \nl
\tablebreak
\cutinhead{Q$1101-264~~z_{\rm abs}=0.359002$}
Total &   $ ( -57,  94)$     & $>13.621$ & & \nl
\cutinhead{Q$1148+384~~z_{\rm abs}=0.553362$}
Total &   $ ( -96, 101)$     & $ 13.433\pm 0.019$ & $<11.077$ & $ 13.186\pm 0.046$ \nl
\cutinhead{Q$1206+459~~z_{\rm abs}=0.927602$}
1 &   $ (-421, -386)$     &    $ 11.910\pm 0.031$ & $<10.262$ &$<11.140$ \nl
2 &   $ (-366, -310)$     &  $ 12.384\pm 0.014$ & $<10.366$ &$<11.242$ \nl
3 &  $(-275, -236)   $  &  $ 12.304\pm 0.015$ & $<10.286$ & $<11.156$ \nl
4 &    $ (-204, -160)$   & $ 12.099\pm 0.022$  & $<10.306$ & $<11.187$\nl
5 &   $ ( -80,  89)$     & $>13.683$ & $ 11.535\pm 0.053$ & $12.870\pm 0.019$ \nl
Total &   $ (-421, 89)$  & $ 13.738\pm 0.001$ & $ 11.627\pm 0.042$ & $ 12.905\pm 0.018$ \nl
\cutinhead{Q$1222+228~~z_{\rm abs}=0.668052$ }
1 &   $ ( -43,  42)$      & $ 13.290\pm 0.010$ & $ 11.477\pm 0.068$ &$ 12.988\pm 0.030$ \nl
2 &   $ ( 213, 253)$     & $ 12.412\pm 0.019$ & $<10.509$ & $12.518\pm 0.055$ \nl 
Total &   $ ( -43, 253)$     &  $ 13.344\pm 0.009$ & $ 11.521\pm0.061$ &  $ 13.115\pm 0.026$ \nl
\cutinhead{Q$1241+176~~z_{\rm abs}=0.550482$}
1 &   $ ( -49,  62)$      & $>13.497$ & $ 11.964\pm 0.042$ & $>13.728$ \nl
2 &   $ ( 133, 161)$     & $ 11.941\pm 0.049$ & $<10.634$ &  $<11.825$\nl
Total &   $ ( -49, 161)$     & $ 13.509\pm 0.001$ & $ 11.984\pm 0.040$& $<13.734$ \nl
\cutinhead{Q$1248+401~~z_{\rm abs}=0.772957$}
1 &   $ ( -52,  69)$      &  $>13.710$ &  $ 11.620\pm 0.058$ & $ 13.544\pm 0.017$ \nl
2 &   $ ( 207, 248)$     &  $ 12.304\pm 0.015$ & $ 11.141\pm 0.088$ & $ 11.707\pm 0.178$ \nl
Total &   $ ( -52, 248)$  & $ 13.726\pm 0.001$ & $ 11.744\pm 0.049$ & $ 13.550\pm 0.017$ \nl
\cutinhead{Q$1254+044~~z_{\rm abs}=0.519389$}
1 &   $ (-277, -244)$     & $ 12.219\pm 0.038$ & $<10.728$ & $<11.881$ \nl
2 &   $ (-173, -143)$     & $ 11.970\pm 0.057$  & $<10.674$ &$<11.832$\nl
3 &   $ ( -33,  40)$      & $>13.625$ & $ 12.313\pm 0.041$ & $>13.948$ \nl
Total &   $ (-407,  40)$  &$ 13.655\pm 0.002$ & $ 12.342\pm 0.038$ & $<13.957$ \nl
\cutinhead{Q$1254+044~~z_{\rm abs}=0.934232$}
Total &   $ ( -61,  32)$ &  $ 13.261\pm 0.024$ & $ 10.585\pm 0.569$ & $ 12.699\pm 0.030$  \nl
\cutinhead{Q$1317+274~~z_{\rm abs}=0.660051$ }
1 &   $ ( -33,  30)$  & $ 12.991\pm 0.008$ & $ 11.341\pm 0.063$ & $ 12.992\pm 0.019$ \nl
2 &   $ (  36,  58)$  & $ 11.398\pm 0.085$ & $<10.247$ &  $<11.218$ \nl
3 &   $ (  62, 118)$  & $ 12.314\pm 0.017$ & $<10.450$ & $ 12.380\pm 0.051$ \nl
4 &   $ ( 126, 158)$  & $ 11.703\pm 0.050$  &  $<10.328$ &  $<11.309$ \nl
Total &   $ ( -33, 158)$ & $ 13.100\pm 0.007$ &  $ 11.457\pm 0.048$ & $ 13.100\pm 0.018$  \nl
\cutinhead{Q$1421+331~~z_{\rm abs}=0.902871$ }
1 &   $ (-118,  46)$      & $>13.827$ & $ 11.631\pm 0.071$ & $ 13.699\pm 0.026$ \nl
2 &   $ (  50, 112)$     &   $ 12.914\pm 0.030$  & $<10.632$ & $ 11.928\pm 0.157$ \nl
Total &   $ (-118, 112)$  &  $ 13.877\pm 0.003$ & $ 11.673\pm 0.065$ &$ 13.706\pm 0.026$ \nl
\tablebreak
\cutinhead{Q$1421+331~~z_{\rm abs}=1.172609$}
Total &   $ ( -60,  53)$  & $ 13.429\pm 0.018$ & $ 11.688\pm 0.061$ &  $ 13.049\pm 0.014$  \nl
\cutinhead{Q$1634+704~~z_{\rm abs}=0.990240$}
Total &   $ ( -48,  59)$      & $ 13.519\pm 0.001$ & $ 11.684\pm0.024$ & $ 13.059\pm 0.010$  \nl
\cutinhead{Q$2128-123~~z_{\rm abs}=0.429735$}
Total &   $ ( -54,  35)$     & $ 13.471\pm 0.003$ & $ 12.248\pm 0.023$& \nl
\cutinhead{Q$2145+064~~z_{\rm abs}=0.790777$}
1 &   $ (-127, -82)$      & $12.860\pm0.006$ & $<10.558$ & $12.688\pm0.035$ \nl
2 &   $ ( -23,  30)$      & $12.681\pm0.011$ & $<10.597$ & \nodata \nl
3 &   $ (  32, 105)$     & $12.845\pm0.009$ & $<10.660$ &$12.401\pm0.149$ \nl
Total &   $ (-127, 105)$   & $13.279\pm0.005$ & $<11.084$ &$12.869\pm0.056$ \nl
\enddata
\label{tab:aodcols}
\end{deluxetable}


\endgroup

\begingroup
\small

\begingroup
\footnotesize
\begin{deluxetable}{llrrrrrrr}
\tablenum{7}
\tablewidth{0pc}
\tablecolumns{9}
\tablecaption{Cloud Properties for {\MgII} Absorbers}
\tablehead
{ 
\multicolumn{2}{c}{System} &
\colhead{Cloud} &
\multicolumn{2}{c}{{\MgII}} &
\multicolumn{2}{c}{{\FeII}} &
\multicolumn{2}{c}{{\MgI}} \\
\cline{4-5} \cline {6-7} \cline{8-9}
\colhead{QSO} &
\colhead{$z_{\rm abs}$} &
\colhead{$\left< v \right>$} &
\colhead{$\log N$} &
\colhead{$b$} &
\colhead{$\log N$} &
\colhead{$b$} &
\colhead{$\log N$} &
\colhead{$b$} \\
 &
 &
\colhead{[{\kms}]} &
\colhead{[{\cmsq}]} &
\colhead{[{\kms}]} &
\colhead{[{\cmsq}]} &
\colhead{[{\kms}]} &
\colhead{[{\cmsq}]} &
\colhead{[{\kms}]} 
}
\startdata
$0002+051$ & 0.85139 
    & $-32.2$ & $13.36\pm0.70$ & $ 7.36\pm1.66$ & $13.13\pm0.00$ & $ 3.70\pm0.01$ & $<10.23$ & \nodata \nl
 &  & $-28.1$ & $13.88\pm0.30$ & $ 6.17\pm4.89$ & $12.74\pm0.01$ & $ 1.23\pm0.01$ & $11.84\pm0.10$ & $11.42\pm2.45$ \nl
 &  & $ 28.5$ & $12.50\pm0.01$ & $ 8.32\pm0.31$ & $12.21\pm0.10$ & $ 3.62\pm0.09$ & $\sim11.6$ & \nodata \nl
 &  & $-13.7$ & $15.85\pm2.67$ & $ 2.34\pm1.27$ & $13.78\pm0.01$ & $ 6.84\pm0.12$ & $11.70\pm0.10$ & $ 3.91\pm1.18$ \nl
 &  & $ -1.6$ & $13.51\pm0.23$ & $ 3.65\pm2.26$ & $<12.28$ & \nodata & $<10.73$ & \nodata \nl
 &  & $  5.4$ & $14.17\pm0.21$ & $ 3.68\pm0.23$ & $13.18\pm0.03$ & $ 3.23\pm0.35$ & $11.71\pm0.13$ & $ 8.37\pm1.39$ \nl
 &  & $142.8$ & $12.92\pm0.02$ & $ 3.03\pm0.17$ & $12.43\pm0.05$ & $ 1.86\pm0.55$ & $<10.77$ & \nodata \nl
 &  & $149.4$ & $12.39\pm0.06$ & $ 9.41\pm0.79$ & $11.39\pm0.52$ & \nodata & $<10.77$ & \nodata \nl
 &  & $172.4$ & $13.42\pm0.03$ & $ 3.44\pm0.14$ & $12.30\pm0.06$ & $ 5.59\pm1.14$ & $<10.78$ & \nodata \nl
 &  & $182.6$ & $13.20\pm0.01$ & $ 4.48\pm0.17$ & $12.42\pm0.04$ & $ 3.58\pm0.56$ & $<10.72$ & \nodata \nl
 &  & $296.0$ & $12.04\pm0.03$ & $ 6.70\pm0.70$ & $<11.74$ & \nodata & $<10.75$ & \nodata \nl
 &  & $407.4$ & $12.05\pm0.03$ & $ 5.23\pm0.54$ & $<11.72$ & \nodata & $<10.80$ & \nodata \nl
\multicolumn{9}{c}{ } \nl
$0117+213$ & 0.57640 
    & $-30.6$ & $13.48\pm0.31$ & $10.53\pm2.20$ & \nodata & \nodata & $11.37\pm0.14$ & $19.02\pm6.54$ \nl
 &  & $ -8.9$ & $13.80\pm0.31$ & $19.49\pm2.81$ & \nodata & \nodata & $11.78\pm0.05$ & $ 8.66\pm0.82$ \nl
 &  & $ 16.3$ & $14.61\pm0.09$ & $14.39\pm0.41$ & \nodata & \nodata & $11.77\pm0.04$ & $23.15\pm2.65$ \nl
 &  & $ 77.2$ & $11.60\pm0.06$ & $ 4.56\pm1.11$ & \nodata & \nodata & $<10.61$ & \nodata \nl
\multicolumn{9}{c}{ } \nl
$0117+213$ & 1.04797 
    & $-97.0$ & $11.55\pm0.09$ & $ 2.98\pm1.21$ & $11.32\pm0.39$ & \nodata & $11.09\pm0.15$ & $17.87\pm8.26$ \nl
 &  & $-85.8$ & $12.46\pm0.01$ & $ 5.28\pm0.26$ & $12.00\pm0.05$ & $ 4.56\pm0.71$ & $<10.35$ & \nodata \nl
 &  & $-23.3$ & $12.20\pm0.06$ & $10.87\pm1.26$ & $<11.32$ & \nodata & $< 10.35$ & \nodata \nl
 &  & $ -0.2$ & $12.74\pm0.13$ & $10.30\pm1.93$ & $12.40\pm0.09$ & $12.54\pm1.96$ & $11.02\pm0.12$ & $14.43\pm5.26$ \nl
 &  & $  6.0$ & $12.21\pm0.28$ & $ 4.44\pm1.26$ & $11.70\pm0.17$ & \nodata & $< 10.35$ & \nodata \nl
 &  & $ 17.4$ & $12.38\pm0.11$ & $ 6.44\pm0.94$ & $11.65\pm0.29$ & $ 4.24\pm2.48$ & $<10.35$ & \nodata \nl
 &  & $ 33.5$ & $11.91\pm0.08$ & $ 9.05\pm1.66$ & $11.84\pm0.44$ & $12.48\pm6.33$ & $<10.35$ & \nodata \nl
\multicolumn{9}{c}{ } \nl
$0420-014$ & 0.63300 
    & $-225.0$ & $12.34\pm0.06$ & $ 4.13\pm0.80$ & $<11.91$ & \nodata & $<10.77$ & \nodata \nl
 &  & $-215.8$ & $11.95\pm0.14$ & $ 3.47\pm1.53$ & $<11.91$ & \nodata & $<10.77$ & \nodata \nl
 &  & $-82.3$ & $11.91\pm0.12$ & $ 4.15\pm1.25$ & $<11.89$ & \nodata & $<10.81$ & \nodata \nl
 &  & $-62.2$ & $12.47\pm0.07$ & $15.05\pm2.96$ & $<11.85$ & \nodata & $<10.82$ & \nodata \nl
 &  & $-39.5$ & $12.62\pm0.01$ & $ 5.99\pm0.60$ & $<11.89$ & \nodata & $<10.83$ & \nodata \nl
 &  & $-22.5$ & $13.05\pm0.04$ & $15.70\pm1.34$ & $<11.88$ & \nodata & $<10.82$ & \nodata \nl
 &  & $  5.4$ & $13.22\pm0.41$ & $ 2.24\pm0.61$ & $12.13\pm0.28$ & $ 6.93\pm6.05$ & $11.01\pm1.16$ & $\sim29.0$ \nl
 &  & $ 16.1$ & $13.37\pm0.05$ & $ 8.83\pm0.70$ & $12.64\pm0.09$ & $ 5.94\pm1.89$ & $11.57\pm0.15$ & $11.05\pm3.37$ \nl
 &  & $ 30.9$ & $12.80\pm0.03$ & $ 3.97\pm0.41$ & $12.20\pm0.15$ & $ 1.79\pm3.27$ & $10.99\pm0.27$ & $ 3.74\pm2.80$ \nl
\multicolumn{9}{c}{ } \nl
$0454+039$ & 0.85957 
    & $-72.2$ & $13.03\pm0.07$ & $ 6.25\pm0.51$ & $12.73\pm0.03$ & $ 3.00\pm0.36$ & $10.34\pm0.46$ & $\sim 7.0$ \nl
 &  & $-53.8$ & $14.50\pm2.91$ & $\sim 6.5$ & $14.51\pm0.05$ & $ 6.50\pm0.33$ & $11.74\pm0.04$ & $ 6.50\pm0.72$ \nl
 &  & $-42.5$ & $\sim13.9$ & $\sim 5.0$ & $13.33\pm0.21$ & $ 5.00\pm1.46$ & $11.19\pm0.16$ & $ 5.00\pm1.81$ \nl
 &  & $-26.9$ & $\sim13.9$ & $\sim 8.8$ & $13.51\pm0.06$ & $ 8.75\pm1.64$ & $11.12\pm0.22$ & $ 8.75\pm5.72$ \nl
 &  & $-10.4$ & $\sim14.2$ & $\sim 5.5$ & $13.77\pm0.19$ & $ 5.25\pm0.99$ & $11.47\pm0.16$ & $ 5.50\pm1.27$ \nl
 &  & $  4.2$ & $\sim14.5$ & $\sim10.8$ & $14.31\pm0.13$ & $10.50\pm3.49$ & $11.75\pm0.17$ & $10.75\pm4.13$ \nl
 &  & $ 21.2$ & $\sim14.6$ & $\sim 8.0$ & $14.73\pm0.14$ & $ 8.00\pm1.62$ & $11.86\pm0.10$ & $ 8.00\pm1.26$ \nl
 &  & $ 39.6$ & $\sim14.0$ & $\sim 6.8$ & $13.88\pm0.14$ & $ 6.75\pm2.21$ & $11.24\pm0.21$ & $ 6.75\pm3.90$ \nl
 &  & $ 53.8$ & $\sim13.9$ & $\sim 5.5$ & $13.51\pm0.54$ & $ 5.50\pm5.30$ & $11.12\pm0.57$ & $\sim 5.5$ \nl
 &  & $ 60.8$ & $\sim14.1$ & $\sim 4.0$ & $13.68\pm0.29$ & $ 4.25\pm0.68$ & $11.30\pm0.30$ & $ 3.50\pm1.40$ \nl
 &  & $ 73.6$ & $12.73\pm0.79$ & $ 7.00\pm4.26$ & $12.73\pm0.04$ & $ 4.50\pm0.54$ & $10.34\pm0.47$ & $\sim 7.5$ \nl
\multicolumn{9}{c}{ } \nl
$0454+039$ & 1.15325 
    & $-96.3$ & $12.00\pm0.02$ & $ 2.86\pm0.36$ & $11.97\pm0.06$ & $ 2.02\pm1.14$ & $<10.48$ & \nodata \nl
 &  & $-11.5$ & $12.57\pm0.08$ & $ 9.22\pm0.53$ & $12.06\pm0.08$ & $ 3.86\pm1.20$ & $11.09\pm1.06$ & $\sim37.9$ \nl
 &  & $ -1.2$ & $13.24\pm0.02$ & $ 8.46\pm0.42$ & $12.54\pm0.03$ & $ 4.07\pm0.50$ & $10.87\pm0.12$ & $ 3.56\pm1.66$ \nl
 &  & $ 12.4$ & $12.58\pm0.04$ & $ 9.35\pm0.49$ & $12.02\pm0.09$ & $ 9.25\pm2.27$ & $\sim10.9$ & $\sim 0.5$ \nl
 &  & $ 41.8$ & $11.63\pm0.06$ & $ 7.42\pm1.37$ & $11.32\pm0.25$ & $\sim 1.2$ & $<10.47$ & \nodata \nl
 &  & $ 67.7$ & $11.62\pm0.05$ & $ 6.35\pm1.13$ & $11.49\pm0.18$ & $ 6.60\pm3.94$ & $<10.21$ & \nodata \nl
 &  & $ 99.0$ & $11.99\pm0.02$ & $ 3.27\pm0.35$ & $<11.54$ & \nodata & $<10.48$ & \nodata \nl
\multicolumn{9}{c}{ } \nl
$0454-220$ & 0.47441 
    & $-81.1$ & $12.48\pm0.07$ & $ 3.72\pm0.54$ & $<12.10$ & \nodata & $10.63\pm0.22$ & $ 5.50\pm2.62$ \nl
 &  & $-70.8$ & $12.33\pm0.27$ & $ 9.77\pm4.30$ & $<12.07$ & \nodata & $< 10.16$ & \nodata \nl
 &  & $-52.4$ & $13.06\pm0.09$ & $10.64\pm1.84$ & $12.81\pm0.53$ & $13.60\pm3.95$ & $<10.14$ & \nodata \nl
 &  & $-40.4$ & $14.59\pm2.53$ & $ 3.65\pm2.59$ & $13.29\pm0.31$ & $14.39\pm7.31$ & $11.11\pm0.12$ & $ 8.12\pm2.27$ \nl
 &  & $-21.9$ & $15.06\pm2.63$ & $ 3.74\pm2.47$ & $12.49\pm0.49$ & $3.85\pm5.03$ & $11.47\pm0.64$ & $ 0.56\pm0.33$ \nl
 &  & $ -4.3$ & $13.90\pm2.67$ & $ 3.60\pm4.14$ & $15.73\pm7.98$ & $5.63\pm8.99$ & $12.31\pm0.02$ & $17.94\pm0.97$ \nl
 &  & $ 18.6$ & $13.90\pm0.06$ & $23.97\pm4.55$ & $14.00\pm0.05$ & $28.40\pm1.91$ & $11.95\pm0.05$ & $12.68\pm0.66$ \nl
 &  & $ 51.4$ & $13.46\pm0.07$ & $13.94\pm0.43$ & $13.61\pm0.29$ & $ 6.11\pm1.54$ & $11.39\pm0.03$ & $10.22\pm0.92$ \nl
\multicolumn{9}{c}{ } \nl
\tablebreak
$0454-220$ & 0.48334 
    & $-46.6$ & $11.81\pm0.05$ & $ 0.92\pm0.08$ & $<11.96$ & \nodata & $10.49\pm0.24$ & $ 6.25\pm5.15$ \nl
 &  & $-33.1$ & $12.32\pm0.01$ & $ 3.88\pm0.23$ & $<12.30$ & \nodata & $<10.37$ & \nodata \nl
 &  & $-20.8$ & $11.95\pm0.03$ & $ 2.19\pm0.68$ & $<12.27$ & \nodata & $<10.20$ & \nodata \nl
 &  & $ -9.6$ & $12.67\pm0.49$ & $ 3.78\pm1.36$ & $<12.24$ & \nodata & $10.82\pm0.25$ & $ 4.71\pm4.25$ \nl
 &  & $ -1.0$ & $13.17\pm0.11$ & $ 5.69\pm3.48$ & $13.41\pm0.10$ & $ 7.75\pm0.52$ & $11.45\pm0.07$ & $ 2.83\pm0.79$ \nl
 &  & $  5.3$ & $13.00\pm0.18$ & $ 5.93\pm0.80$ & $<12.28$ & \nodata & $11.58\pm0.04$ & $ 2.57\pm0.74$ \nl
 &  & $ 21.0$ & $12.37\pm0.01$ & $ 4.55\pm0.15$ & $12.47\pm0.14$ & $ 4.00\pm2.44$ & $10.47\pm0.17$ & $2.57\pm3.27$ \nl
\multicolumn{9}{c}{ } \nl
$0823-223$ & 0.91102 
    & $-139.2$ & $12.49\pm0.01$ & $ 1.53\pm0.12$ & $12.21\pm0.00$ & $ 1.97\pm0.68$ & $10.76\pm5.56$ & $ 5.64\pm4.65$ \nl
 &  & $-119.3$ & $12.35\pm0.00$ & $ 4.23\pm0.29$ & $12.22\pm0.01$ & $ 4.23\pm0.77$ & $10.92\pm5.67$ & $11.87\pm22.52$ \nl
 &  & $-117.7$ & $12.44\pm0.01$ & $27.54\pm2.05$ & $<10.95$ & \nodata & $<10.57$ & \nodata \nl
 &  & $-67.5$ & $13.25\pm0.01$ & $ 7.17\pm0.15$ & $13.04\pm0.00$ & $ 6.56\pm0.20$ & $11.93\pm0.01$ & $ 6.90\pm0.44$ \nl
 &  & $-52.4$ & $12.80\pm0.01$ & $ 3.51\pm0.32$ & $12.33\pm0.01$ & $ 1.02\pm0.17$ & $11.47\pm0.04$ & $ 3.32\pm0.72$ \nl
 &  & $-43.2$ & $12.27\pm0.02$ & $ 3.00\pm0.50$ & $11.98\pm0.01$ & $ 3.74\pm1.16$ & $<10.91$ & \nodata \nl
 &  & $-29.3$ & $12.02\pm0.01$ & $ 5.01\pm0.54$ & $11.53\pm0.01$ & $ 1.09\pm0.87$ & $10.69\pm0.46$ & $11.27\pm10.67$ \nl
 &  & $ -2.2$ & $12.27\pm0.01$ & $ 0.95\pm0.20$ & $11.92\pm0.02$ & $ 0.85\pm0.21$ & $<10.90$ & \nodata \nl
 &  & $  6.3$ & $12.62\pm0.04$ & $ 2.58\pm0.44$ & $12.55\pm0.01$ & $ 0.96\pm0.22$ & $11.32\pm0.34$ & $11.53\pm8.76$ \nl
 &  & $  8.2$ & $12.98\pm0.00$ & $17.78\pm0.65$ & $12.81\pm5.98$ & $16.70\pm1.27$ & $<10.55$ & \nodata \nl
 &  & $ 13.8$ & $12.49\pm0.01$ & $ 3.22\pm0.60$ & $12.36\pm0.01$ & $ 3.57\pm0.69$ & $<10.56$ & \nodata \nl
 &  & $ 46.2$ & $12.53\pm0.00$ & $ 9.92\pm0.40$ & $12.43\pm0.00$ & $10.33\pm0.97$ & $<10.18$ & \nodata \nl
 &  & $ 62.9$ & $11.45\pm0.03$ & $ 0.77\pm0.14$ & $<11.53$ & \nodata & $<10.22$ & \nodata \nl
 &  & $131.2$ & $12.48\pm0.08$ & $ 9.78\pm1.08$ & $12.37\pm0.09$ & $ 8.26\pm1.70$ & $11.09\pm0.33$ & $18.87\pm12.10$ \nl
 &  & $140.3$ & $12.40\pm0.11$ & $ 3.53\pm0.89$ & $12.31\pm0.11$ & $ 1.96\pm1.15$ & $<10.55$ & \nodata \nl
 &  & $150.1$ & $12.68\pm0.09$ & $10.27\pm2.05$ & $12.62\pm0.09$ & $ 8.84\pm2.03$ & $11.13\pm0.47$ & $22.59\pm15.07$ \nl
 &  & $166.1$ & $12.30\pm0.05$ & $ 3.93\pm0.43$ & $12.19\pm0.05$ & $ 1.88\pm0.87$ & $<10.54$ & \nodata \nl
 &  & $216.5$ & $11.79\pm0.03$ & $ 3.64\pm0.59$ & $12.02\pm0.99$ & $ 2.43\pm1.99$ & $<10.54$ & \nodata \nl
\multicolumn{9}{c}{ } \nl
$1101-264$ & 0.35900 
    & $-37.4$ & $11.91\pm0.05$ & $10.37\pm1.75$ & \nodata & \nodata & \nodata & \nodata \nl
 &  & $ -8.0$ & $13.48\pm0.04$ & $ 7.92\pm0.30$ & \nodata & \nodata & \nodata & \nodata \nl
 &  & $ 14.0$ & $12.18\pm0.05$ & $ 5.03\pm1.03$ & \nodata & \nodata & \nodata & \nodata \nl
 &  & $ 30.2$ & $12.98\pm0.02$ & $ 6.26\pm0.40$ & \nodata & \nodata & \nodata & \nodata \nl
 &  & $ 46.6$ & $12.42\pm0.03$ & $ 5.40\pm0.48$ & \nodata & \nodata & \nodata & \nodata \nl
 &  & $ 71.4$ & $11.57\pm0.07$ & $ 3.99\pm1.28$ & \nodata & \nodata & \nodata & \nodata \nl
\multicolumn{9}{c}{ } \nl
$1148+384$ & 0.55336 
    & $-79.8$ & $12.56\pm0.04$ & $ 2.90\pm0.27$ & $13.23\pm1.90$ & $1.88\pm0.97$ & $<10.92$ & \nodata \nl
 &  & $-49.4$ & $12.78\pm0.02$ & $ 6.02\pm0.24$ & $12.61\pm0.07$ & $ 5.29\pm1.34$ & $<10.91$ & \nodata \nl
 &  & $ -6.0$ & $12.64\pm0.04$ & $ 7.15\pm0.74$ & $12.32\pm0.15$ & $ 5.85\pm3.01$ & $<10.92$ & \nodata \nl
 &  & $  6.6$ & $12.91\pm0.04$ & $ 4.11\pm0.37$ & $12.62\pm0.08$ & $ 3.16\pm1.21$ & $<10.93$ & \nodata \nl
 &  & $ 25.6$ & $12.41\pm0.03$ & $ 3.62\pm0.34$ & $11.88\pm0.29$ & $ 3.99\pm5.13$ & $<10.94$ & \nodata \nl
 &  & $ 49.7$ & $12.43\pm0.03$ & $ 3.79\pm0.41$ & $11.93\pm0.24$ & $ 2.91\pm4.45$ & $<10.95$ & \nodata \nl
 &  & $ 63.6$ & $12.09\pm0.04$ & $ 4.94\pm0.83$ & $<12.00$ & \nodata & $<10.95$ & \nodata \nl
 &  & $ 85.5$ & $11.85\pm0.05$ & $ 3.52\pm0.90$ & $<12.02$ & \nodata & $<10.96$ & \nodata \nl
\multicolumn{9}{c}{ } \nl
$1206+459$ & 0.92760 
    & $-403.5$ & $11.92\pm0.03$ & $ 5.98\pm0.55$ & $<11.47$ & \nodata & $<10.40$ & \nodata \nl
 &  & $-350.7$ & $12.16\pm0.02$ & $ 2.89\pm0.26$ & $<11.48$ & \nodata & $<10.41$ & \nodata \nl
 &  & $-327.7$ & $12.12\pm0.02$ & $ 2.98\pm0.28$ & $<11.48$ & \nodata & $<10.40$ & \nodata \nl
 &  & $-256.8$ & $12.35\pm0.01$ & $ 4.77\pm0.20$ & $<11.48$ & \nodata & $<10.41$ & \nodata \nl
 &  & $-188.0$ & $12.01\pm0.02$ & $ 3.10\pm0.37$ & $<11.49$ & \nodata & $<10.40$ & \nodata \nl
 &  & $-174.2$ & $11.60\pm0.05$ & $ 3.69\pm0.89$ & $<11.49$ & \nodata & $<10.41$ & \nodata \nl
 &  & $-62.8$ & $11.72\pm0.07$ & $16.19\pm3.54$ & $<11.47$ & \nodata & $< 10.40$ & \nodata \nl
 &  & $-29.0$ & $13.44\pm0.04$ & $ 5.67\pm0.15$ & $12.75\pm0.02$ & $ 6.19\pm0.27$ & $11.03\pm0.10$ & $ 5.09\pm1.51$ \nl
 &  & $-13.3$ & $11.53\pm0.08$ & $ 1.07\pm0.26$ & $<11.50$ & \nodata & $<10.43$ & \nodata \nl
 &  & $  6.0$ & $13.29\pm0.01$ & $ 7.38\pm0.14$ & $12.21\pm0.06$ & $ 8.52\pm1.26$ & $11.30\pm0.10$ & $10.42\pm1.89$ \nl
 &  & $ 30.4$ & $12.60\pm0.01$ & $12.70\pm0.49$ & $<11.45$ & \nodata & $<10.41$ & \nodata \nl
 &  & $ 66.0$ & $12.82\pm0.01$ & $ 5.08\pm0.10$ & $12.07\pm0.05$ & $ 2.85\pm0.76$ & $10.81\pm0.29$ & $12.34\pm6.50$ \nl
\multicolumn{9}{c}{ } \nl
$1222+228$ & 0.66805 
    & $-28.2$ & $12.05\pm0.10$ & $ 4.79\pm1.22$ & $11.89\pm0.15$ & $1.94\pm3.20$ & $<10.60$ & \nodata \nl
 &  & $-15.4$ & $12.73\pm0.07$ & $ 5.51\pm1.12$ & $12.11\pm0.21$ & $ 5.32\pm3.74$ & $10.91\pm0.49$ & $8.80\pm13.8$ \nl
 &  & $ -4.9$ & $12.70\pm0.07$ & $ 3.83\pm1.02$ & $12.58\pm0.07$ & $ 4.35\pm1.20$ & $10.66\pm0.56$ & $1.98\pm5.81$ \nl
 &  & $  5.2$ & $12.62\pm0.04$ & $ 3.65\pm0.66$ & $12.65\pm0.14$ & $ 1.52\pm0.57$ & $10.93\pm0.18$ & $ 4.70\pm3.03$ \nl
 &  & $ 19.7$ & $12.68\pm0.02$ & $ 6.50\pm0.41$ & $12.27\pm0.08$ & $ 6.68\pm1.89$ & $10.94\pm0.17$ & $ 7.64\pm4.01$ \nl
 &  & $233.4$ & $12.40\pm0.02$ & $ 5.19\pm0.31$ & $<14.94$ & \nodata & $11.02\pm0.22$ & $ 8.06\pm5.2$ \nl
\multicolumn{9}{c}{ } \nl
$1241+176$ & 0.55048 & 
      $ -8.4$ & $13.07\pm0.02$ & $19.41\pm0.58$ & $<12.92$ & \nodata & $11.29\pm0.21$ & $17.79\pm8.42$ \nl
 &  & $  1.6$ & $13.52\pm0.15$ & $ 5.90\pm0.57$ & $13.57\pm0.05$ & $ 7.59\pm0.68$ & $11.74\pm0.06$ & $ 5.79\pm0.93$ \nl
 &  & $ 40.5$ & $12.31\pm0.03$ & $ 8.05\pm0.65$ & $12.19\pm0.18$ & $ 4.05\pm3.34$ & $11.42\pm0.24$ & $26.65\pm18.87$ \nl
 &  & $145.2$ & $11.96\pm0.05$ & $ 6.54\pm1.13$ & $<12.07$ & \nodata & $<10.85$ & \nodata \nl
\multicolumn{9}{c}{ } \nl
\tablebreak
$1248+401$ & 0.77296 
    & $-29.6$ & $12.66\pm0.02$ & $ 8.48\pm0.43$ & $12.27\pm0.07$ & $ 9.81\pm2.05$ & $<10.72$ & \nodata \nl
 &  & $-13.1$ & $12.80\pm0.16$ & $ 5.66\pm1.27$ & $12.62\pm0.15$ & $ 5.31\pm1.47$ & $10.91\pm0.48$ & $ 7.77\pm7.24$ \nl
 &  & $ -5.2$ & $13.33\pm0.04$ & $ 7.41\pm0.34$ & $13.54\pm0.03$ & $ 3.39\pm0.20$ & $11.45\pm0.17$ & $ 1.12\pm0.64$ \nl
 &  & $ 12.6$ & $12.31\pm0.02$ & $ 3.54\pm0.40$ & $12.04\pm0.11$ & $ 8.32\pm3.06$ & $<10.71$ & \nodata \nl
 &  & $ 23.8$ & $11.70\pm0.07$ & $ 1.35\pm0.88$ & $<10.44$ & \nodata & $<10.74$ & \nodata \nl
 &  & $ 39.7$ & $13.26\pm0.02$ & $ 6.51\pm0.23$ & $12.87\pm0.02$ & $ 4.22\pm0.33$ & $11.23\pm0.48$ & $ 5.80\pm2.92$ \nl
 &  & $ 52.2$ & $11.89\pm0.15$ & $ 5.34\pm1.31$ & $11.70\pm0.18$ & $ 3.85\pm3.43$ & $<10.69$ & \nodata \nl
 &  & $225.2$ & $12.30\pm0.01$ & $ 5.53\pm0.24$ & $11.98\pm0.10$ & $ 4.15\pm1.98$ & $11.16\pm0.07$ & $ 7.66\pm1.89$ \nl
\multicolumn{9}{c}{ } \nl
$1254+044$ & 0.51939 
    & $-262.4$ & $12.25\pm0.04$ & $ 3.31\pm0.52$ & $<12.17$       & \nodata        & $<10.95$ & \nodata \nl
 &  & $-159.4$ & $12.05\pm0.05$ & $ 6.28\pm1.12$ & $<12.15$       & \nodata        & $<10.94$ & \nodata \nl
 &  & $  -9.7$ & $15.75\pm0.10$ & $ 3.18\pm0.22$ & $13.81\pm0.15$ & $ 4.73\pm0.58$ & $12.18\pm0.04$ & $ 6.36\pm0.64$ \nl
 &  & $   4.1$ & $12.86\pm0.17$ & $ 4.77\pm2.83$ & $13.37\pm0.22$ & $ 9.47\pm6.24$ & $11.17\pm0.17$ & $ 2.13\pm3.30$ \nl
 &  & $  16.2$ & $13.00\pm0.09$ & $ 7.95\pm1.00$ & $12.60\pm0.64$ & $ 7.36\pm3.72$ & $11.52\pm0.12$ & $ 12.90\pm4.23$ \nl
\multicolumn{9}{c}{ } \nl
$1254+044$ & 0.93423 
    & $-37.6$ & $12.72\pm0.01$ & $ 9.68\pm0.27$ & $11.90\pm0.12$ & $11.53\pm4.10$ & $<10.67$ & \nodata \nl
 &  & $  4.3$ & $13.07\pm0.01$ & $ 7.57\pm0.16$ & $12.66\pm0.02$ & $ 4.41\pm0.34$ & $11.00\pm0.11$ & $ 4.77\pm2.01$ \nl
\multicolumn{9}{c}{ } \nl
$1317+274$ & 0.66005 
    & $-10.4$ & $12.63\pm0.04$ & $ 7.29\pm0.49$ & $12.60\pm0.04$ & $ 6.53\pm0.90$ & $<10.54$ & \nodata \nl
 &  & $  0.0$ & $12.69\pm0.04$ & $ 4.55\pm0.36$ & $12.81\pm0.04$ & $ 2.49\pm0.37$ & $11.27\pm0.07$ & $ 4.48\pm1.17$ \nl
 &  & $ 11.6$ & $12.02\pm0.05$ & $ 6.02\pm0.77$ & $11.99\pm0.06$ & $ 2.05\pm1.50$ & $<10.52$ & \nodata \nl
 &  & $ 47.5$ & $11.50\pm0.07$ & $ 3.68\pm1.21$ & $<11.63$ & \nodata & $<10.54$ & \nodata \nl
 &  & $ 77.6$ & $11.81\pm0.11$ & $ 6.98\pm1.89$ & $<11.63$ & \nodata & $<10.52$ & \nodata \nl
 &  & $ 89.7$ & $11.92\pm0.08$ & $ 4.65\pm1.04$ & $12.00\pm0.15$ & $ 5.04\pm2.52$ & $<10.53$ & \nodata \nl
 &  & $100.2$ & $11.77\pm0.08$ & $10.24\pm2.19$ & $11.88\pm0.11$ & $ 1.57\pm3.20$ & $<10.52$ & \nodata \nl
 &  & $140.4$ & $11.75\pm0.05$ & $ 9.01\pm1.48$ & $<11.64$ & \nodata & $<10.52$ & \nodata \nl
\multicolumn{9}{c}{ } \nl
$1421+331$ & 0.90287 
   & $-103.6$ & $11.96\pm0.14$ & $ 7.57\pm1.65$ & $11.61\pm0.18$ & $\sim 2.5$ & $<10.69$ & \nodata \nl
 &  & $-87.8$ & $12.77\pm0.08$ & $ 9.23\pm1.71$ & $11.93\pm0.17$ & $ 6.77\pm3.75$ & $< 9.74$ & \nodata \nl
 &  & $-77.0$ & $12.52\pm0.11$ & $ 3.56\pm0.84$ & $11.75\pm0.22$ & $ 4.90\pm3.37$ & $<10.69$ & \nodata \nl
 &  & $-59.7$ & $12.65\pm0.01$ & $10.34\pm0.45$ & $12.13\pm0.07$ & $ 5.48\pm1.36$ & $<11.58$ & \nodata \nl
 &  & $-17.6$ & $12.79\pm0.04$ & $ 3.37\pm0.32$ & $12.86\pm0.10$ & $ 0.76\pm0.07$ & $<10.70$ & \nodata \nl
 &  & $ -8.2$ & $12.93\pm0.08$ & $ 3.46\pm0.85$ & $12.48\pm0.05$ & $ 3.89\pm0.79$ & $<10.32$ & \nodata \nl
 &  & $  7.6$ & $13.29\pm0.06$ & $10.23\pm2.03$ & $12.77\pm0.05$ & $ 7.74\pm1.35$ & $10.74\pm0.33$ & $ 4.60\pm4.43$ \nl
 &  & $ 23.4$ & $14.44\pm0.85$ & $ 4.03\pm0.94$ & $13.60\pm0.02$ & $ 4.99\pm0.18$ & $11.58\pm0.06$ & $ 7.47\pm1.23$ \nl
 &  & $ 65.2$ & $11.62\pm0.06$ & $ 2.45\pm1.24$ & $<11.26$ & \nodata & $<10.72$ & \nodata \nl
 &  & $ 83.1$ & $12.99\pm0.03$ & $ 4.41\pm0.17$ & $12.27\pm0.07$ & $ 5.49\pm1.27$ & $<10.70$ & \nodata \nl
\multicolumn{9}{c}{ } \nl
$1421+331$ & 1.17261
    & $-43.4$ & $12.46\pm0.03$ & $ 4.08\pm0.42$ & $12.20\pm0.06$ & $ 1.91\pm1.08$ & $10.85\pm0.38$ & $ 6.83\pm5.38$ \nl
 &  & $-36.2$ & $11.94\pm0.10$ & $ 4.88\pm1.57$ & $11.58\pm0.16$ & $ 0.93\pm0.63$ & $<10.27$ & \nodata \nl
 &  & $-17.9$ & $12.57\pm0.03$ & $ 6.71\pm0.63$ & $12.32\pm0.04$ & $ 6.36\pm0.99$ & $<10.14$ & \nodata \nl
 &  & $  0.6$ & $13.11\pm0.02$ & $ 7.23\pm0.36$ & $12.75\pm0.02$ & $ 6.32\pm0.45$ & $11.53\pm0.07$ & $10.59\pm2.62$ \nl
 &  & $ 20.2$ & $12.06\pm0.06$ & $ 4.09\pm1.03$ & $\sim12.3$ & $\sim 0.2$ & $<10.08$ & \nodata \nl
 &  & $ 28.9$ & $11.96\pm0.22$ & $ 2.26\pm1.76$ & $<10.95$ & \nodata & $10.72\pm0.60$ & $\sim 4.5$ \nl
 &  & $ 35.8$ & $12.75\pm0.07$ & $ 2.95\pm0.49$ & $12.32\pm0.09$ & $ 4.86\pm1.08$ & $10.81\pm0.32$ & $\sim 3.1$ \nl
\multicolumn{9}{c}{ } \nl
$1634+706$ & 0.99024 
    & $-21.9$ & $13.11\pm0.01$ & $ 7.39\pm0.15$ & $12.77\pm0.02$ & $ 5.19\pm0.27$ & $11.35\pm0.04$ & $ 6.85\pm0.74$ \nl
 &  & $-10.0$ & $12.17\pm0.11$ & $ 2.15\pm0.81$ & $12.15\pm0.09$ & $ 4.33\pm1.57$ & $< 9.97$ & \nodata \nl
 &  & $  1.7$ & $12.82\pm0.16$ & $ 6.46\pm0.86$ & $12.38\pm0.13$ & $ 4.20\pm1.12$ & $10.85\pm0.48$ & $ 7.65\pm5.33$ \nl
 &  & $ 12.9$ & $13.10\pm0.09$ & $16.68\pm2.40$ & $12.33\pm0.22$ & $12.47\pm5.54$ & $11.30\pm0.30$ & $17.59\pm5.69$ \nl
 &  & $ 24.7$ & $12.09\pm0.20$ & $ 5.03\pm1.34$ & $<11.49$ & \nodata & $<10.39$ & \nodata \nl
\multicolumn{9}{c}{ } \nl
$2128-123$ & 0.42973 
    & $-39.8$ & $12.38\pm0.04$ & $ 7.66\pm0.91$ & \nodata & \nodata & $<10.47$ & \nodata \nl
 &  & $-25.7$ & $12.01\pm0.08$ & $ 3.48\pm1.23$ & \nodata & \nodata & $11.37\pm0.14$ & $ 9.03\pm4.18$ \nl
 &  & $ -0.6$ & $15.24\pm0.51$ & $ 4.45\pm0.46$ & \nodata & \nodata & $12.12\pm0.03$ & $ 4.98\pm0.48$ \nl
 &  & $ 12.2$ & $13.49\pm0.07$ & $ 5.58\pm0.47$ & \nodata & \nodata & $11.70\pm0.08$ & $ 6.39\pm1.52$ \nl
\multicolumn{9}{c}{ } \nl
$2145+064$ & 0.79078 
    & $-109.5$ & $12.55\pm0.00$ & $ 2.04\pm0.01$ & $\sim10.1$ & $\sim24.8$ & $\sim 0.0$ & \nodata \nl
 &  & $-101.7$ & $13.38\pm0.06$ & $ 2.42\pm0.08$ & $12.82\pm0.02$ & $ 3.86\pm0.36$ & $<10.69$ & \nodata \nl
 &  & $ -3.4$ & $12.56\pm0.15$ & $ 1.04\pm0.17$ & $<11.82$ & \nodata & $<10.78$ & \nodata \nl
 &  & $  3.6$ & $12.57\pm0.02$ & $14.12\pm0.58$ & $<11.79$ & \nodata & $<10.71$ & \nodata \nl
 &  & $ 63.3$ & $12.76\pm0.01$ & $ 9.91\pm0.32$ & $12.39\pm0.09$ & $ 5.28\pm1.65$ & $<10.69$ & \nodata \nl
 &  & $ 86.6$ & $12.18\pm0.04$ & $ 8.86\pm1.00$ & $11.86\pm0.23$ & $\sim 1.6$ & $<10.72$ & \nodata \nl
\enddata
\label{tab:vptab}
\end{deluxetable}
\endgroup


\endgroup


\begin{figure*}[th]
\figurenum{2}
\plotone{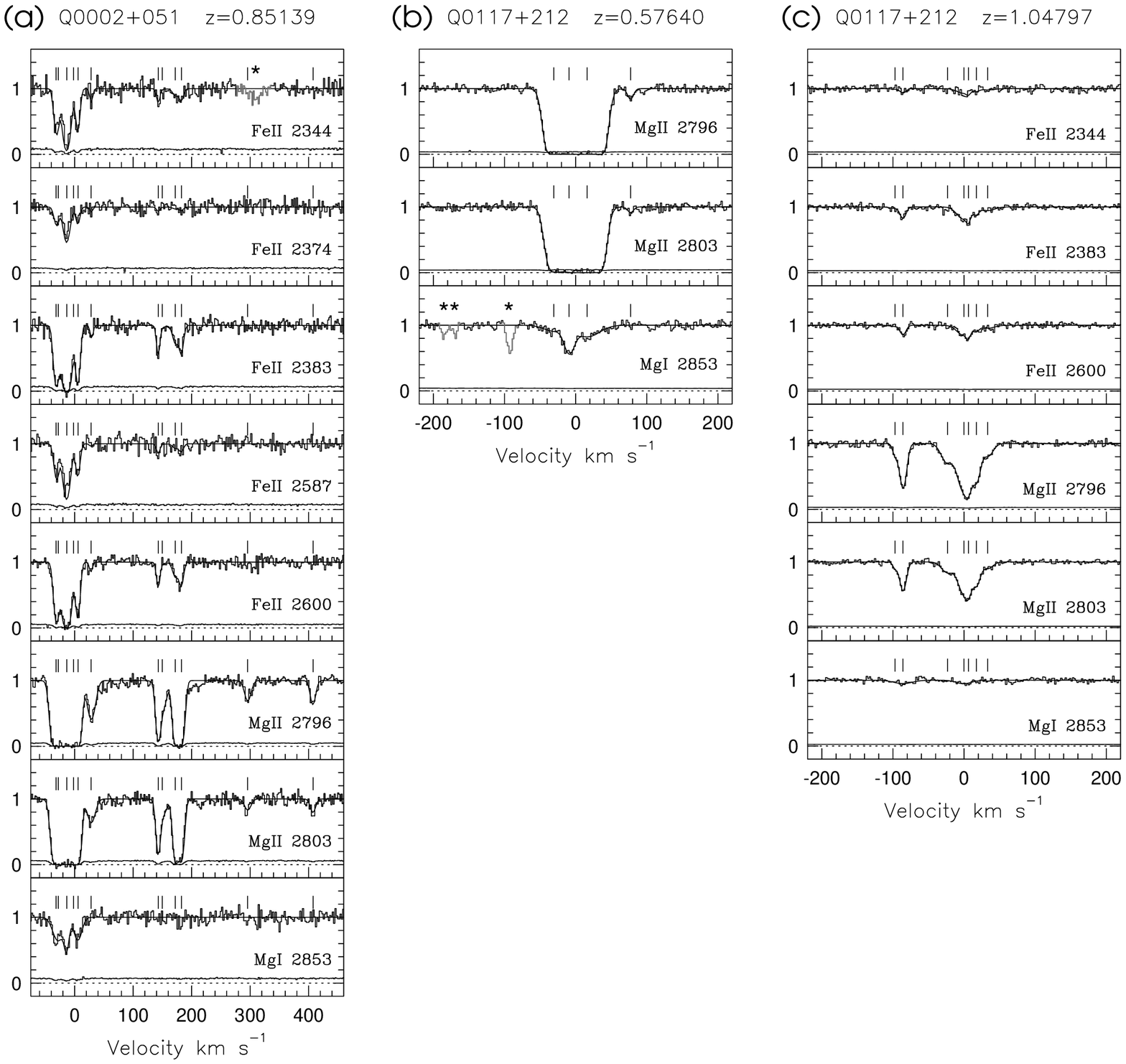}
\caption[fig2abc.eps]   {The   HIRES/Keck   profiles  of   detected
transitions for systems in the $0002+051$ and $0117+212$ spectra.  The
data are aligned in  rest--frame velocity.  Ticks above the normalized
spectra are the centroids of  the VP components.  The VP model spectra
(the fits) are shown as solid curves superimposed upon the data.
\label{fig:data}}
\end{figure*}

\begin{figure*}[th]
\figurenum{2. cont}
\plotone{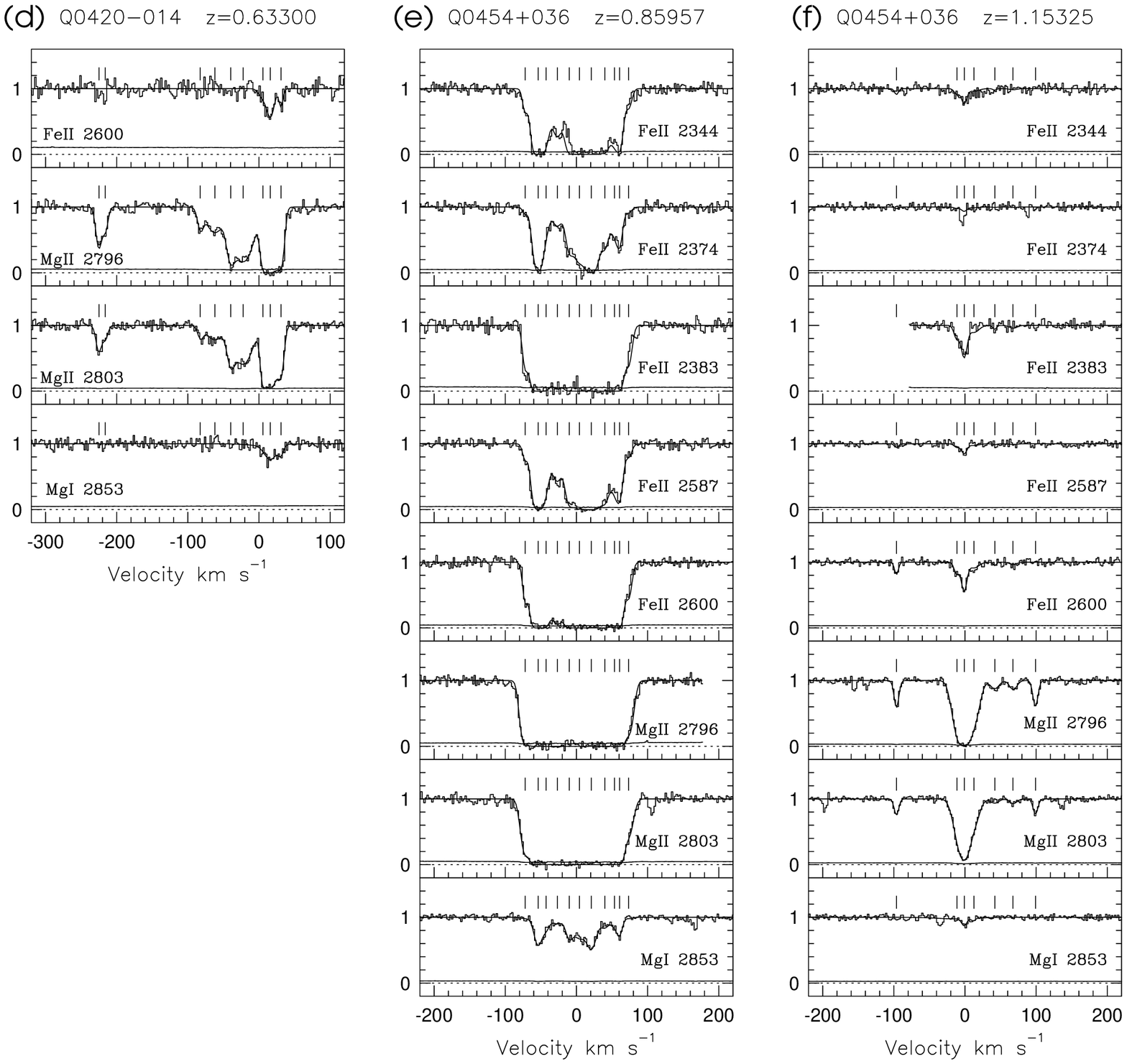}
\caption[fig2def.eps] { Same as for Figures~\ref{fig:data}$a$--$c$,
but for $0420-014$ and $0454+036$.  }
\end{figure*}

\begin{figure*}[th]
\figurenum{2. cont}
\plotone{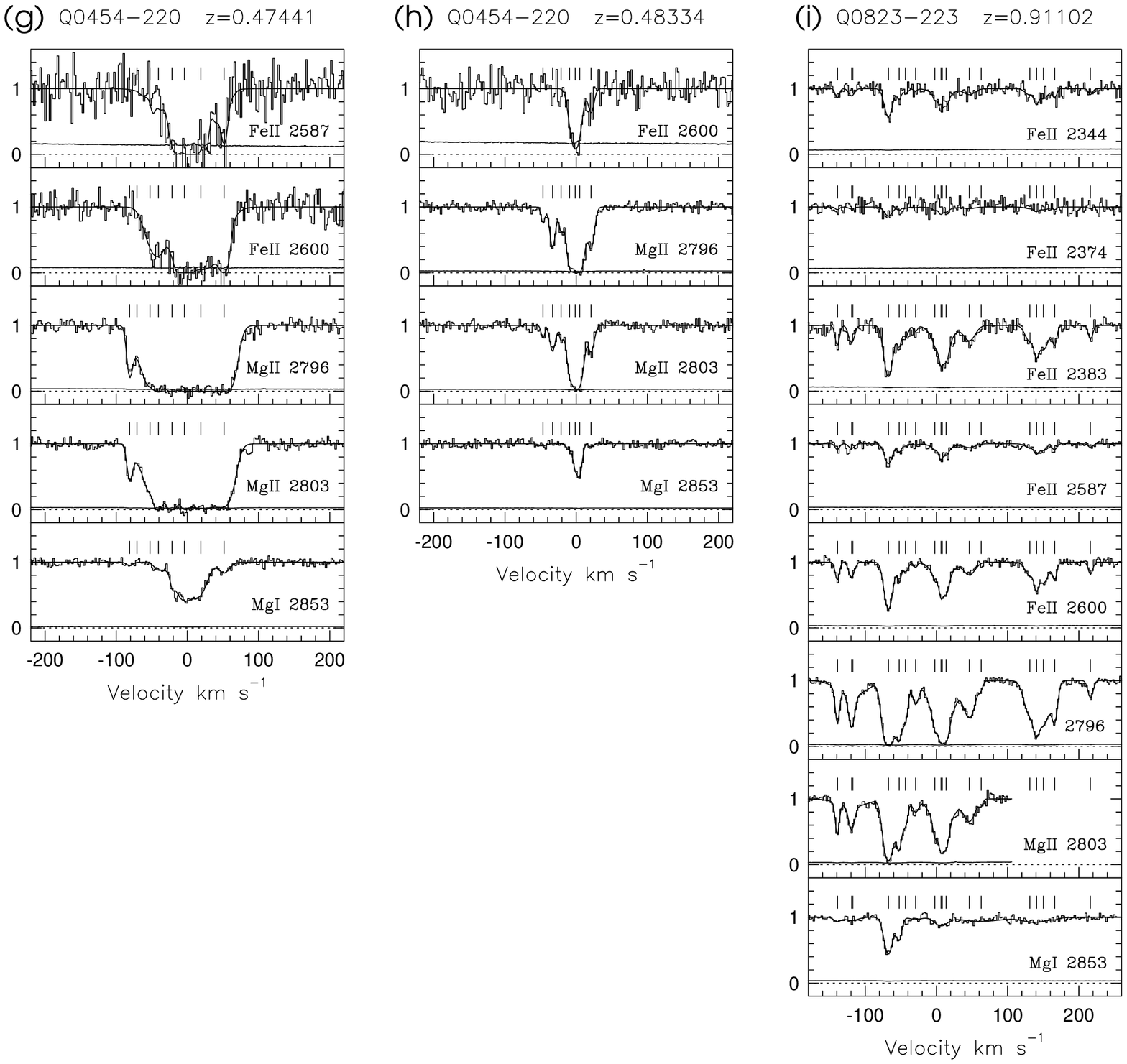}
\caption[fig2ghi.eps] { Same as for Figures~\ref{fig:data}$a$--$c$,
but for $0454-220$ and $0823-223$.  }
\end{figure*}

\begin{figure*}[th]
\figurenum{2.  cont} 
\plotone{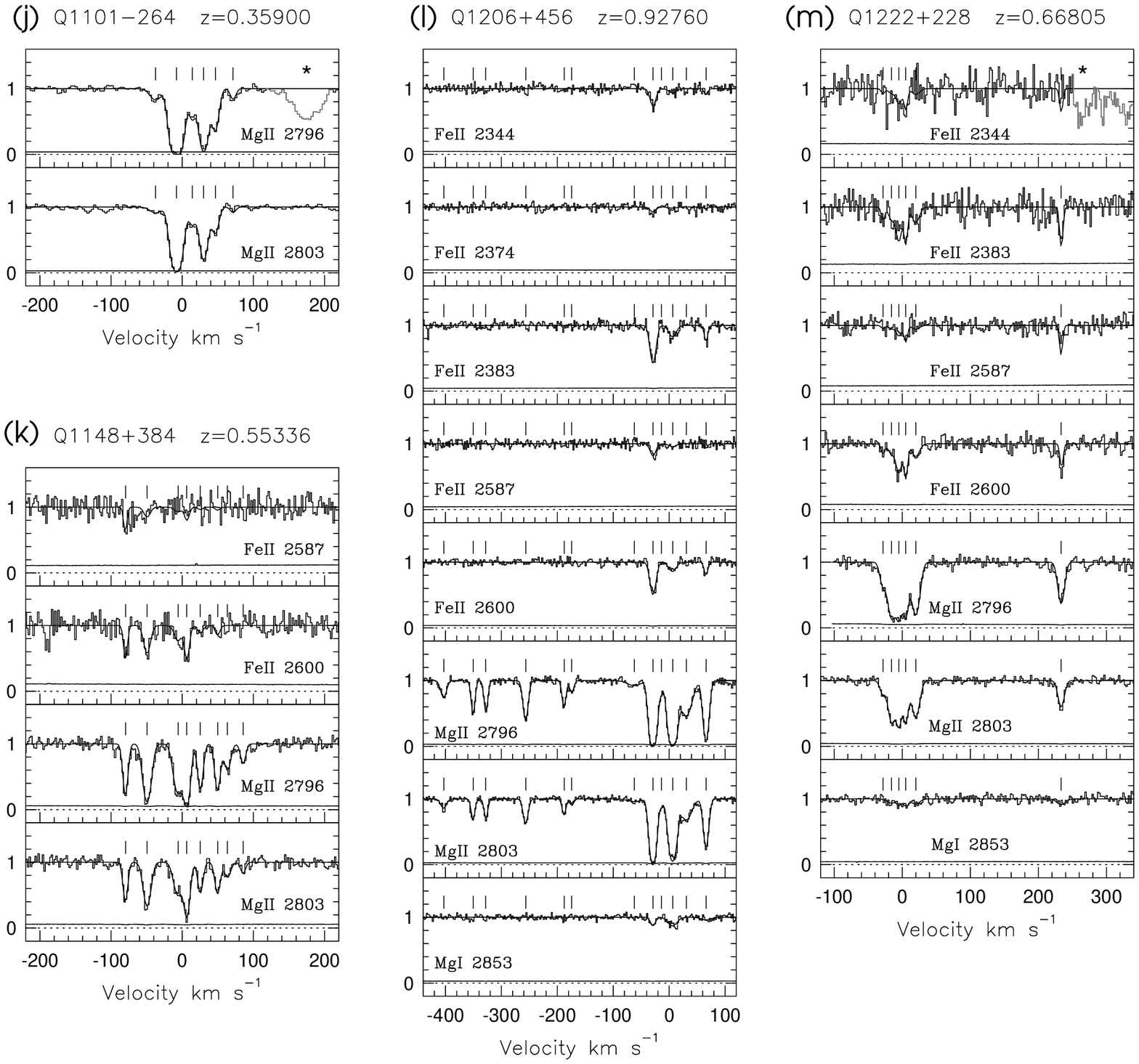}
\caption[fig2jklm.eps] {Same as for Figures~\ref{fig:data}$a$--$c$,
but for $1101-264$, $1148+384$, $1206+459$, and $1222+228$.  }
\end{figure*}

\begin{figure*}[th]
\figurenum{2. cont}
\plotone{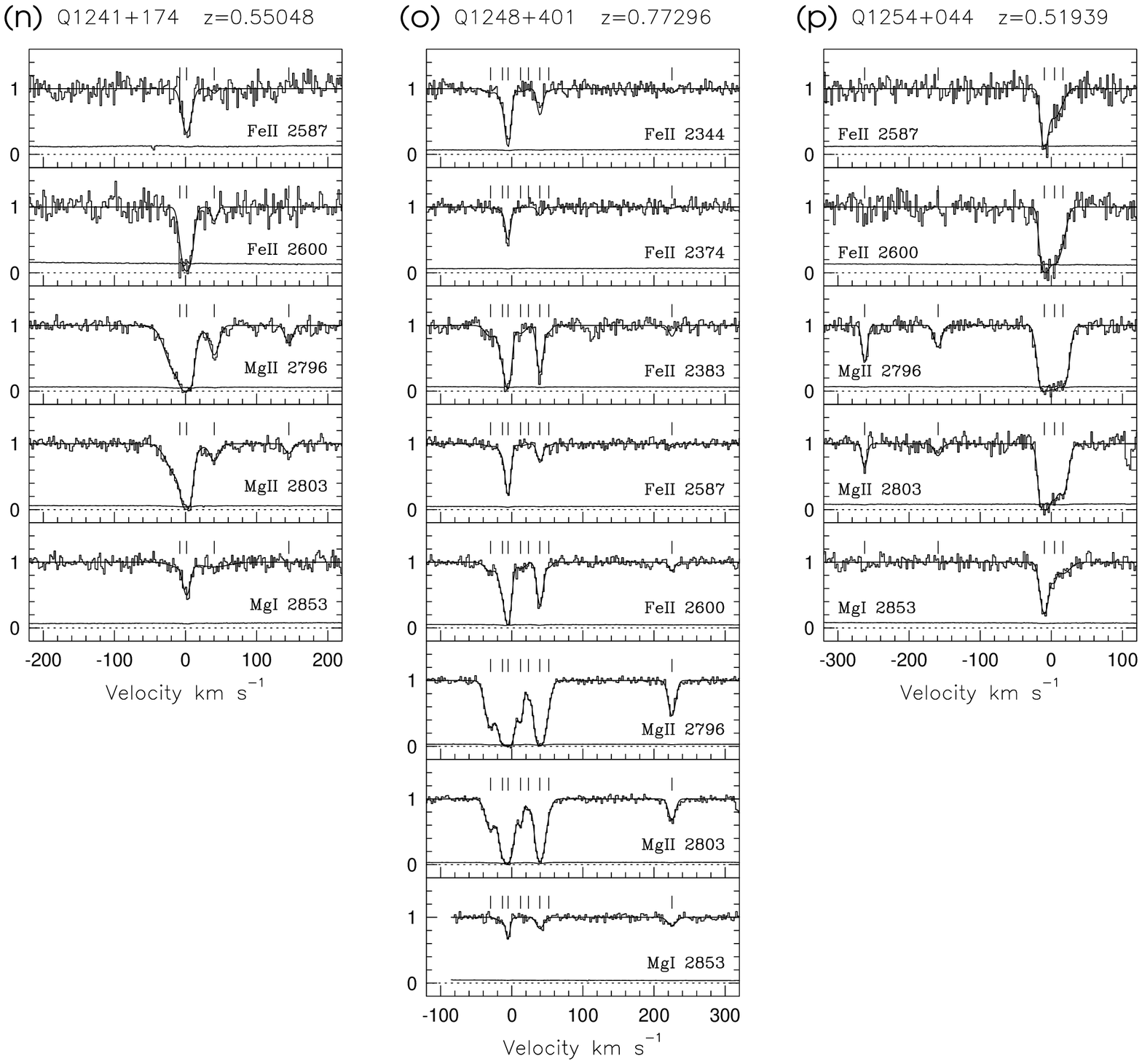}
\caption[fig2nop.eps] { Same as for Figures~\ref{fig:data}$a$--$c$,
but for $1241+174$, $1248+401$, and $1254+044$.  }
\end{figure*}

\begin{figure*}[th]
\figurenum{2. cont}
\plotone{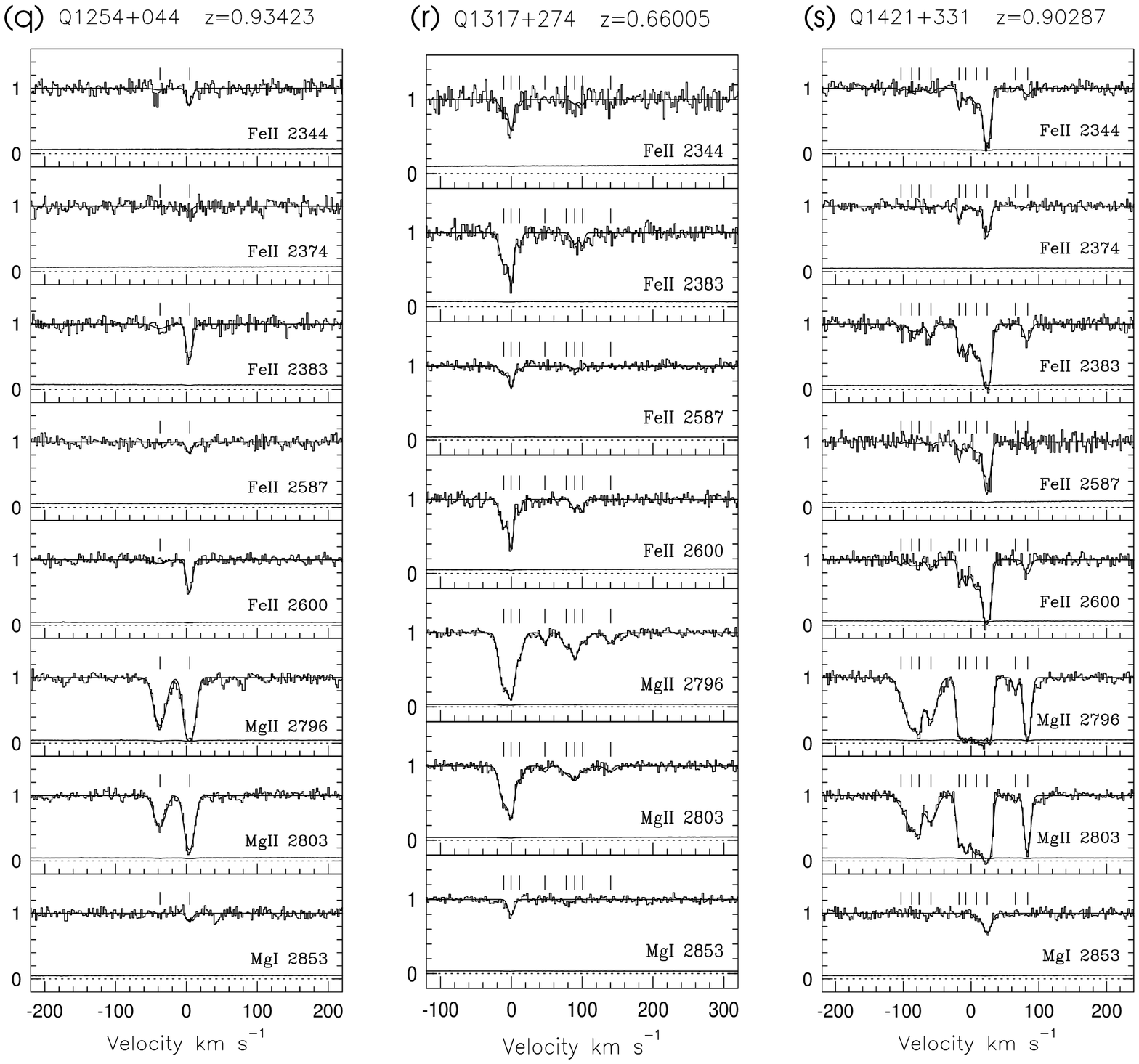}
\caption[fig2qrs.eps] { Same as for Figures~\ref{fig:data}$a$--$c$,
but for $1254+044$, $1317+274$, and $1421+331$.  }
\end{figure*}

\begin{figure*}[th]
\figurenum{2. cont}
\plotone{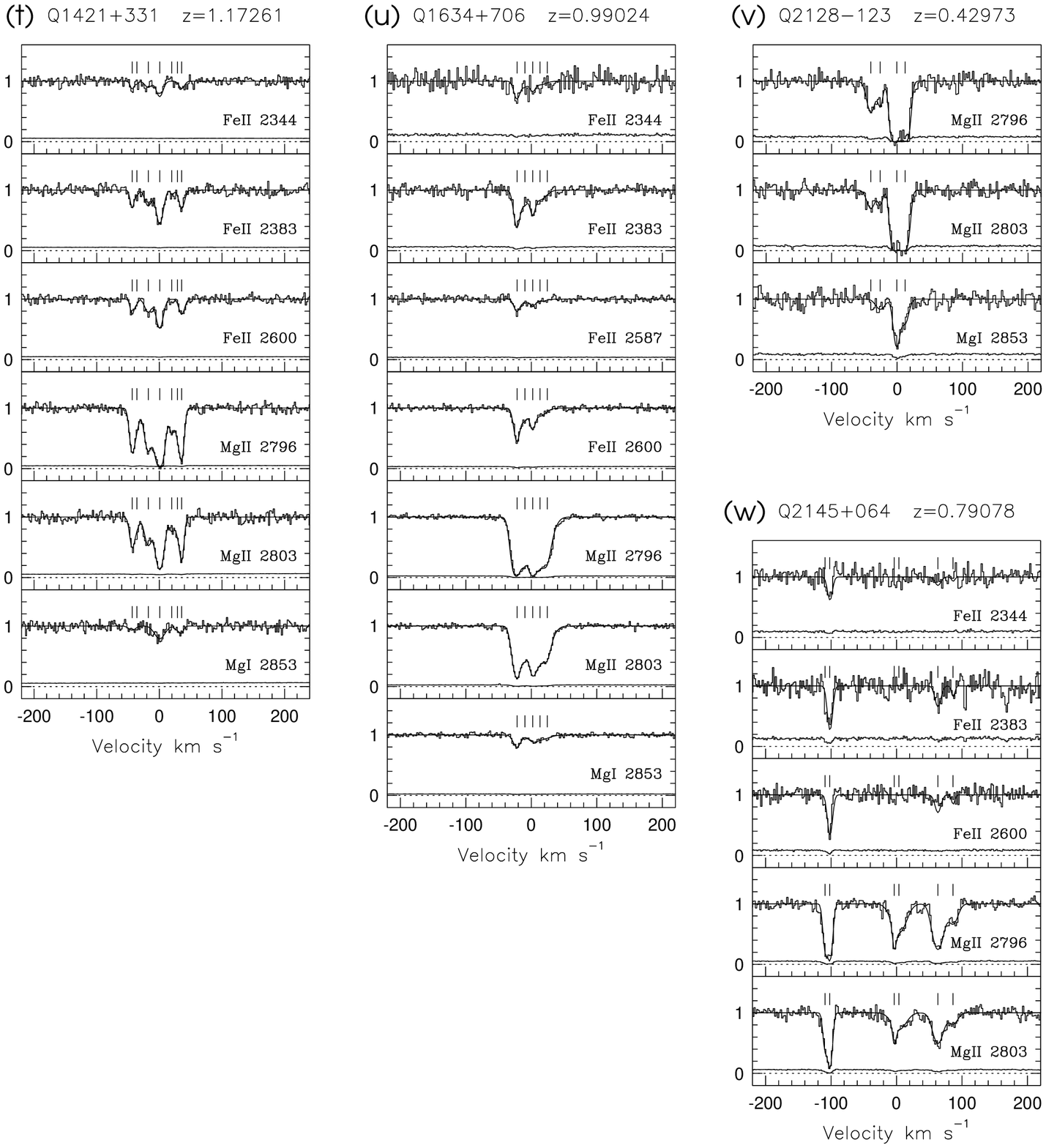}
\caption[fig2tuvw.eps] {Same as for Figures~\ref{fig:data}$a$--$c$,
but for $1421+331$, $1634+706$, $2128-123$, and $2145+064$.  }
\end{figure*}


\end{document}